\documentclass{aa}
\usepackage{natbib}
\bibpunct{(}{)}{;}{a}{}{,} 
\usepackage{amsmath}
\usepackage{wasysym}
\usepackage{marvosym}
\usepackage{txfonts}

\usepackage{hyperref} 
\hypersetup{pdfauthor=Esther Linder}
\hypersetup{backref=true, pagebackref=true, hyperindex=true, breaklinks=true,colorlinks=true,urlcolor=blue, linkcolor=blue,  citecolor=blue,pagecolor=red, bookmarks=true, bookmarksopen=true}

\usepackage{siunitx}

\usepackage{lscape}

\def\mearth{M_\oplus}
\def\rearth{R_\oplus}
\def\msun{M_\odot}

\def\mcore{M_{\rm core}}
\def\menve{M_{\rm enve}}

\def\f1{f_{\rm I}}
\def\fice{f_{\rm ice}}

\def\beq{\begin{equation}}
\def\eeq{\end{equation}}

\def\t2{\tau_{\rm II}}

\def\mtot{M_{\rm tot}}
\def\sigmas0{\Sigma_{\rm s,0}}
\def\mj{M_{\textrm{\tiny \jupiter }}}

\newcommand{\muran}{M_{\textrm{\tiny \uranus}}}

\newcommand{\luran}{L_{\textrm{\tiny \uranus}}}
\newcommand{\ruran}{R_{\textrm{\tiny \uranus}}}
\newcommand{\mnept}{M_{\textrm{\tiny \neptune}}}

\newcommand{\lnept}{L_{\textrm{\tiny \neptune}}}
\newcommand{\rnept}{R_{\textrm{\tiny \neptune}}}

\newcommand{\lj}{L_{\textrm{\tiny \jupiter}}}
\newcommand{\rj}{R_{\textrm{\tiny \jupiter}}}

\def\s0{S_0}

\newcommand{\lsun}{L_{\odot}}
\newcommand{\msat}{M_{\textrm{\tiny \saturn}}}

\newcommand{\kB}{\text{k}_{\rm B}}

\defcitealias{marleyfortney2007}{M07}
\defcitealias{spiegelburrows2012}{SB12}

\def\apjl{ApJ}                
\def\apjs{ApJS}               
\def\ao{Appl.Optics}          
\def\aap{A\&A}                

\def\({\left(}
\def\){\right)}
\def\<{\left<}
\def\>{\right>}

\usepackage{etoolbox}

\makeatletter

\patchcmd{\NAT@citex}
  {\@citea\NAT@hyper@{%
     \NAT@nmfmt{\NAT@nm}%
     \hyper@natlinkbreak{\NAT@aysep\NAT@spacechar}{\@citeb\@extra@b@citeb}%
     \NAT@date}}
  {\@citea\NAT@nmfmt{\NAT@nm}%
   \NAT@aysep\NAT@spacechar\NAT@hyper@{\NAT@date}}{}{}

\patchcmd{\NAT@citex}
  {\@citea\NAT@hyper@{%
     \NAT@nmfmt{\NAT@nm}%
     \hyper@natlinkbreak{\NAT@spacechar\NAT@@open\if*#1*\else#1\NAT@spacechar\fi}%
       {\@citeb\@extra@b@citeb}%
     \NAT@date}}
  {\@citea\NAT@nmfmt{\NAT@nm}%
   \NAT@spacechar\NAT@@open\if*#1*\else#1\NAT@spacechar\fi\NAT@hyper@{\NAT@date}}
  {}{}

\makeatother

\begin{document}

\title{Evolutionary models of cold and low-mass planets: \\ Cooling curves, magnitudes, and detectability} 
\subtitle{}
\author{Esther F. Linder\inst{1}, Christoph Mordasini\inst{1}, Paul Molli\`{e}re\inst{2}, Gabriel-Dominique Marleau\inst{3,1}, Matej Malik\inst{4}, \\
Sascha P. Quanz\inst{5}, Michael R. Meyer\inst{6}}
\institute{Physikalisches Institut, University of Bern, Gesellschaftsstrasse 6, 3012 Bern, Switzerland \and Sterrewacht Leiden, Huygens Laboratory, Niels Bohrweg 2, 2333 CA Leiden, The Netherlands \and Institut f\"ur Astronomie und Astrophysik, Eberhard Karls Universit\"at T\"ubingen, Auf der Morgenstelle 10, 72076 T\"ubingen, Germany \and Center for Space and Habitability, University of Bern, Gesellschaftsstrasse 6, CH-3012, Bern, Switzerland \and Institute for Particle Physics and Astrophysics, ETH Zurich, Wolfgang-Pauli-Strasse 27, CH-8093 Zurich, Switzerland  \and Department of Astronomy, University of Michigan, 1085 S. University, Ann Arbor MI 48109, United States } 
\offprints{Esther Linder, \email{esther.linder@space.unibe.ch}}
\date{Received 16.07.2018 / Accepted 03.12.18}

\abstract
{Future instruments like the Near Infrared Camera (NIRCam) and the Mid Infrared Instrument (MIRI) on the James Webb Space Telescope ({JWST}) or  the Mid-Infrared E-ELT Imager and Spectrograph ({METIS}) at the  European Extremely Large Telescope {(E-ELT)} will be able to image exoplanets that are too faint (because they have a low mass, and hence a small size or low effective temperature) for current direct imaging instruments. On the theoretical side, core accretion formation models predict a significant population of  low-mass and/or cool planets at orbital distances of $\sim10$--100  { {au}}.}{ Evolutionary models predicting the planetary intrinsic luminosity as a function of time have traditionally concentrated on gas-dominated giant planets. We extend these cooling curves to Saturnian and Neptunian planets.}{We simulated  the cooling of isolated core-dominated and gas giant planets with masses of 5 $\mearth$ to 2 $\mj$. The planets consist of a core made of iron, silicates, and ices surrounded by a H/He envelope, similar to the ice giants in the solar system. The luminosity includes the contribution from the cooling and contraction of the core and of the H/He envelope, as well as radiogenic decay. For the atmosphere we used grey, \texttt{AMES-Cond}, \texttt{petitCODE}, and \texttt{HELIOS} models. We considered solar and non-solar metallicities as well as cloud-free and cloudy atmospheres. The most important initial conditions, namely the core-to-envelope ratio and the initial (i.e. post formation) luminosity are taken from planet formation simulations based on the core accretion paradigm.}{We first compare our cooling curves for Uranus, Neptune, Jupiter, Saturn, GJ 436b, and a 5 $\mearth$ planet with a 1\% H/He envelope with other evolutionary models. We then present the temporal evolution of planets with masses between 5 $\mearth$ and 2 $\mj$ in terms of their luminosity, effective temperature, radius, and entropy. We discuss the impact of different post formation entropies. For  the different atmosphere types and initial conditions, magnitudes in various filter bands   between 0.9 and 30 micrometer wavelength are provided.}
{Using blackbody fluxes and non-grey spectra, we estimate the detectability of such planets with { {JWST}}. We found that a 20 (100) $\mearth$ planet can be detected with JWST in the background limit up to an age of about 10 (100) Myr with NIRCam and MIRI, respectively.}


\titlerunning{Evolutionary models of low-mass planets}
\authorrunning{E. F. Linder et al.}

\maketitle

\section{Introduction}
During the last few years, the Kepler satellite has detected numerous exoplanets of which many are in the sub-Neptunian or super-Earth mass range (e.g. \citealt{batalharowe2012,burkebryson2014,fressintorres2013a,petiguramarcy2013}). Different from the solar system, especially in close-in orbits, sub-Neptunian planets seem to be very abundant in the solar neighbourhood \citep{howardmarcy2012}. Various studies on sub-Neptunians and super-Earths  have been conducted. For example, \citet{bodenheimerlissauer2014} studied the origin and evolution of  low-density planets in the mass range from 1-10~$\mearth$, which are within 0.5~ { {au}} from their star. In particular, they wanted to find out if these planets formed in situ or further out and then moved inwards. Another analysis conducted by \citet{chenrogers2016} was dedicated to computing mass-radius-composition-age relations for low-mass planets and how these depend on the evolution history of the planets. Finally, \citet{jinmordasini2017} studied whether through photoevaporation a certain planetary composition is revealed. 

In the literature, evolutionary calculations for gas giants are abundant (e.g. \citealt{burrowsmarley1997,baraffechabrier2003,podolakweizman1995,nettelmannhelled2013, fortneynettelmann2010,fortneyikoma2011}).  However,  studies of the thermodynamic evolution of low-mass planets have been scarce so far (e.g. \citealt{howeburrows2015,nettelmannhelled2013,fortneyikoma2011,beichmankrist2010}). In this paper we want to extend calculations of  the thermodynamic evolution and cooling curves to lower mass planets in a small parameter study. An important initial condition for planetary cooling calculations  is the post formation entropy (e.g. \citealt{marleyfortney2007,marleauklahr2017,mordasinimarleau2017}). For low-mass planets, the core-envelope mass ratios are also important. We study non-irradiated planets in a mass range of 5-636~$\mearth$ (2~$\mj$) and provide magnitudes corresponding to filter bands from various instruments and systems. While non-irradiated planets are simpler in the sense that the three-dimensional redistribution of insolation energy through the planetary atmosphere does not have to be modelled, they still present a unique set of challenges when trying to model their atmospheric structures and spectra. For one, we expect the atmospheres of the  objects studied here to be  heavily enriched in metals, because the  bulk enrichment of a planet appears to be a function of its mass \citep{millerfortney2011,thorngrenfortney2016}, and so may the atmospheric enrichment  \citep[see Fig. 3 in][]{mordasinivanboekel2016}. The degree to which the the bulk enrichment of a planet is visible in its atmosphere  is not straightforward to assess and hence remains an open challenge \citep[see the discussion in Sect. 2.4.4 in][]{mordasinivanboekel2016}. In addition, the question of when and how clouds form in self-luminous planets is far from being understood. The challenge lies in understanding and quantifying the multitude of microphysical processes that lead to cloud formation and evolution \citep{rossow1978}. Because of this, most cloud models currently in use are heavily simplified or parametrized \citep{tsujiohnaka1996,ackermanmarley2001,allardhauschild2001,allardguillot2003,zsomkaltenegger2012,mollierevanboekel2017}, and remove certain cloud species as a function of temperature in an ad hoc fashion in order to mimic the settling of these species below the planet's photosphere \citep{morleyfortney2012,mollierevanboekel2017}. Even with the use of a sophisticated micro-physical model, the cloud properties depend strongly on the unknown vertical mixing and the detailed atmospheric chemistry and hence remain a priori under-determined without further observational constraints \citep{gaomarley2018, ohnookuzumi2018}. Moreover, the recovery of the optical properties of such a vast variety of potential condensates is still an ongoing endeavour \citep{kitzmannheng2018}. Finally, since some self-luminous sub-stellar objects cannot be reasonably fitted with current cloud models, an altogether different process has been suggested to affect the atmospheres and spectra of planets \citep{tremblinchabrier2017}.

In terms of imaging observations of exoplanets, the soon-to-be-launched  James Webb Space Telescope ({JWST)} as well as the next generation of ground-based optical and near-infrared telescopes with 30-40 m primary mirrors will probe currently uncharted parameter space in terms of exoplanet mass and luminosity. The Near Infrared Camera (NIRCam) and the Mid Infrared Instrument (MIRI) on the JWST will provide unprecedented sensitivity to cool and/or low-mass objects at near- and mid-infrared wavelengths, respectively, and instruments like the { {Mid-infrared E-ELT Imager and Spectrograph (METIS)}} \citep{brandlagocs2016}, to be installed at European Southern Observatory's (ESO's) 39-m { {European Extremely Large Telescope (E-ELT)}}, with its unparalleled spatial resolution and superior sensitivity compared to current ground-based instruments, will be able to search for low-luminosity objects in the solar vicinity \citep[e.g.][]{crossfield2013,quanzcrossfield2015}. Therefore, evolutionary models extending to smaller masses (ice giants, super-Earths) are needed to inform these future observations and interpret possible detections.

The structure of this paper is as follows. In Sect. \ref{sect:model}, the model and the improvements made in the planetary evolution code for calculating the evolution of (low-mass) planets are described. Following in Sect. \ref{sect:atmgrids} is the description of the atmospheres used in this work. Section \ref{sect:benchmarking} shows various example and benchmark calculations, where  the results obtained in this work are compared to measurements and earlier theoretical evolution calculations. The initial conditions for the final calculations are presented in Sect. \ref{sect:intialcond}.  After this, the results and discussion are presented in Sect. \ref{sect:results}. Section \ref{sect:conclusions}  summarizes the findings and major conclusions.

\section{Evolutionary and internal model}\label{sect:model}

The evolutionary calculations presented here were obtained with the evolutionary model described in \citet{jinmordasini2014}, which is itself based on the  model of  planetary evolution of \citet{mordasinialibert2012b,mordasinialibert2012c}.  This model describes the planets as consisting of three distinct homogeneous layers, namely a H/He envelope (using the equation of state (EoS) of \citealt{saumonchabrier1995}), an ice layer (for planets which have accreted outside of the iceline), and a rocky core, which itself consists of silicates and iron. To address the cooling  and contraction of very low-mass planets, we have extended the model in regard to two aspects.

In our previous simulations, the source of luminosity of the planets were the cooling and contraction of the H/He envelope,  the radiogenic luminosity due to long-lived radionuclides, and the luminosity generated from the contraction of the solid core when the external pressure exerted by the envelope changes. The contributions resulting from the non-zero temperature of the core were, in contrast, not considered. Therefore, the contraction of the core due to a change of its mean density, because of a decrease in its mean temperature,  as well as the decrease of the core's   internal energy, were neglected. While the contribution of the core to the total luminosity is negligible for H/He dominated giant planets (e.g. \citealt{baraffechabrier2008}), neglecting it for core-dominated low-mass planets leads to inaccurate cooling sequences, as demonstrated by \citet{baraffechabrier2008} and \citet{lopezfortney2014}. We have therefore added a first order temperature correction of the mean core density to take into account the temperature dependency of the core radius, which is described in more detail in Appendix \ref{app:coretemp}. We also take into account additional terms in the energy equation that is used to calculate the temporal evolution and thus the luminosity of the whole planet. This addition is described below.


As described in \citet{mordasinialibert2012b}, the calculation of evolutionary sequences in our model is based on the fundamental relation between the change of the total energy of the planet, and its luminosity because of  energy conservation, $d E_{\rm tot}/dt=-L$ (for other energy based approaches, see \citealt{lecontechabrier2013,pisoyoudin2014}). In previous versions of our planet evolution model (except for \citealt{lindermordasini2016}), the thermal energy of the core was, however, neglected. As the second modification of the code, we include it here, considering both the isothermal and adiabatic case.

The total energy $E_{\rm tot}$ of the planet consists of four terms,
\beq
E_{\rm tot}=E_{\rm grav,e}+E_{\rm int,e}+E_{\rm grav,c}+E_{\rm int,c}
,\eeq
which are the gravitational potential energy of the gaseous envelope $E_{\rm grav,e}$,  its thermal (internal) energy $E_{\rm int,e}$, 
the potential energy of the core $E_{\rm grav,c}$, and finally its thermal  energy  $E_{\rm int,c}$. Additional sources of luminosity we include are the radiogenic decay (calculated as in \citealt{mordasinialibert2012c}, assuming a chondritic abundance of radionucleides), which is important for  low-mass planets, as well as deuterium burning (\citealt{mollieremordasini2012,bodenheimerdangelo2013}), which is in contrast negligible for the planets we study here.

The potential energy of the gaseous envelope is found as
\beq
E_{\rm grav,e}=-\int_{\mcore}^M \frac{G m}{r}\ d m
,\eeq
where $\mcore$ is the core mass, $M$ the total planet mass, $r$ the distance from the planet's centre, and $G$ is the gravitational constant. The internal energy is 
\beq
E_{\rm int,e}=\int_{\mcore}^M u\ d m
,\eeq
where $u$ is the specific internal energy of the H/He gas, which is directly given by the Saumon, Chabrier, and van Horn (SCvH) EoS \citep{saumonchabrier1995}.

For the  core's gravitational energy, we assume for simplicity a (mean) density that is constant within the core, as density contrasts in the cores are smaller than in the gaseous envelope, even if this is strictly speaking not self-consistent with the internal structure model of the core. We then have for the potential energy of the core
\beq
E_{\rm grav,c}=-\frac{3 G \mcore^{2}}{5 R_{\rm core}}
,\eeq
where $R_{\rm c}$ is the core's radius that is found as described in the previous section.  {To relate the pressure to the density in the core, we use the modified polytropic EoS of \cite{seagerkuchner2007} for iron, rock (perovskite: MgSiO$_3$), and ice.}

Finally, for the core's internal energy we consider two cases, reflecting the uncertainty in the heat transport mechanism in the core \citep{baraffechabrier2008}. First, as in \citet{lopezfortney2014}, we consider  the isothermal case, where the internal energy is given as
\beq
E_{\rm int,c,iso}=c_{\rm v} \mcore T_{\rm ceb}
,\eeq
where $c_{v}$ is the specific heat capacity that is set to  $10^{7}$ erg g$^{-1}$ K$^{-1}$ for rocky material \citep{guillotchabrier1995}. As noted by \citet{baraffechabrier2008}, this value is compatible with the predictions by the Analytic Equations of State (ANEOS) in the relevant pressure and temperature range. For (water) ice we assume  $c_{\rm v}=6\times10^{7}$ erg g$^{-1}$ K$^{-1}$. For cores consisting of both rocky material and ice, we use the mass weighted average.

We also consider an adiabatic thermal structure of the core to estimate the core's thermal energy content. Under the (rough) approximation that in the core the density $\rho$, heat capacity at constant pressure $c_{\rm p}$, and thermal expansion coefficient $\alpha$ are constant with radius, one can find the temperature as a function of radius $r$ from the adiabatic temperature gradient
\beq
\frac{d T}{dP}=\frac{T \alpha}{\rho  c_{\rm p}}
\eeq
and the pressure as a function of radius 
\beq
P(r)=P_{\rm ceb}+\frac{3 G M_{\rm core}^{2}}{8 \pi R_{\rm core}^{6}}(R_{\rm core}^{2}-r^{2})
,\eeq
where $P_{\rm ceb}$ is the pressure at the core-envelope boundary, that is, at $R_{\rm core}$. Integrating the first equation and replacing the pressure using the second one yields the temperature structure as
\beq
T(r)=T_{\rm ceb} \exp\left[\frac{\alpha G M_{\rm core}}{2 c_{\rm p} R_{\rm core}^{3}}(R_{\rm core}^{2}-r^{2})\right].
\eeq
This temperature structure can be integrated to find $E_{\rm int,c,adia}$ via
\beq
E_{\rm int,c,adia}=\int_{0}^{R_{\rm core}} 4 \pi r^{2} \rho c_{\rm v} T(r) dr.
\eeq
This integral is evaluated to 
\beq
\begin{aligned}
E_{\rm int,c,adia}= 
4 \pi  c_{\rm v}  T_{\rm ceb} \rho \left[\sqrt{\frac{\pi c_{\rm p}^{3} R_{\rm core}^{9}}{2 G^{3} M_{\rm core}^{3} \alpha^{3}}}\exp\left(\frac{G M_{\rm core} \alpha}{2 c_{\rm p} R_{\rm core}}\right) \times  \right. \\ 
\left. \mathrm{erf}\left(\sqrt{\frac{G M_{\rm core} \alpha}{2 c_{\rm p} R_{\rm core}}}\right)-\frac{c_{\rm p} R_{\rm core}^{4}}{G M_{\rm core} \alpha}  \vphantom{\sqrt{\frac{\pi c_{\rm p}^{3} R_{\rm core}^{9}}{2 G^{3} M_{\rm core}^{3} \alpha^{3}}}} \right].\label{eq:adienergy}
 \end{aligned}
\eeq
In the simulations presented below, this is used as the nominal expression for the core thermal energy, since the core's energy transport is assumed to be convective. However, as already noted by \citet{baraffechabrier2008} and \citet{lindermordasini2016}, we find that using  $E_{\rm int,c,adia}$ instead of $E_{\rm int,c,iso}$ has a much smaller impact on the results compared to neglecting the temperature dependence of the core material or its energy contribution in general. The influence on the luminosity of an isothermal versus an adiabatic core is biggest for the 5, 10, and 20 $\mearth$ planets. The simulations with an isothermal core are up to 21\% smaller in luminosity, whereas for simulations where the core is excluded the luminosity is up to 86\% smaller than in the adiabatic case. For the 50 $\mearth$ planet, the luminosity with an isothermal core is 11\% smaller compared to the luminosity with an adiabatic core, and the luminosity without any core contribution is 57\% smaller than in the adiabatic case. For higher masses the differences become  smaller, but the general trend that  simulations without any core contribution included have a much smaller luminosity compared to those with an isothermal or adiabatic core stays the same. 

In both the isothermal and adiabatic cases, we assume that the temperature is continuous across the core-envelope boundary. As discussed by \citet{lopezfortney2014}, this should be the case for planets with an envelope sufficiently massive  that the surface of the rocky core is partially or completely molten ($T_{\rm ceb}\gtrsim$ 2000 K), allowing an efficient heat transfer, as also discussed by \citet{ginzburgsari2016} in the context of icy cores. For the smallest planetary masses studied in this work for ages younger than 1 Gyr, $T_{\rm ceb}$ is always  above 2000 K. For the heavier masses this is true for even later times. These  times are much later than the time when the planets could be detected, such  that this does not pose a problem.
In this paper we deal with planets at large orbital distances, therefore a mechanism that is in principle also included in the evolution model, namely atmospheric escape \citep{jinmordasini2017}, can be neglected.

\section{Atmospheric models}\label{sect:atmgrids}

\begin{figure}
\begin{center}
\begin{minipage}[t]{0.49\textwidth}
\centering\includegraphics[width=1\linewidth]{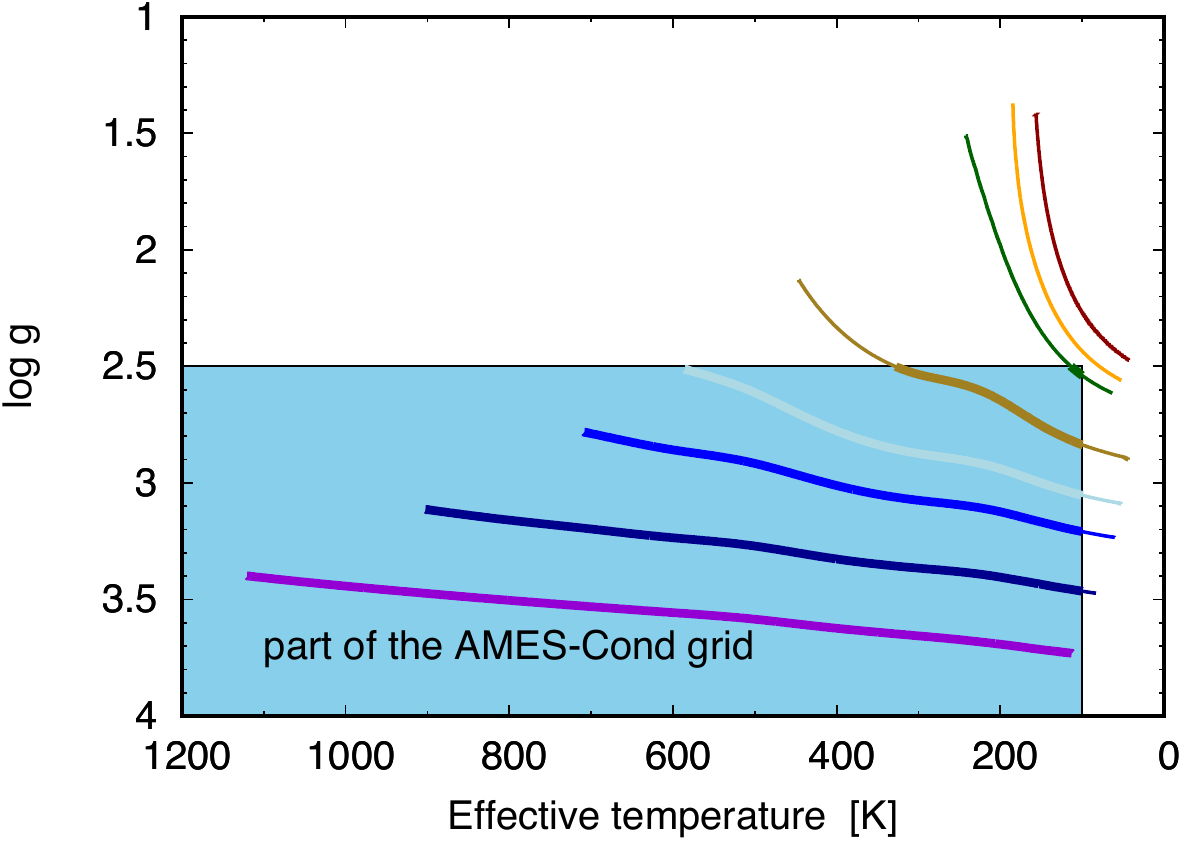}
\end{minipage}
\begin{minipage}[t]{0.49\textwidth}
\centering\includegraphics[width=1\linewidth]{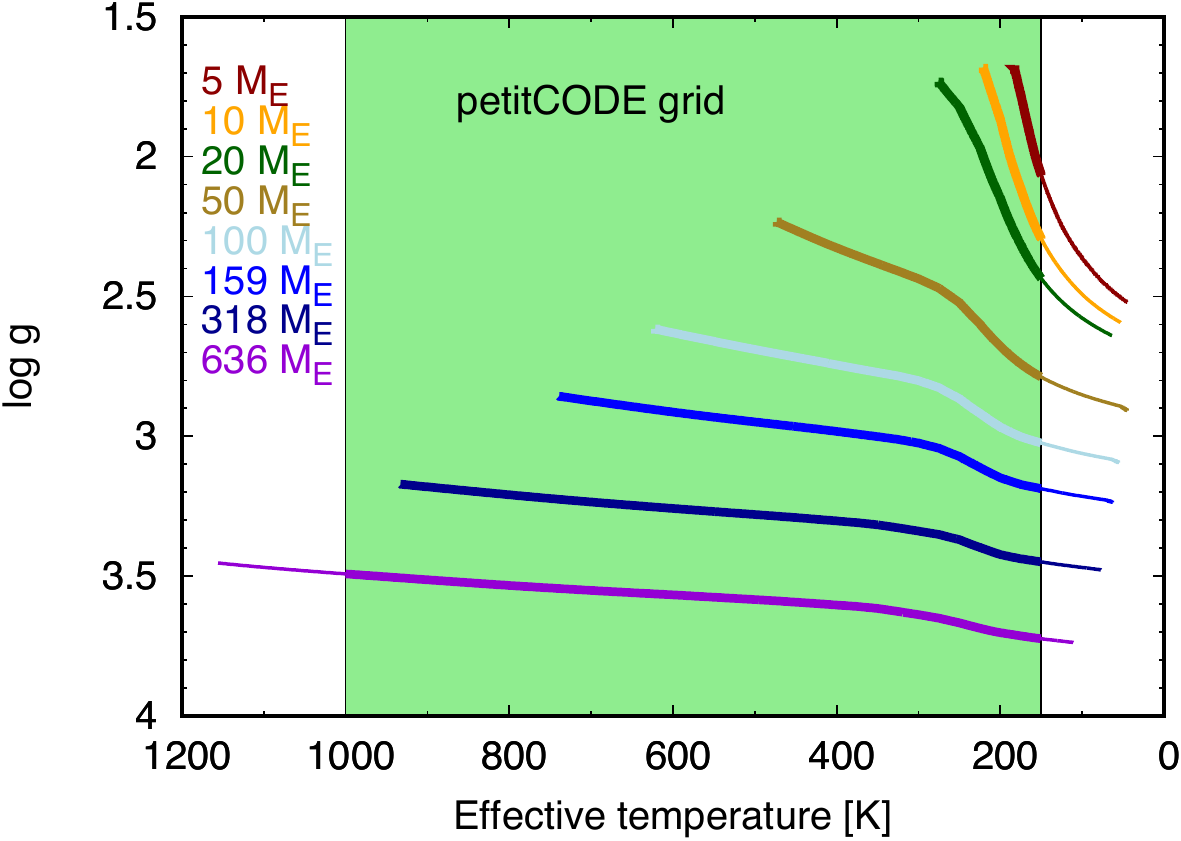}
\end{minipage}
\begin{minipage}[t]{0.49\textwidth}
\centering\includegraphics[width=1\linewidth]{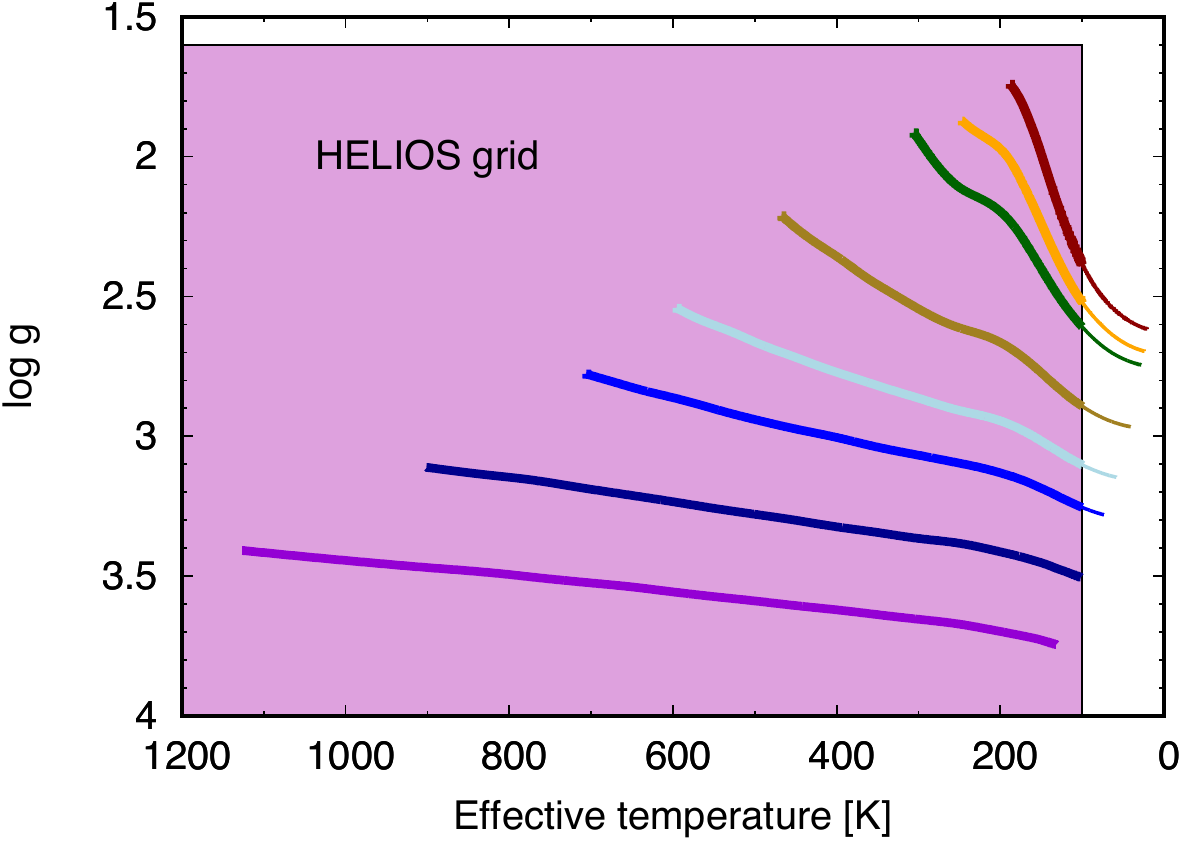}
\end{minipage}
\caption{Evolution of the planets in the $\log g$--$T_{\rm eff}$ space, together with the coverage by the atmosphere grids given as rectangles. The colour code for the masses is given in the top right panel. Thick lines indicate cooling curves that are in the atmosphere grid and thin lines those that are outside of the grid. The {top} panel shows the evolution in the \texttt{AMES-Cond} grid. In the {middle} panel the evolution in the \texttt{petitCODE} grid for the cloudy ($f_{\rm sed}$=0.5) and solar metallicity atmosphere is shown. This is representative also for other types of \texttt{petitCODE} atmospheres. The {bottom} panel finally shows the evolution in the \texttt{HELIOS} grid for a cloud-free atmosphere with a [M/H]=0.6 and is representative also for the solar  {-metallicity} evolution paths. }\label{fig:loggTeff}
\end{center}
\end{figure}

The atmospheric models provide boundary conditions for the cooling of the planet and also determine the spectral appearance of the planet.
In this work we use the approach of \citet{chabrierbaraffe1997} to couple externally calculated atmospheres to the interior calculation. Briefly, for a given $\log g$ and $T_{\rm eff}$ reached in the course of evolution, this simply means looking up (interpolating) the pressure and temperature in the convective very deep layers of the atmospheric model, and using this $(P,T)$ pair as the starting point for the inward structure integration (see \citealp{mordasinialibert2012b} for the structure calculation). Further details will be given in Marleau et al., in prep. For \texttt{petitCODE} and \texttt{HELIOS} we use a pressure of 50~bar as the connecting level but we verified that taking another pressure does not change the cooling curves and that the error in the radius from neglecting the layers above 50~bar is negligible ({at roughly the percent level}, smaller than the effects of model uncertainties such as cloud types or metallicities). For the \texttt{AMES-Cond} model, we took the structures available on F.~Allard's website\footnote{\url{https://phoenix.ens-lyon.fr/Grids/AMES-Cond/STRUCTURES/}.} and extracted the layer at $\tau_{\rm std}\approx100$, where all models are convective (this is almost but not quite the case at $\tau_{\rm std}\approx30$), where the index ``std'' stands for ``standard'' and refers to 2.15~$\mu$m (F.~Allard 2014, priv.\ comm.).

Additionally we also calculate evolutionary tracks using so-called Eddington atmospheres (Eddington atmospheres as always simply use $T=T_{\rm eff}$ at $ {P=P_{\rm phot}=(2\,g)/(3\kappa)}$; \citealp{mordasinialibert2012b}.) Since Eddington atmospheres as relevant for this work have been described in \citet{mordasinialibert2012b}, we only summarize the main features of \texttt{AMES-Cond} and briefly describe the more recent models \texttt{petitCODE} and \texttt{HELIOS}.

When the planet leaves the atmospheric grid, we extrapolate linearly from the last two grid points for the \texttt{AMES-Cond} grid. For the \texttt{petitCODE} and \texttt{HELIOS} grid, we extrapolate using  splines when the planet leaves the atmospheric grid. We have verified that this extrapolation is reasonable and similar to a linear extrapolation.

\subsection{\texttt{AMES-Cond} grid}

The \texttt{AMES-Cond} grid \citep{allardhauschild2001} consists of cloud-free atmosphere calculations that have been obtained with the \texttt{PHOENIX} code \citep{hauschildtbaron1999}. The models are calculated assuming radiative-convective and chemical equilibrium. While clouds are not included in these calculations, the sequestration of elements into condensates is treated.  {The \texttt{AMES-Cond} models treat condensation in strictly local chemical equilibrium, which means that the condensated particles do not rain out and are therefore still available for chemical reactions \citep{allardhauschild2001}.} The models contain opacities important for young, high temperature brown dwarfs (e.g. oxides such as TiO, VO, as well as hydrides such as FeH and MgH). Molecules that are important at intermediate to low temperatures, such as H$_2$O and CH$_4$, are also included, where the CH$_4$ line list with 47,415 lines is likely to be very incomplete when compared to modern CH$_4$ line lists with $\sim 10^{10}$ lines \citep{yurchenkoatennyson2014}. Convection in the \texttt{AMES-Cond} models is treated using mixing length theory. \texttt{AMES-Cond} models have been widely used in the literature for the calculation of planet evolutionary tracks, but also for the spectral fitting of brown dwarf and dwarf star atmospheres \citep[see e.g.][]{baraffechabrier2003}.

\subsection{\texttt{petitCODE} grid}\label{subsect:petitgrid}

{We calculated another grid using the} \texttt{petitCODE}, a 1D model that self-consistently calculates atmospheric structures and spectra of exoplanets. {The m}ain assumptions of the code are radiative-convective and chemical equilibrium.
 {\texttt{petitCODE} treats condensation of solids in chemical equilibrium. This means that gas phase chemistry will be affected by the depletion of elements into solids. Rain-out is neglected, however. {The choice to neglect feldspar condensation in the equilibrium chemistry calculations effectively mimics the rain-out of silicon atoms into silicates (which are included). Hence the alkalis, which would otherwise condense into feldspars, will stay in the atmosphere until Na2S and KCl condense, which seems to be confirmed by observations \citep{linemarley2017}. Indeed \texttt{petitCODE}  spectra and structures agree with calculations including rain-out \citep{baudinomolliere2017}.} The equilibrium condensate mass fractions are used as an input to the Ackerman \& Marley cloud model, as described in \cite{mollierevanboekel2017}. This cloud model includes rain-out of the cloud particles, but this is not coupled back to the chemical equilibrium calculations.}

  {Convection is modelled applying adiabatic adjustment: \texttt{petitCODE} solves the radiative temperature structure of the atmosphere from top to bottom. If, during that process, the temperature gradient from one layer to the next is found to be steeper than the adiabatic temperature gradient, the temperature of the bottom layer is corrected to follow the adiabatic temperature gradient instead. This process is repeated until the bottom of the atmosphere is reached, and the full atmospheric temperature structure has been found. Further details can be found in \cite{molliereboekel2015} on how the adiabatic adjustment is implemented.} The radiative transfer implementation treats absorption, emission, and scattering. Clouds can be added self-consistently, making use of the \citet{ackermanmarley2001} model, or a model that simply parametrizes the cloud particle size and maximum cloud mass density \citep{mollierevanboekel2017}. In the calculations presented here, we used the \citet{ackermanmarley2001} model. The gas absorbtion of the following species is considered: CH$_4$, HCN \citep[ExoMol, see][]{tennyson:2012aa}, H$_2$O, CO, CO$_2$, OH \citep[HITEMP, see][]{rothman:2010aa}, H$_2$, H$_2$S, C$_2$H$_2$, NH$_3$, PH$_3$ \citep[HITRAN, see][]{rothman2013}, and Na , K \citep[VALD3, see][]{piskunov1995}. Ultraviolet electronic transitions are included for H$_2$ and CO \citep{kurucz1993}. The code also includes collision induced absorption (CIA) of H$_2$--H$_2$ and H$_2$--He \citep{borysowfrommhold1989a,borysowfrommhold1989b,richardgordon2012}. Lastly, Rayleigh scattering is included arising from H$_2$, He, CO$_2$, CO, CH$_4$ , and H$_2$O. The cross sections are taken from \citet{dalgarnowilliams1962} (H$_2$),
\citet{chandalgarno1965} (He), \citet{sneepubachs2005} (CO$_2$, CO, CH$_4$), and \citet{harveygallagher1998} (H$_2$O). \texttt{petitCODE} is described in \citet{molliereboekel2015,mollierevanboekel2017}. Recently, \texttt{petitCODE} was benchmarked against the \texttt{ATMO} and \texttt{Exo-REM} codes \citep{baudinomolliere2017}.

The grid presented here is an extension of the grid of self-luminous atmospheres calculated for spectral fitting of {51 Eri b} \citep{samlandmolliere2017}. Identical to \citet{samlandmolliere2017}, we assume the clouds to consist of Na$_2$S and KCl. The following parameter values were considered, spanning the grid in a rectangular fashion:
$T_{\rm eff}$ = [150, 1000], $\Delta T_{\rm eff}$ = 50;
$\log_{10}$(g) (cgs) = [1.5, 4.0], $\Delta \log_{10}$(g) = 0.5;
[Fe/H] = [-0.4, 1.4], $\Delta$[Fe/H] = 0.2; 
$f_{\rm sed}$ = [0.5, 3.0], $\Delta f_{\rm sed}$ = 0.5, cloud-free.  As described in \citet{ackermanmarley2001}
$f_{\rm sed}$ is the cloud settling parameter. In contrast to the original \citet{ackermanmarley2001} model, the \citet{mollierevanboekel2017} implementation assumes the cloud mixing length to be always equal to the pressure scale height. This effectively lowers the $f_{\rm sed}$ value when compared to \citet{ackermanmarley2001}, as described in \citet{mollierevanboekel2017} and \citet{samlandmolliere2017}.

\subsection{ {Absence of water clouds in atmosphere models}}

 {The major limitation of the \texttt{petitCODE} cloudy grid for the temperature range to which it is applied in this work is the absence of water clouds. Water clouds are expected to form at temperatures from 300-400 K (see e.g. \cite{morleyfortney2012}, \cite{morleymarley2014}). This is especially relevant for planets with masses smaller than 20 $ {\mearth}$, which start their evolutions at temperature below 300 K. Water clouds can heavily impact the spectra by absorbing flux in the 4.5 micron region at higher temperatures (300 K) and heavily absorbing across the full spectral range for even cooler planets \citep{morleymarley2014}. 
Thus the models are not very realistic at low temperatures, until very low temperatures are reached where the water cloud has disappeared below the photosphere and cloudless models become relevant again (until methane and ammonia condense). This has a potentially large effect on the surface boundary conditions as well as the predicted magnitudes. }

\subsection{\texttt{HELIOS} grid}

 {A third atmosphere} grid is produced with the \texttt{HELIOS} code \citep{malik17}, which is an open-source 1D radiative transfer code developed specifically for exoplanetary atmospheres. As  {in the \texttt{petitCODE},}  \texttt{HELIOS} self-consistently calculates the temperature structures and resulting emission spectra in radiative-convective equilibrium through radiative iteration and convective adjustment. The chemistry model \texttt{FASTCHEM} is used \citep{stockkitzmann2018}, which includes 550 gas-phase reactants to calculate the atmospheric abundances in chemical equilibrium.  {In the current version of \texttt{FASTCHEM,} condensation is not taken into account for the gas phase abundances. This may lead to an overestimation of the gas absorption in the cool temperature regime, for example of water vapour for $ {T \lesssim 300}$ K. T}he following opacities are included from these line lists --- EXOMOL: H$_2$O \citep{barber2006}, CH$_4$ \citep{yurchenkoatennyson2014}, NH$_3$ \citep{yurchenko2011}, HCN \citep{harris2006}, H$_2$S \citep{azzam2016}; -- HITEMP \citep{rothman:2010aa}: CO$_2$, CO;  -- HITRAN \citep{rothman2013}: C$_2$H$_2$. Also added are the resonance lines for Na, K as described in \cite{heng2015} and \cite{heng2016} and CIA H$_2$-H$_2$ H$_2$-He absorption \citep{richardgordon2012}. Also included is isotropic Rayleigh scattering of H$_2$ \citep{sneepubachs2005}. With \texttt{HELIOS} we calculated a grid of cloud-free, self-luminous atmospheres with the following parameters: [Fe/H]=0, 0.6; $T_{\rm eff}$ = [100, 1200], $\Delta T_{\rm eff}$ = 50; $\log_{10}$(g) (cgs) = [1.6, 4.0], $\Delta \log_{10}$(g) = 0.1.

\begin{figure*}
\begin{center}
\begin{minipage}[t]{0.48\textwidth}
\centering\includegraphics[width=0.99\linewidth]{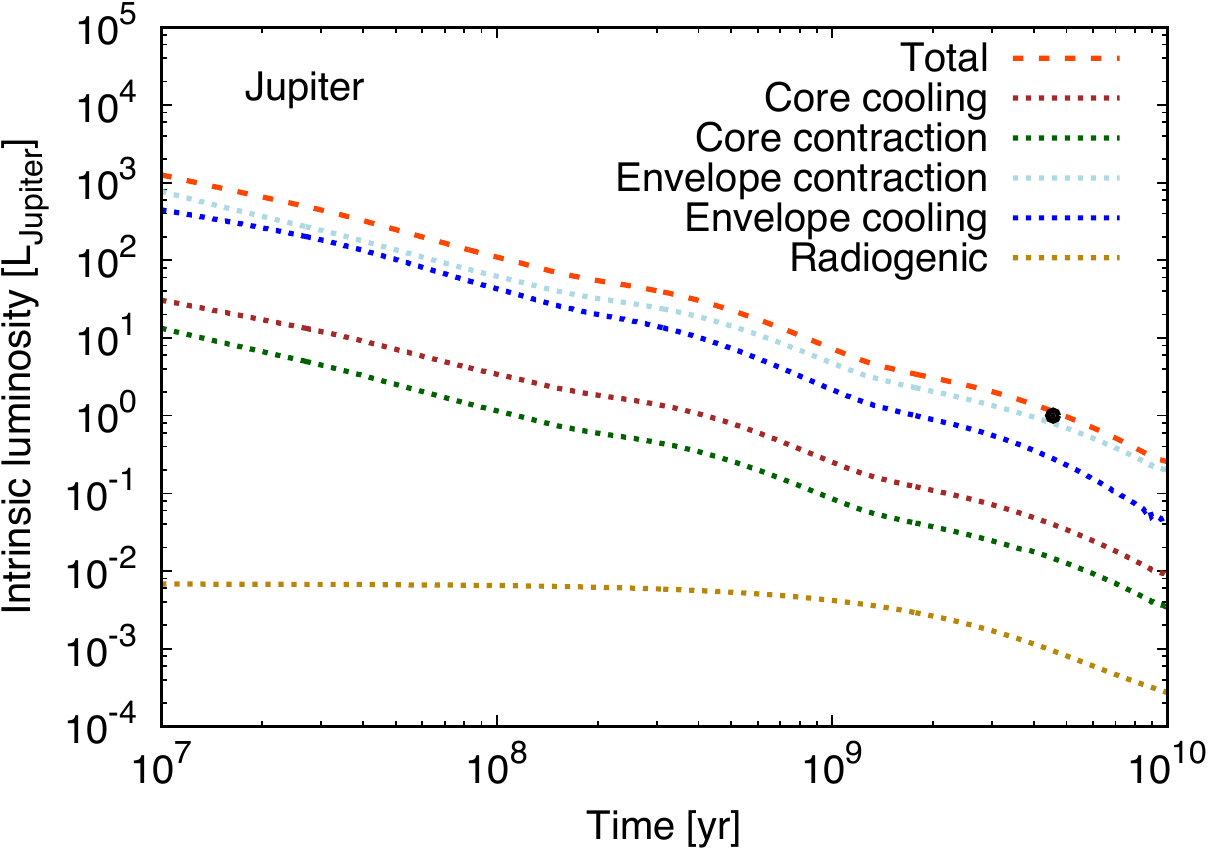}
\end{minipage}
\begin{minipage}[t]{0.48\textwidth}
\centering\includegraphics[width=0.99\linewidth]{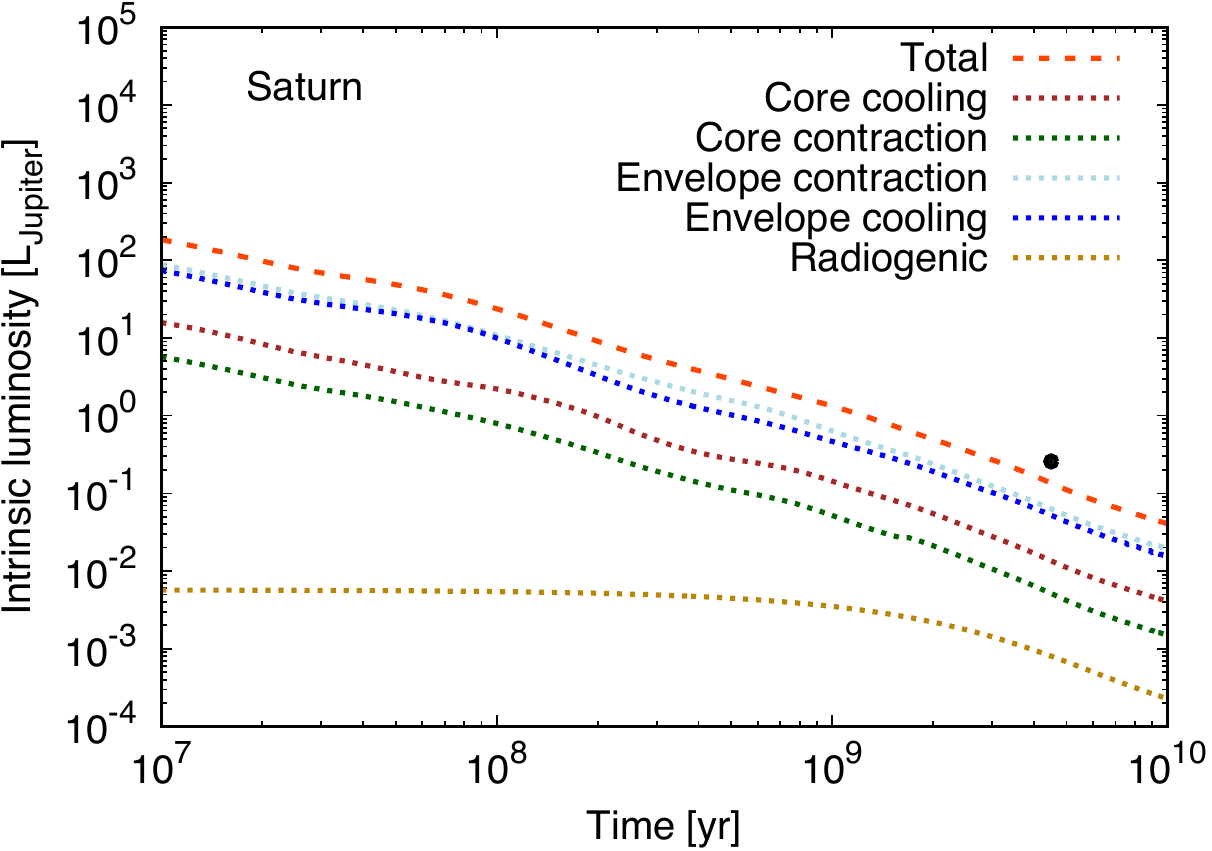}
\end{minipage}
\begin{minipage}[t]{0.48\textwidth}
\centering\includegraphics[width=0.99\linewidth]{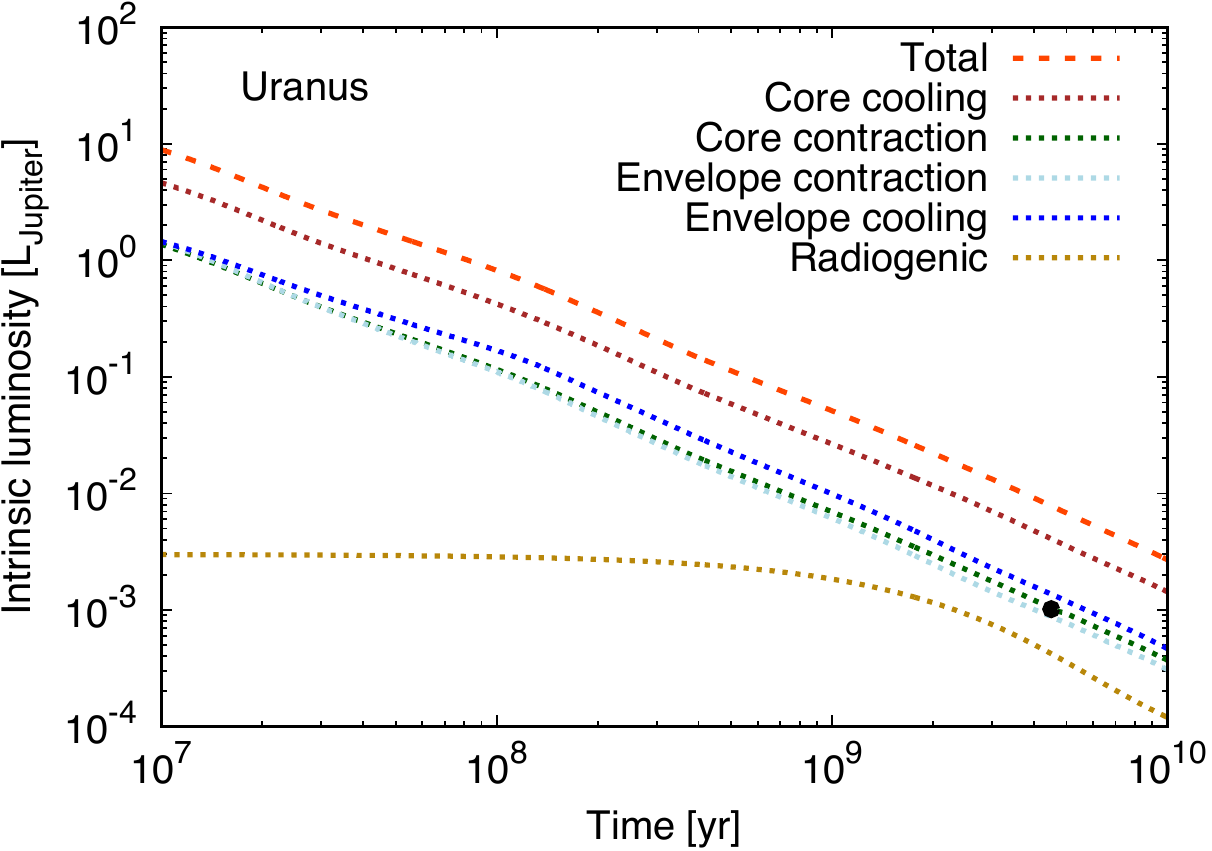}
\end{minipage}
\begin{minipage}[t]{0.48\textwidth}
\centering\includegraphics[width=0.99\linewidth]{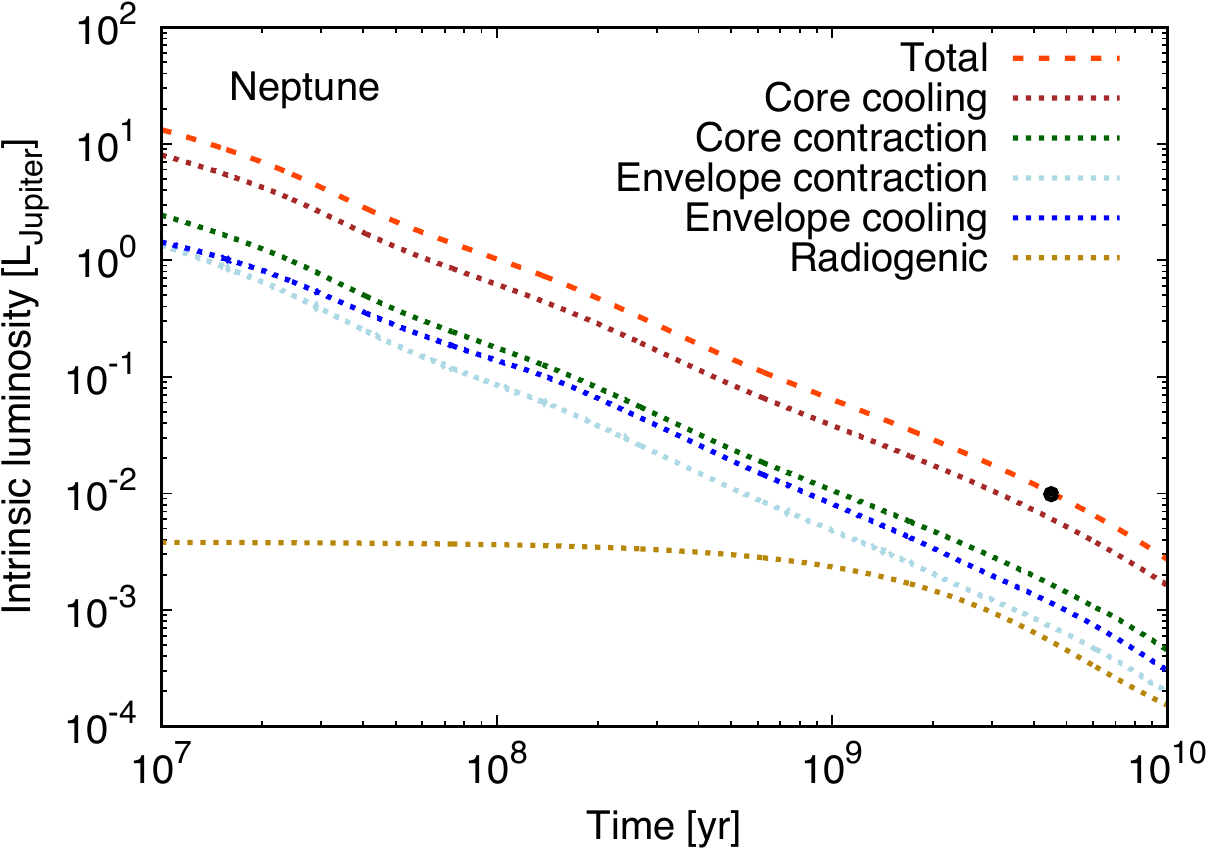}
\end{minipage}
\caption{Temporal evolution of the luminosity of the simplified solar system gas giants. The different contributions to the total luminosity are indicated with colours, the luminosity of today is shown as a black dot.}\label{fig:solsystlumis}
\end{center}
\end{figure*}

\begin{figure}
\begin{center}
\includegraphics[width=\columnwidth]{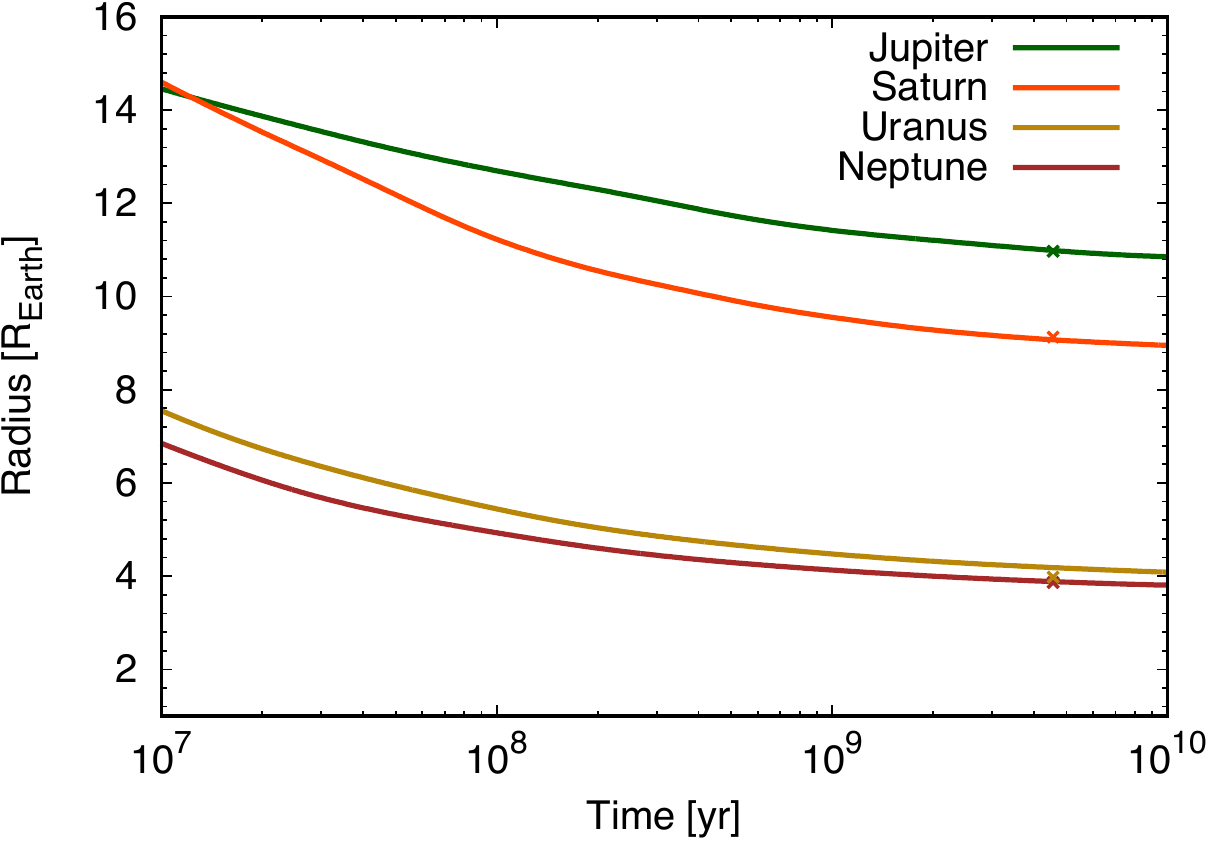}
\caption{Temporal evolution of the radius of the simplified solar system gas giants.  The radius of today is given in the respective colour of the planet as a  {cross. On this scale, the symbols for the radii of Uranus and Neptune partially overlap.}}\label{fig:solsystradis}
\end{center}
\end{figure}

\begin{table*}
\caption{Overview of the simulated solar system planets. For all of them, a  {h}elium abundance of 0.27 and an ice mass fraction in the core of 0.5 was assumed. The [M/H] metallicities, the radii $R_{\rm meas. }$ , as well as the luminosities $L_{\rm meas.}$ of the planets today   were taken from \citet{guillotgautier2014}. }
\begin{center}
\begin{tabular}{lrrrrrrrrr}
\hline\hline
Planet  &       [M/H]   &       $R_{\rm meas. }$ [$\rearth$]    &       $R_{\rm simu. }$ [$\rearth$] & $L_{\rm meas. }$  [$\lj$] &$L_{\rm simu. }$  [$\lj$]&$\mcore$ [$\mearth$]&$\menve$ [$\mearth$]& $ {\mtot}$ [$ {\mearth}$]\\
\hline
Jupiter &       0.5                     &       10.97   &       10.99   &1                                                      &1.13   &       27.5    &290.3  & {317.8}\\
Saturn  &       1.0                     &       9.14    &       9.07    &$257.6\times10^{-3}$           &$135.01\times10^{-3}$  &       23.0    &72.2   & {95.2}\\
Uranus  &       1.8                     &       3.98    &       4.18    &$\le 1.01 \times10^{-3}$     &$      7.84\times10^{-3}$&     12.1    &2.5 &  {14.6}\\
Neptun{e}       &       1.8                     &       3.87    &       3.88    &$9.85\times10^{-3}$            &$      10.13\times10^{-3}$&    15.3    &1.8 & {17.1}\\
\hline
\end{tabular}
\end{center}
\label{tab:solsystoverview}
\end{table*}

\section{Examples of evolutionary calculations}\label{sect:benchmarking}

In this section  {examples of}  cooling curves of simple models for a Neptune-, Uranus-, Jupiter-, and Saturn-like planet, for a planet like GJ 436 b, and for a close-in, core-dominated, sub-Neptunian planet are presented. The results are compared with other thermal evolution models to validate our evolutionary model.

The thermal evolution of the planets is modelled with a three layer interior of the planet, namely an iron/silicate core, potentially an ice layer, and a (pure) H/He envelope, as was described in more detail in Sect. \ref{sect:model}. This is similar to but simpler than the approach in \citet{fortneynettelmann2010}, who include water mixed into the H/He layer, and H/He mixed into the water layer  above the rock core. {The assumption is that a mixed envelope composition is favoured} from a planet formation point of view, because planetesimals might get dissolved in the envelope of the accreting planet (e.g. \citealt{podolakpollack1988,mordasinialibert2006}). In contrast to the calculations further down, a grey atmosphere is assumed here as boundary condition for the inward integration, as introduced in Sect. \ref{sect:atmgrids}. The metallicity [M/H] enters the opacity calculation, using the \citet{freedmanmarley2014} Rosseland mean opacity. Our aim is not to present detailed models for the evolution of the giant planets of the solar system for which numerous observational constraints exist. Rather, we want to understand how our simplified model that is used for exoplanets compares to existing more detailed simulations. For exoplanets we usually only have little observational constraint, for example a rough age and magnitudes in some bands. This makes a simplified approach appropriate, as more complexity would in any case remain unconstrained. 

The evolution of the planet is calculated in the following way. 
First, with static interior structure calculations, the envelope-to-core mass ratio is determined for the planet today, given its measured mass, luminosity, and radius.
Then, with the derived core-to-envelope mass ratio and a starting luminosity that is several orders of magnitudes higher than the one measured today, the planets' temporal evolution is calculated. Since the Kelvin-Helmholtz timescale is  short in the beginning, the exact starting luminosity respectively starting entropy is no longer important at the present time. The evolution is then calculated by taking into account the contraction of the envelope, as well as its cooling, and the contraction, cooling and radioactive energy production in the core, as explained in Sect. \ref{sect:model}.

\subsection{Neptune and Uranus}\label{subsect:neptuneuranus}

In Fig. \ref{fig:solsystlumis} the bottom right panel shows the luminosity as a function of time for our simplified Neptune model. The planet 
has a total mass of $1~\mnept=17.147~\mearth$. The atmospheric opacity  {corresponds} to a  [M/H]=1.8 \citep{guillotgautier2014}. An ice mass fraction of 50\% in the core is assumed for all four giant planets in the solar system. A similar value is  expected for a condensation of water ice in the solar nebula \citep{mindullemond2011, lodders2003} and is also motivated from planetary formation models \citep{guillotgautier2014}. Much higher ice mass fractions have sometimes been used in interior structure models, which is difficult to understand from a formation point of view \citep{guillotgautier2014}. For this fixed ice mass fraction, the core and envelope mass is determined with static interior structure calculations  so that the planet has for its observed intrinsic luminosity $\lnept=9.85 \times10^{-3}~\lj$  \citep{guillotgautier2007} a radius of $1~\rnept  = 3.87 ~\rearth$. We find a composition with a H/He envelope mass of $1.80~\mearth$. The central part, built of an ice layer wrapped around an iron/silicate core, thus has a mass of $15.347 ~\mearth$. An overview of these values is given in Table \ref{tab:solsystoverview}. This composition can be compared with the models of \citet{podolakweizman1995}. Their Neptune model 1 (2) has a H/He envelope of $2.2~(0.9)~\mearth$, and a total heavy element mass of $14.9 ~(16.2)~\mearth$. These values bracket the ones found in our model. The bulk mass of heavy elements in Neptune found in \citet{nettelmannhelled2013}  is $14-14.5 ~ \mearth$, which is also similar to our value.

We then followed the thermal evolution of this planet, starting with a high luminosity of $50~ \lj$. The total luminosity is split in the contributions coming from the core cooling, core contraction, envelope cooling and contraction, as well as radiogenic heating. The total luminosity at the age of the solar system agrees well (difference of $3~\%$) with the observed value, which is shown as a black dot. Our cooling curve for Neptune overlaps especially at later times with the one presented in \citet{fortneyikoma2011}. In  Fig. \ref{fig:solsystradis}, the change of the radius in time is shown, where the measured radius of today of the respective planet is given as a coloured dot. Neptune's measured radius of today is by construction well reproduced (within $0.5~\%$) by our simulations.
The change in intrinsic luminosity over time of Neptune (bottom right panel in Fig. \ref{fig:solsystlumis}) follows a $t^{-4/3}$ slope, which is analytically expected for the cooling of a planet where the dominant energy source is the thermal cooling of the core \citep{ginzburgsari2016}.

We also simulated the cooling of an Uranus-like object. It is well known that Uranus is much fainter than expected from fully convective cooling models (for example \citealt{podolakhubbard1991}; \citealt{podolakweizman1995}; \citealt{fortneyikoma2011}; \citealt{nettelmannhelled2013}). We therefore expect that it is not possible to find a correct cooling age with our model. This is indeed the case:
for the simulation of a simple Uranus model, we assume that the planet 
has a total mass of $1~\muran  = 14.536~ \mearth$ with an opacity corresponding to  [M/H]=1.8 \citep{guillotgautier2014}. In order to reproduce the radius of Uranus today, $\ruran  = 3.98~\rearth$ with the current upper limit of its intrinsic luminosity of $\luran=1.01 \times10^{-3}~\lj$ (one order of magnitude less than Neptune, \citealt{guillotgautier2014}), we need an interior consisting of a H/He envelope of $2.470~\mearth$, and a solid part of $12.066~\mearth$, which is split into 50\% ice and 50\% silicate/iron (see Table \ref{tab:solsystoverview}). These values are similar to the Uranus model of \citet{podolakweizman1995} with $1.5~ \mearth$ for the envelope and $13~ \mearth$ for the ice and silicate/iron part, as well as to the Uranus model computed in \citet{nettelmannhelled2013}, who find a bulk composition of heavy elements of $12.5~ \mearth$. 

With this interior composition of Uranus, we simulated the cooling of the planet, starting from a high initial luminosity ($50 ~\lj$ as for Neptune, but the precise value is not important as long as it is high). Our evolution calculation of Uranus agrees at later times very well with the one  in \citet{fortneyikoma2011}. We find that at the current age of the solar system the simulated planet has a luminosity that is eight times too high relative to the upper limit of the observations, which can also be seen in the bottom left panel of Fig. \ref{fig:solsystlumis}. We therefore recover the result (e.g. \citealt{fortneyikoma2011,nettelmannhelled2013}) that standard fully convective cooling models fail to explain Uranus's very low luminosity. This is also mirrored in the fact that the simulated radius is $5~\%$ too big compared to the measured radius of today (see Fig. \ref{fig:solsystradis}).  {So even though we matched a static model of the planet  {with the present measured luminosity and radius of the planet, these may not be reproduced} by the modelled evolution of the planet at 4.5 Gyr. }

With the improved gravity field data from the Voyager fly-by of Uranus and Neptune and modified rotation period and shape of the planets, \citet{nettelmannhelled2013} compute adiabatic three-layer structures for the two planets and find that Uranus and Neptune might differ in their atmospheric enrichment within an observational{ly} significant amount. This could be due to a stable stratification in the interior of Uranus and originate from a giant impact \citep{nettelmannhelled2013}. We conclude that Uranus and Neptune might have very different internal structures and/or thermodynamic states despite their similar masses and radii.

\subsection{Jupiter and Saturn}\label{subsect:jupisat}

Jupiter is simulated with a total mass of $\mj=317.83~\mearth$. We assumed again an ice fraction of 50\% in the core as for Uranus and Neptune and an opacity corresponding to [M/H]=0.5 \citep{guillotgautier2014}. As before, we matched today's given radius to the observed present-day luminosity by varying the core-envelope mass ratio. With static interior calculations, we obtained the radius of the planet today ($\rj=10.97~\rearth$) for a model with a central part containing iron, silicate, and ice of $27.50~\mearth$ and a H/He envelope mass of $290.33~\mearth$ (Table \ref{tab:solsystoverview}). \citet{fortneynettelmann2010} conclude that the current Jupiter models show a range in the core mass of $0-18~\mearth$ and a heavy element mass in the envelope of $0-37~\mearth$. The more recent analysis of \citet{wahlhubbard2017} based on Juno data finds core masses between 6 to 24 $\mearth$, and total heavy element masses (core and metals mixed in the envelope) of 24 to 46 $\mearth$, depending on assumptions concerning the core's state and the EoS. Thus, keeping in mind that the core mass in our model rather represents the bulk heavy element in the planet, our results lie in a similar interval for the heavy element content of Jupiter.

We then followed the evolution of this planet. {The evolution of the luminosity can be seen in the top left panel of Fig. \ref{fig:solsystlumis}, and the evolution of the radius in Fig. \ref{fig:solsystradis}}. Our modelled Jupiter is slightly too bright compared to the measured luminosity (difference of $13~\%$), as was already found for example by \citet{fortneyikoma2011}. For the Neptune model, the biggest contribution to the total luminosity came from the core cooling, as is expected for core-dominated planets \citep{baraffechabrier2008}. However, for our Jupiter model, the biggest contribution comes from the envelope contraction. Our modelled Jupiter is cooling too slowly compared to the real planet, reaching the observed luminosity at 4.91~Gyr. Therefore, also the modelled radius of the planet is slightly too big ($1.0014\times\rj$) at the present age, which can be seen in Fig. \ref{fig:solsystradis}. 

We also modelled the evolution of Saturn. The planet was simulated with a mass of $~1\msat=95.159~\mearth$. As for the other solar system planets, we assumed an ice mass fraction of 50\% in the core, but an opacity corresponding to [M/H]=1 \citep{guillotgautier2014}. We find a solution for the static interior model with an envelope mass of $72.159~\mearth$ and a central part of $23~\mearth$ (Table \ref{tab:solsystoverview}). Following the evolution of the planet over time, we expect a cooling time for our homogeneous adiabatic models of 2-3 Gyr, as in \citet{puestownettelmann2016}. Our modelled Saturn reaches the luminosity of today at 3.1 Gyr, while at 4.6 Gyr the simulated Saturn is $48 \%$ too dim (note the logarithmic scale). The difference between today's simulated luminosity and the measured one for Saturn is much larger than for Jupiter, by a factor of 3.7. Also the radius of the simulated planet is thus smaller than  the measured radius. The modelled Saturn is therefore cooling too fast. Hence, we reproduce the common result in the literature that adiabatic homogeneous models underpredict Saturn's current luminosity, or in other words that Saturn exhibits a strong excess luminosity \citep{stevensonsalpeter1977a}. This is commonly attributed to H/He demixing (e.g. \citealt{stevensonsalpeter1977a, fortneyhubbard2003,puestownettelmann2016}), but other explanations exist as well \citep{lecontechabrier2013}. We note that for a Saturnian-mass planet, the demixing sets in  at an age of 1-2 Gyr \citep{puestownettelmann2016}, so for objects younger than that it should not pose a big problem.

\subsection{GJ 436 b: Comparison with Baraffe et al. (2008)}\label{subsect:GJ436b}

The transiting planet GJ 436 b \citep{gillondemory2007a} orbits a $0.44~\msun$ M-star at 0.028~ {  {au}}. The mass of the planet is $M_{\rm GJ436b }= 22.6 \pm 1.9~\mearth$ and its radius is determined to be $0.386\pm0.016~\rj$ (\citealt{gillondemory2007a}; \citealt{demingharrington2007}). The age of the system is unconstrained by  observations. \citet{baraffechabrier2008} assume a system age of 1-5 Gyr and do not take  irradiation  on the planet into account because of the low luminosity of the parent star. They find a good match with the observed radius within the  {uncertainty} of the age of the system for a $22.6~\mearth$  planet with a $21~\mearth$ water core surrounded by  a $1.6~\mearth$   H/He envelope. {We want to test whether we can match the observations of this planet with the same composition which is what we find.}

\begin{figure}
\begin{center}
\begin{minipage}[t]{0.48\textwidth}
\centering
\includegraphics[angle=0,width=0.99\linewidth]{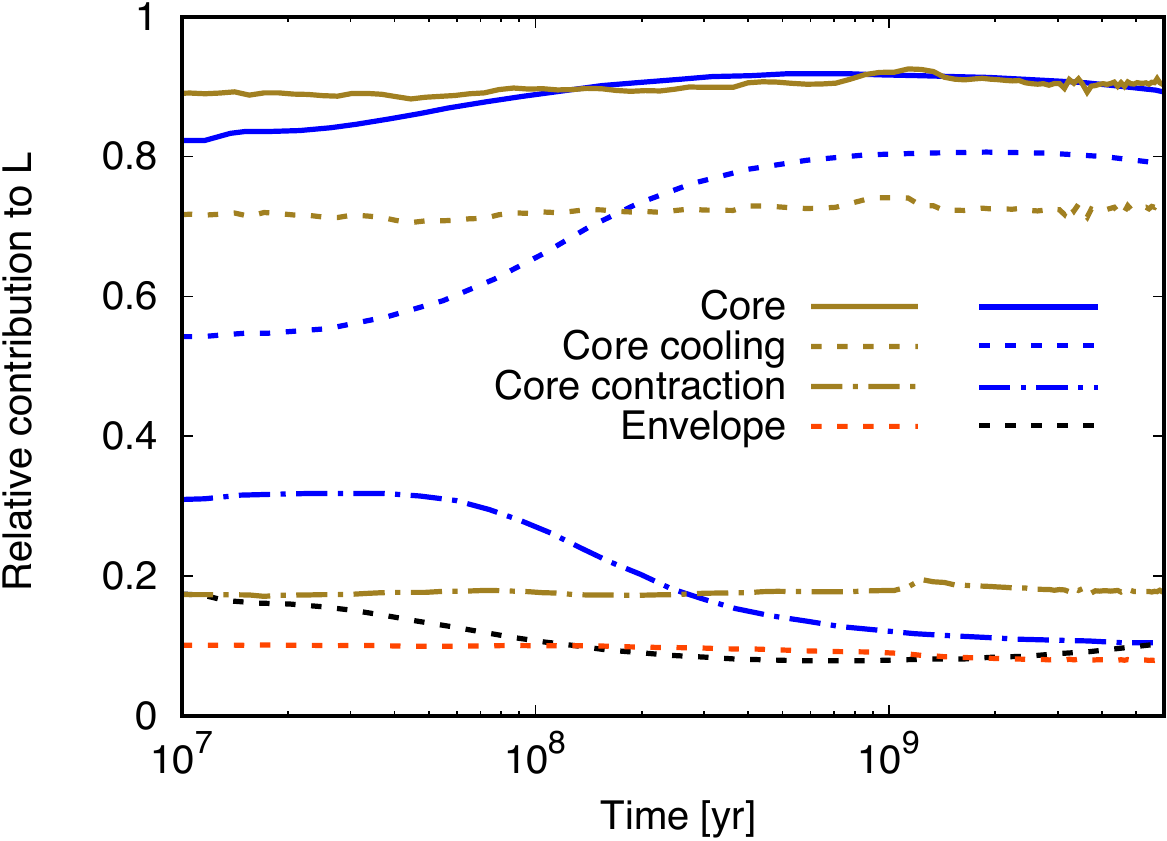}
\end{minipage}
\begin{minipage}[t]{0.48\textwidth}
\centering
\end{minipage}
\caption{Relative contributions to the total luminosity $L$ over time for the water core  simulation of GJ 436 b. The blue and black lines are from Baraffe 2015 (personal communication), the  {red} and  {brown} lines are from this work and represent the following: {Solid blue/ {brown} line:} energy release from the core{, which is the sum of core cooling and core contraction}; {dashed blue/ {brown} line:} core cooling; {dash-dotted blue/ {brown} line:} core contraction; {dashed black/ {red} line:} envelope cooling and contraction. A discussion can be found in Sect. \ref{subsect:GJ436b}. }\label{fig:GJ436benergies}
\end{center}
\end{figure}

Due to its core-dominated nature, GJ 436 b is a good example to study the contributions of the core and envelope to the planet's temporal evolution. 
The contributions to the gravothermal energy release for the water core look qualitatively similar if  the structure of the planet is calculated with the EoS SESAME or ANEOS  (\citet{baraffechabrier2005}, Baraffe 2015 personal communication). Figure \ref{fig:GJ436benergies} shows the luminosity from the core cooling and contraction as a sum and separate relative to the total luminosity $L$ over time, as well as the  contribution from the envelope relative to the total $L$. Shown are the luminosities predicted by our model assuming a water core  together with the results from Baraffe (2015, personal communication). 
 
 {In constrast} to Fig. 10 in \cite{baraffechabrier2008},  {which used SESAME, the black and blue lines in Fig. \ref{fig:GJ436benergies}}
  {were} obtained using the ANEOS EoS on which we base our expressions for the heat capacities and the thermal expansion coefficient  {as well}.  Looking  {in more detail} at the calculation employing ANEOS, which is  shown in Fig. \ref{fig:GJ436benergies} (blue and black lines), the biggest contribution to the gravothermal energy release comes from the core with a fraction of $\sim$~0.85  from the total luminosity averaged over time. The total contribution from the core is split into a bigger part coming from the thermal cooling of the core with a fraction of $\sim$~0.55 between 10 to 100 Myr and then rising  to reach  $\sim$~0.8 around 1 Gyr. The smaller part of the total core contribution  comes from the contraction of the core with a fraction of $\sim$~0.3 from 10 to 100 Myr and then sinking to $\sim$~0.1 at 1 Gyr. The envelope contribution to the total luminosity of the planet is the smallest with a fraction of $\sim$~0.15.  
  
{The relative contributions do not strictly sum up to 1, as is expected. This is due to the fact that the  {total luminosity is} calculated using the entropy that is given by the EoS, {} $ {   {L}_\text{tot}} = - \int T   {(}dS/dt) \,   { dm} $. If the applied EOS were thermodynamically coherent, one would have $ {-TdS/dt = - P d   {V}/dt - d   {U}/dt }$. However, this is not exactly the case with ANEOS (from Baraffe 2015, personal communication).}  {Therefore, the sum of the volume work and the internal energy terms, which are shown separately in Fig. \ref{fig:GJ436benergies}, do not sum up to the total luminosity obtained from the entropy change.}

The gravothermal energy contributions from this work are similar (shown in Fig. \ref{fig:GJ436benergies} as well, olive and orange lines), namely a fraction of $\sim$~0.9 for the biggest contribution to the total energy from the core which is split into a contribution of $\sim$~0.73 from the cooling of the core and $\sim$~0.17 from the contraction of the core.The smallest contribution to the total luminosity originates again from the envelope (fraction of $\sim$~0.1).  {Different to the calculation by Baraffe 2015, the energy contributions sum up to 1.}

 {In this calculation, we aim to reproduce the model of \cite{baraffechabrier2008} and use the same model parameters. Accordingly, the core is composed of  pure water and thus there is no radiogenic contribution to the luminosity}{, as the radioactive material content of the planet is assumed to be proportional to the rock mass fraction of the planet. Simulating a planet with the same core and envelope mass fraction as above, but containing 0.8 $\mearth$ of iron and rock in the core (corresponding to an ice mass fraction in the core of 50\%), so  containing 0.53 $\mearth$ rock, the radiogenic luminosity contributions can be estimated. For example, it accounts for   0.01\% (0.5\%) at 1 (100) Myr of the total luminosity. } 
 
{As we discussed above, the overall agreement between the two simulations regarding the luminosity contributions  in the planet is good, as expected. The agreement especially concerns the fact that the core makes the dominant energy contribution to the total luminosity budget of the planet. Therefore, neglecting the core's contribution in the evolution calculation of such a  planet has a significant impact. }

\begin{figure}
\begin{center}
\includegraphics[width=\columnwidth]{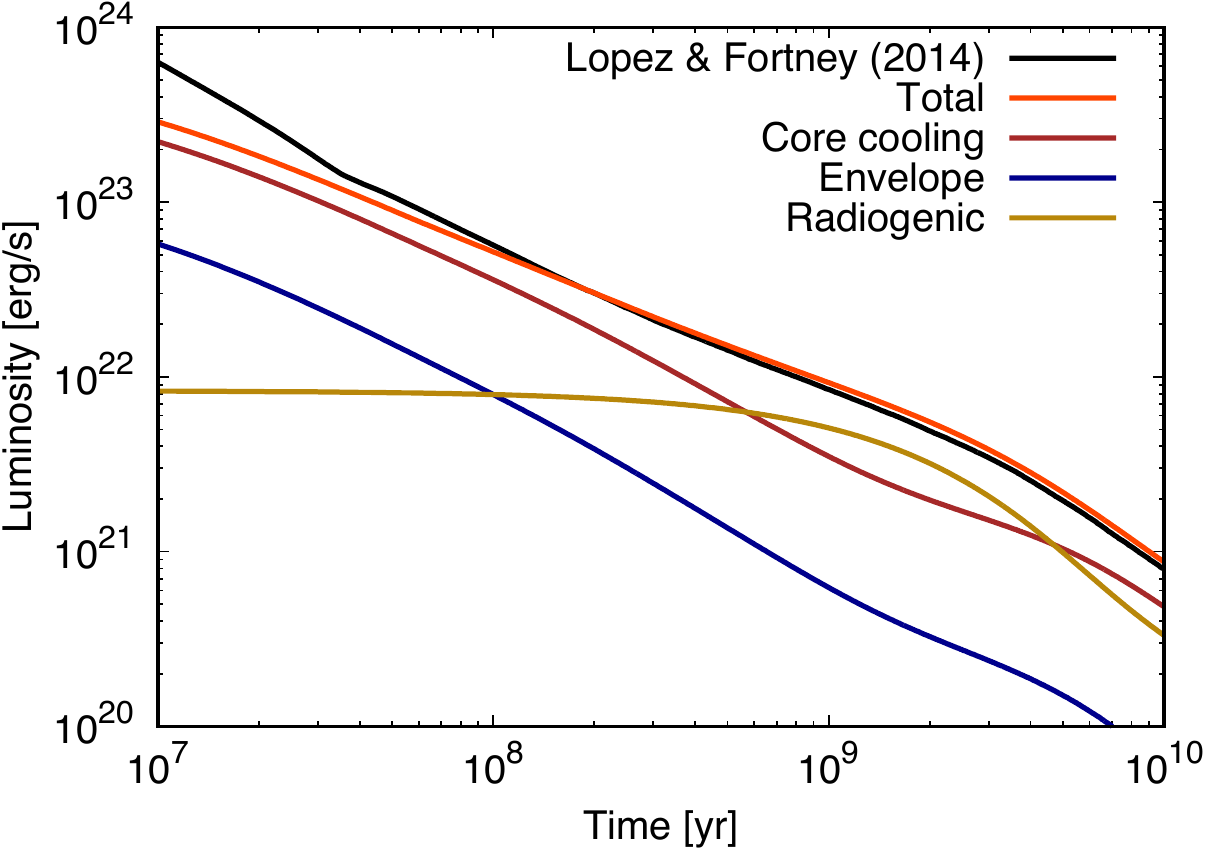}
\caption{Luminosity as a function of time for a 5 $\mearth$ planet with a 1\% H/He envelope at a distance of 0.1~ {  {au}} from a 1 $\msun$ star. The core is assumed to be isothermal. The different contributions are indicated in the plot. The solid black line  shows the total luminosity as found by \citet{lopezfortney2014}.}\label{fig:lopezfortney}
\end{center}
\end{figure}

\subsection{A 5 $\mearth$ planet with a 1\% H/He envelope: Comparison with Lopez \& Fortney (2014)}\label{subsect:vglLF}

As an example of the evolution of a strongly core-dominated planet, and for comparison with another  independent evolutionary model, we present  {in Fig. \ref{fig:lopezfortney}}  the cooling curve of a planet with the same properties as the one simulated in Fig. 3 of \citet{lopezfortney2014}.
It is a $5~\mearth$ planet with a rocky core and a 1\% H/He envelope. The planet is located at 0.1~ {  {au}} from a solar-like star. Such close-in sub-Neptunian planets have been found in high numbers by the Kepler satellite (e.g. \citealt{fressintorres2013a,petiguramarcy2013}). The envelope opacity corresponds to a 50-times-solar heavy element enrichment ([M/H]=1.7). In contrast to our simpler model, \citet{lopezfortney2014} directly use the ANEOS \citep{thompson1990} and SESAME \citep{lyonjohnson1992} equations of state to model the interior of the solid core, and they employ fully non-grey atmospheric models.

Different from all other simulations in the present paper, we assume for this comparison  that (i) the core is isothermal, (ii) that the heat capacity of the core is $7.5\times10^{6} ~\textrm{ergK}^{-1} \textrm{g}^{-1}$, (iii) that the core does not shrink due to the reduction of its temperature, which means that the release of internal energy ($du/dt$), but not the associated release of gravitational potential energy ($pdV/dt$) of the core, contributes to its luminosity. All these three settings mimic the ones made by \citet{lopezfortney2014}. The adiabatic model for the core (Eq. \ref{eq:adienergy}) leads in this case to total luminosities that are about 66\%  higher compared to the isothermal core. This comes from the fact that the thermal energy of the core is more than two times higher for the adiabatic core model than for the isothermal core model. For planets with a less extreme core-envelope mass ratio, the impact is correspondingly smaller, as  discussed in Sect. \ref{sect:model}. For giant planets, the difference is completely negligible.

A comparison of the two simulations in Fig. \ref{fig:lopezfortney} shows that the overall agreement is very good, with a tendency of our model to predict a somewhat dimmer planet at earlier times ( $\leq$100~Myr). After 300~Myr, the luminosity found in our model is  slightly higher than in \citet{lopezfortney2014}, with a difference of about 9\% at 10~Gyr. A  feature that is shared in both models is the dominance of the radiogenic heating mainly caused by $^{40}$K decay \citep{mordasinialibert2012c} over the cooling of the core at intermediately late times (0.7-5~Gyr). Further comparison shows that our model predicts a somewhat higher contribution of the core at later times, whereas it is the opposite at earlier times. The radiogenic luminosity is virtually identical in the two models.

\section{Initial conditions}\label{sect:intialcond}

\begin{figure*}[ht]
\begin{center}
\begin{minipage}{0.5\textwidth}
\centering
 \includegraphics[width=0.9\textwidth]{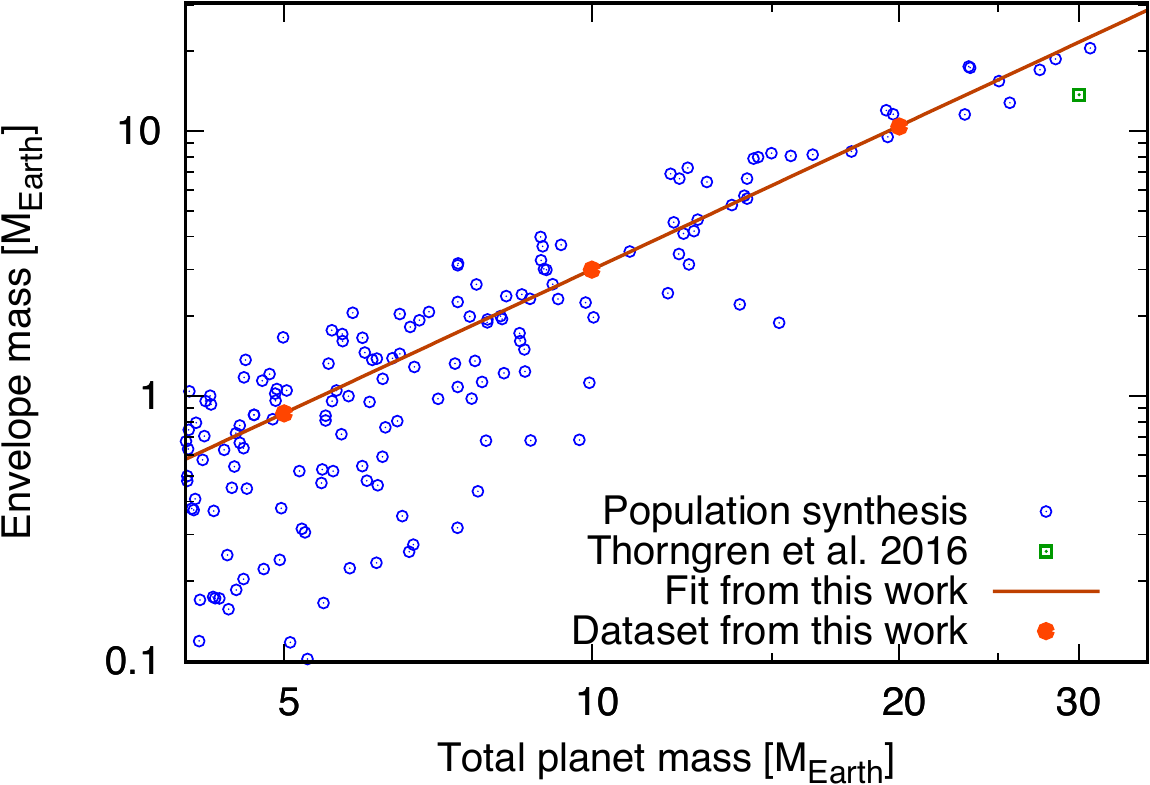}
\end{minipage}\hfill
\begin{minipage}{0.5\textwidth}
\centering
\includegraphics[width=0.95\textwidth]{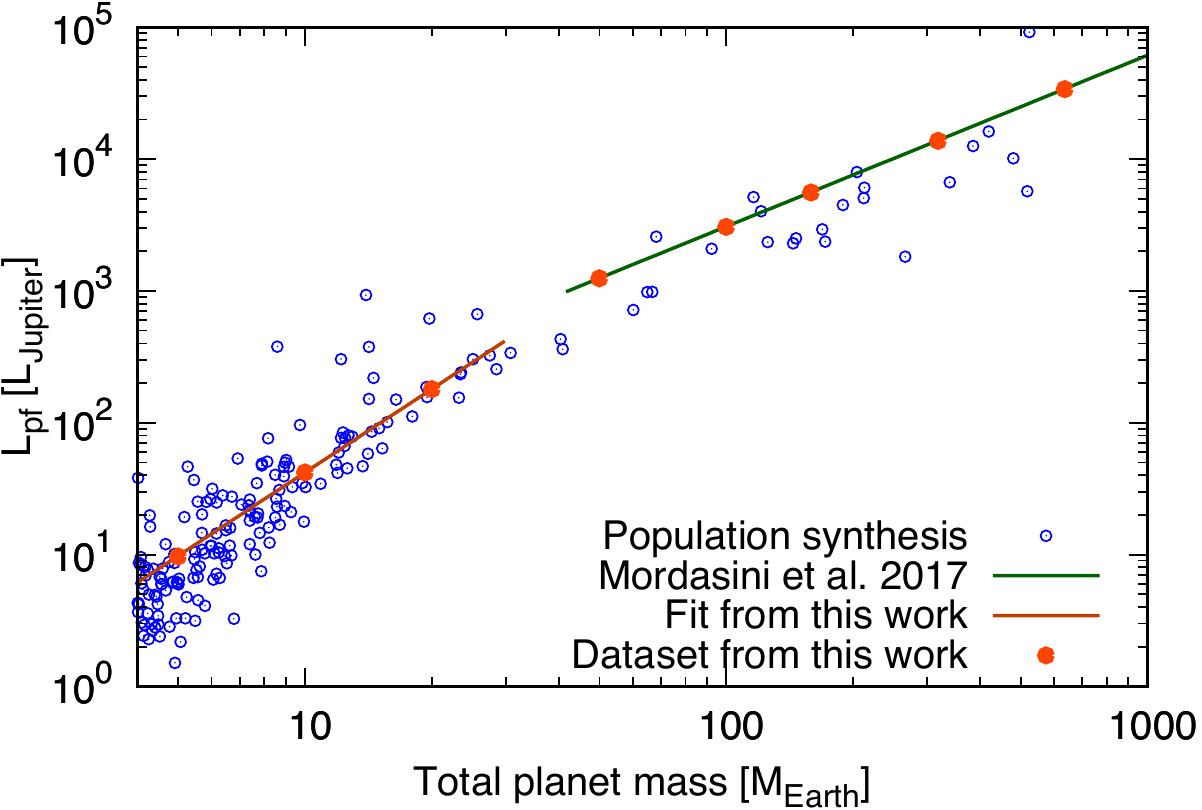}
\end{minipage}
\caption{Results from population synthesis calculations {(blue circles)} of combined planet formation and evolution that are used for the initial conditions. {Orange dots} give our final dataset. {Left panel:} Envelope mass as function of total mass with the fit from this work ({orange-red}, Eq. \ref{eq:fitMtotMenv}). For more massive planets the envelope mass as calculated from the relationship found by \cite{thorngrenfortney2016} (Eq. \ref{eq:fitThorngren}) is shown for the 30~$\mearth$ planet with a {green square}. {Right panel:} Post formation luminosity as a function of total planetary mass, together with the fit from \citet{mordasinimarleau2017} ({dark-green}, Eq. \ref{eq:fitMordasini}) and from this work  ({orange-red}, Eq. \ref{eq:fitML}). }\label{fig:initialcond}
\end{center}
\end{figure*}

\begin{table*}
\caption{Total, envelope, and core mass, specific post formation entropy, as well as luminosity for the eight planets modelled here, after 3 Myr of the formation simulation.}
\begin{center}
\begin{tabular}{rrrrr}
\hline\hline
Planet mass [$\mearth$]  &      $\menve$         [$\mearth$]   &        $\mcore$          [$\mearth$] &  $S_{\text{pf}}$         [$\kB\, {baryon}^{-1}$]&         $L_{\text{pf}}$ [$\lj$] \\
\hline
5                       &       0.9             &4.1            &7.61           &       9.7\\
10              &       3.0             &7.0            &       7.94            &       42\\
20                      &       10.4            &9.6            &8.26   &       180\\
50                      & 31.3          &18.7   &       8.83    &       1240\\ 
100                     &               71.4&28.6       &       9.07    &       3060\\ 
159             &121.1          &37.9   &       9.14    &        5590\\
318             &       260.1   &57.9   &       9.20& 13800\\
636             &        547.6  &88.4   &       9.27    & 33900\\
\hline
\end{tabular}
\end{center}
\label{tab:fitexamples}
\end{table*}    

For the simulation of the thermodynamic evolution of the planets and the calculation of their magnitudes we need to specify the post formation properties of these planets. As exemplified by the large impact of cold versus hot starts for giant planets (e.g. \citealt{marleyfortney2007,mordasinimarleau2017}), the post formation properties can have a very strong influence on the predicted detectability and observational characteristics. To find such initial conditions for the simulations,  the output of planetary population syntheses based on the core accretion paradigm (see e.g. \citealt{mordasinimarleau2017}) was studied. Population synthesis is the attempt  to model planet formation globally. Therefore, a model containing many sub-models coming from specialized studies on one aspect of planet formation (for example gas and dust disc dynamics, type I \& II migration, ice-line behaviour, planetary accretion) is constructed. By simplifying and putting these specialized models together, a global planet formation model can be built. The initial conditions, such as the total disc (gas) mass, the dust-to-gas-ratio, and the lifetime of the disc  are sampled in a Monte-Carlo way from probability distributions that are derived as closely as possible from observations. In the particular population synthesis used for this work, ten planetary embryo were inserted into a disc around a solar-like star. More model settings used in the population synthesis are described in \citet{mordasinialibert2012c} and \citet{mordasinimarleau2017}. Formally, the synthesis was conducted under the assumption of cold gas accretion. However, for low-mass planets studied here, there is no hot versus cold accretion difference in the same way as for giant planets. For giant planets, the dominant fraction of their mass is accreted after detachment from the protoplanetary nebula (at $\sim$~100 $\mearth$) through a potentially entropy-reducing shock. Low-mass planets only detach from the nebula when the nebula itself has already almost completely dissipated. Because of this, only a very small amount of gas is accreted after detachment through a potentially entropy-reducing shock. Different formation histories of the individual planets (e.g. moment when gas and solids are accreted, surrounding disc conditions), however, still induce a diversity in post formation properties. Additional physical mechanisms such as solid accretion at late times, giant impacts, enriched envelope composition, or semi-convection might  change the outcome of our planet formation model, but are currently not considered. 

From the planet population synthesis,  the planet's core and H/He envelope mass fractions as well as the luminosity at the end of formation was estimated.  The output from the population synthesis was studied at an age of 3~Myr, the typical life-time of the synthetic discs. We do not include planets that are still undergoing strong planetesimal accretion. For the case of a hot protoplanet after collisional afterglows, the reader is referred to for example \citet{schaeferfegley2009} and \citet{miller-riccimeyer2009}.

The population synthesis results for envelope mass as a function of total planet masses from 5-30~$\mearth$ are shown in the left panel of Fig. \ref{fig:initialcond} as blue circles. The data were fit by eye, which is given by Eq. \ref{eq:fitMtotMenv}:

\begin{equation}\label{eq:fitMtotMenv}
\frac{\menve}{\mearth}= 3 \, \left( \dfrac{\mtot}{10 \mearth}  \right)^{1.8}, 
\text{ for $\mtot~$\textless$~30~\mearth$},
\end{equation}

and shown as an orange-red line in Fig. \ref{fig:initialcond}, together with the three lowest mass planets simulated in  this work indicated as orange dots. We note that this fit relation represents mainly planets with a high $\menve$ for a given total mass (see Fig. \ref{fig:initialcond}), and that studies of the composition of close-in, low-mass sub-Neptunes indicate rather lower envelope mass fractions (e.g. \citealt{wolfganglopez2015}). However, the envelope mass of many of these planets was potentially reduced by atmospheric escape (e.g. \citealt{owenwu2013,jinmordasini2017}), and the planets we consider here are at larger semi-major axes.

For heavier planets, the gas accretion rate changes from being limited by the planet's Kelvin-Helmholtz contraction to the disc-limited regime (e.g. \citealt{dangelodurisen2010}). \citet{thorngrenfortney2016} derive from the mass-radius relation of observed planets  a fit for the total heavy element content as a function of the total planetary masses, which is given in Eq. \ref{eq:fitThorngren}:

\begin{equation}\label{eq:fitThorngren}
\frac{M_{\text{z}}}{\mearth}= (57.9 \pm 7.03) \, \left( \dfrac{\mtot}{  {318\mearth}} \right)^{0.61\pm0.08}, 
\text{ for $\mtot~$\textgreater$~30~\mearth.$}
\end{equation}

 For planets that are more massive than 30~$\mearth$, their fit was used. In the left panel of Fig. \ref{fig:initialcond} for reference the envelope mass of a 30 $\mearth$ is given (green square) as computed with their expression. A satisfyingly smooth transition between the two fits is found.

From the population syntheses, also the post formation luminosity ($L_{\text{pf}}$) and thus entropy ($S_{\text{pf}}$) was obtained. The $L_{\text{pf}}$ are shown as blue circles in the right panel of Fig. \ref{fig:initialcond}. For planets between 5-30~$\mearth$, the population syntheses output was  fit  by eye, which is given in Eq. \ref{eq:fitML}:

\begin{equation}\label{eq:fitML}
\dfrac{L} {\lj}= 42 \, \left( \dfrac{\mtot}{10\mearth}\right)^{2.1}, 
\text{ for $\mtot~$\textless$~40~\mearth,$}
\end{equation}

and shown as an orange-red line in the right panel of Fig. \ref{fig:initialcond}. 

For giant planets, $L_{\text{pf}}$ increases more slowly with $\mtot$ than for $\mtot~\le~100~\mearth$ because giant planets go through an entropy-reducing shock, therefore the slope becomes shallower. In \citet{mordasinimarleau2017} a fit for the luminosity of giant planets (M~$\ge~100~\mearth$) depending on the total planetary mass was provided. In this work, the fit for the cold-nominal planets is applied and given also in Eq. \ref{eq:fitMordasini}:

\begin{equation}\label{eq:fitMordasini} 
\dfrac{L} { {\lj}}=  {1.378 \times 10^{4}} \, \left( \dfrac{\mtot}{ {318\mearth}}\right)^{1.3}, 
\text{ for $\mtot~\ge~50~\mearth. $}
\end{equation}

This fit is shown as a dark-green line in the right panel of Fig. \ref{fig:initialcond} and used for the heavier planets in our dataset (orange dots in the same figure).

Table \ref{tab:fitexamples} finally gives an overview of the properties of the eight planets the evolution of which we simulate in this study, spanning a wide mass range from 5 $\mearth$ to 2 $\mj$. The post formation entropy $S_{\rm pf}$ is calculated at the bottom of the convective zone with a grey atmosphere after the bulk composition and luminosity of the planet is given.

 \begin{figure*}
\begin{center}
\begin{minipage}[t]{0.49\textwidth}
\centering\includegraphics[width=1\linewidth]{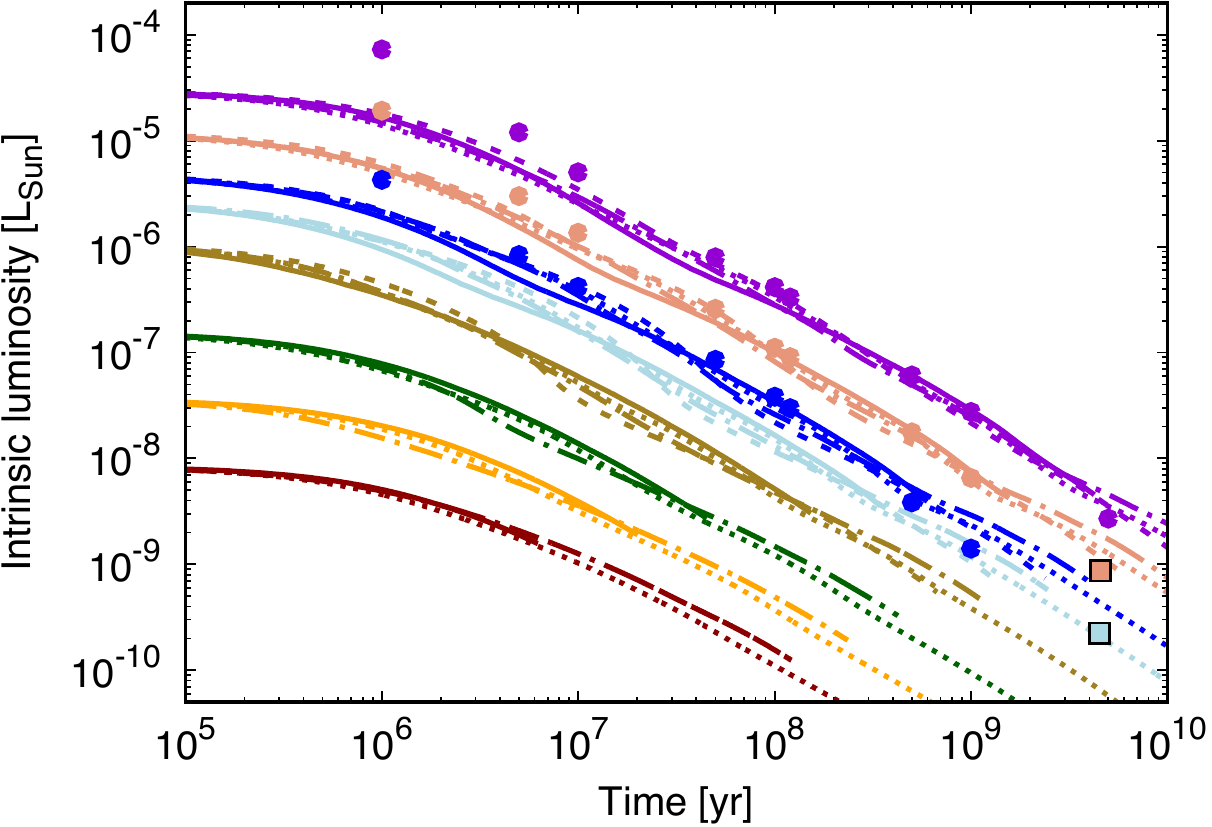}
\end{minipage}
\begin{minipage}[t]{0.49\textwidth}
\centering\includegraphics[width=1\linewidth]{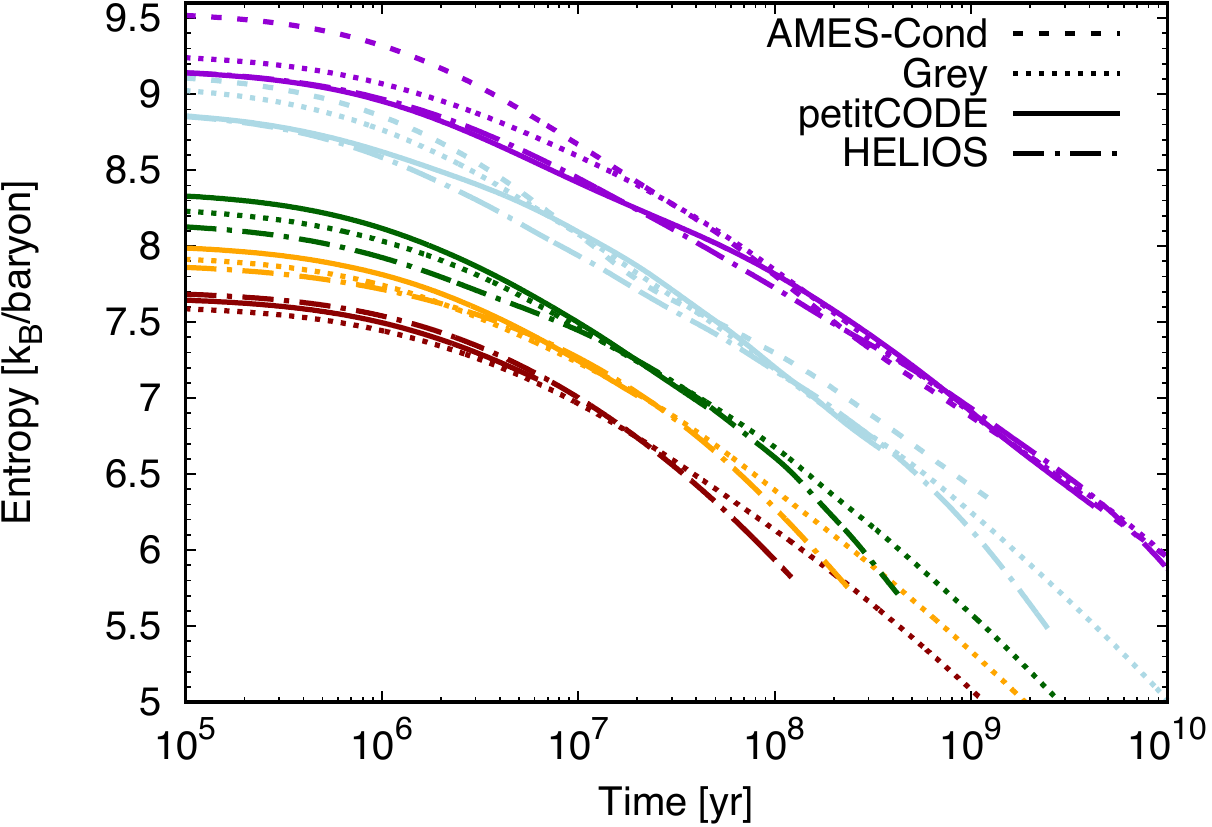}
\end{minipage}
\begin{minipage}[t]{0.49\textwidth}
\centering\includegraphics[width=1\linewidth]{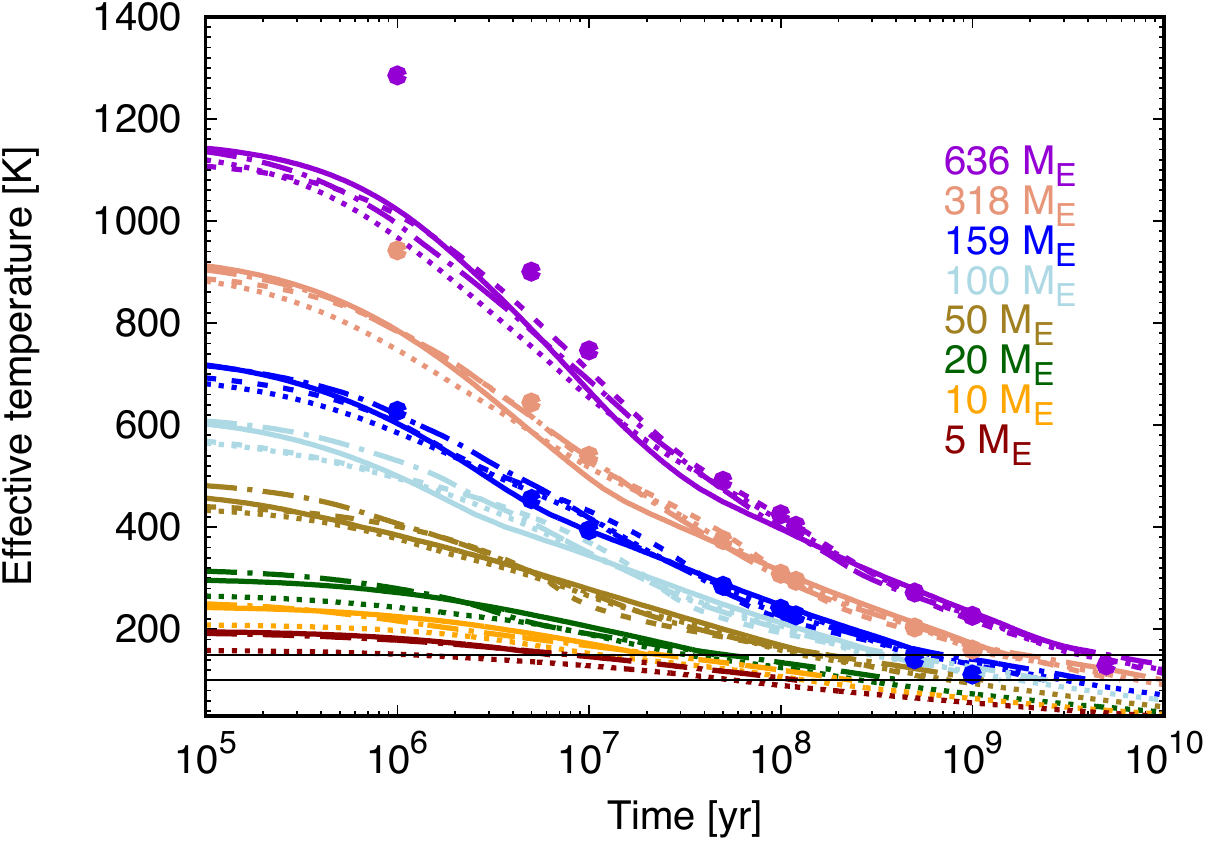}
\end{minipage}
\begin{minipage}[t]{0.49\textwidth}
\centering\includegraphics[width=1\linewidth]{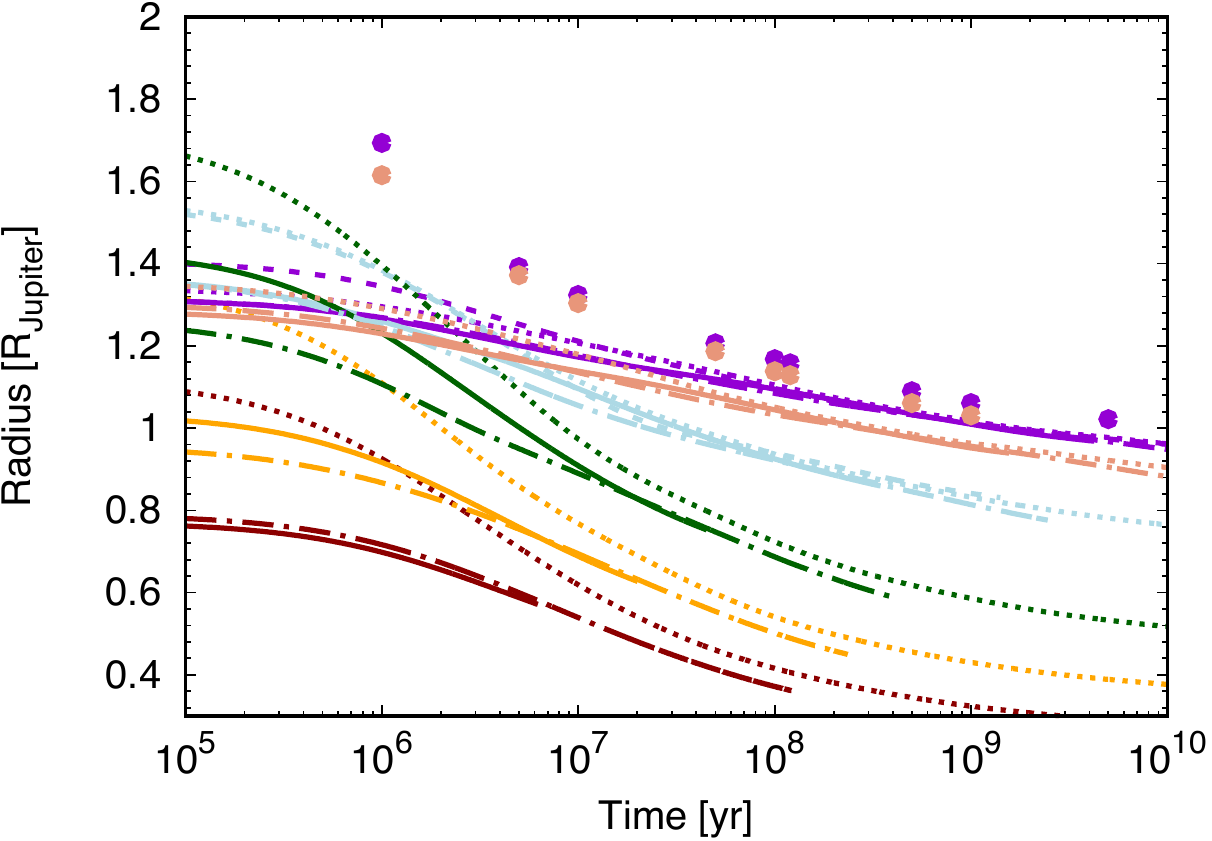}
\end{minipage}
\caption{Temporal evolution of fundamental properties of the eight simulated planets with initial conditions given in Table \ref{tab:fitexamples}. The change in intrinsic luminosity, entropy, temperature, and radius is shown. The colour code for the  masses is given in the  {bottom} left panel. The luminosities of Jupiter and Saturn are indicated as  squares in the top left panel. The dots  {in the luminosity, effective temperature, and radius} panel show the results from \citet{baraffechabrier2003}. {Solid lines} show the evolution with a cloud-free atmosphere with solar metallicity from the \texttt{petitCODE} grid, {dashed lines} show it with an \texttt{AMES-Cond} atmosphere, {dash-dotted lines} show it with a solar metallicity, cloud-free atmosphere from the \texttt{HELIOS} grid, and {dotted lines} finally show the evolution calculated with a grey atmosphere.  {Since the three smallest masses are evolving outside of the \texttt{AMES-Cond} grid (see Fig. \ref{fig:loggTeff}), these masses are not plotted for an \texttt{AMES-Cond} atmosphere. For the \texttt{AMES-Cond, petitCODE} and \texttt{HELIOS} atmospheres, the tracks are shown as long as the planet is evolving on the atmospheric grid.} For better visibility, three {resp. two } evolution curves are not shown in the entropy {resp. }radius plot. The horizontal lines at 100 and 150 K in the $T_{\rm eff}$ plot show the lower limit of the \texttt{HELIOS} and \texttt{petitCODE} grid, respectively.} \label{fig:resultssolarclear}
\end{center}
\end{figure*}

\section{Results and discussion}\label{sect:results}

In this section, we show cooling curves resulting from our evolution code using the initial conditions given in Table \ref{tab:fitexamples} for different planetary atmosphere models and metallicites. Following this, the impact of different post formation entropies on the planet's evolution is studied. {Finally, we calculate magnitudes for various typical filters that can be
found for instance in the Nasmyth Adaptive Optics System  Near-Infrared Imager and Spectrograph  (NACO) on the Very Large Telescope (VLT), in the VLT Imager and Spectrometer for mid Infrared (VISIR), in the Polarimetric High-contrast Exoplanet Research (SPHERE), or on JWST. The influences on{ {JWST}} magnitudes from different atmospheric models or post formation entropies are estimated.} Also, we carry out a comparison with the { {JWST}} sensitivity limits for the simulated planets during their evolution.

For all planets, the ice mass fraction in the core was set to 0.5 as direct imaging is sensitive to planets at large orbital distances. Because of this, stellar irradiation was neglected. It is important to note that in the simulations of this work time zero is when  the gas disc disappears. The stellar age could thus be up to $\sim10$ Myr higher, depending on the specific disc lifetime.

\subsection{Cloud-free solar metallicity models}\label{subsect:solarclear}

At first, the evolution with cloud-free solar metallicity model atmospheres was studied. Figure \ref{fig:resultssolarclear} shows the evolution for eight planetary masses from 5 $\mearth$ to 636 $\mearth$ (2 $\mj$) and four different atmosphere types, from 0.1 Myr to 10 Gyr for the heavier planets.  The luminosity of Jupiter and Saturn is given for reference as  squares  {in the colour corresponding  (roughly, for Saturn) to their masses}.  {Planets with $ {\mtot \textless 20 \mearth}$ are not plotted with an \texttt{AMES-Cond} atmosphere, as these masses evolve outside of the atmospheric grid (see Fig. \ref{fig:loggTeff}). For the \texttt{AMES-Cond, petitCODE,} and \texttt{HELIOS} atmospheres, the tracks are shown as long as the planet is evolving on the atmospheric grid.} The dots represent  the luminosities{, temperatures, and radii} from \citet{baraffechabrier2003}. 

The differences between \citet{baraffechabrier2003} and our cooling curves at early times   {in the luminosity and temperature panel} simply reflect the choice of different initial luminosities, and  {at later times}, the two different cooling calculations agree  well.  {For example, at 50 Myr the difference for the 636 $\mearth$ planet between the luminosity curve with an \texttt{AMES-Cond} atmosphere versus a \texttt{HELIOS} (\texttt{petitCODE}) atmosphere is  14\% (29\%).}   {The mean difference is around 25\%. The mean was obtained by averaging over the maximal procentual differences in bolometric luminosity for the different atmospheric models at 1, 10, and 100 Myr for the four heaviest planets. These  differences are smaller than the error bars  today for measured luminosities (e.g.  48\% for  Eri b in \citealt{macintoshgraham2015}). With future more precise measurements, it could be possible to distinguish between different atmospheric models. We conclude that t}he choice of the atmosphere has, most of the time, only a  {limited} impact on the evolution of the bolometric luminosity, as also found for example in \citet{burrowsliebert1993}, \citet{burrowshubbard2001}, \citet{ baraffechabrier2002}, \citet{ saumonmarley2008},  and \citet{ dupuyliu2015}.  {A posteriori, it is therefore justified to use a grey atmosphere for the comparison calculations in Sect. \ref{sect:benchmarking}.}

In the limit of core-dominated planets with very low $\menve$, where the luminosity is dominated by the core cooling, and in the approximation of a constant radius, the analytical model of \cite{ginzburgsari2016} predicts $\text{L} \propto t^{-4/3}$ for a constant mass. However, the $\menve$ even of the 5 $\mearth$ planet in this work is sufficiently massive so that the radius change cannot be neglected, and numerically a decrease rather like $t^{-1}$ is found for a fixed mass, even for the low-mass planets. This is similar to what was found in \citet{burrowsliebert1993}, and is valid for grey and non-grey atmospheres.  This can be understood from the fact that at the radiative-convective boundary (RCB), the bottleneck for the transport of the luminosity occurs, independent of the atmospheric model, at the high optical depth of the RCB  \citep{arrasbildsten2006,leechiang2018}. 
Radiative transport occurs there by diffusion, meaning that only the Rosseland mean opacity matters, and not the specific wavelength-dependent opacity of the different atmospheric models.

In the top right panel in Fig. \ref{fig:resultssolarclear}, the decline of the specific entropy in the inner convective zone with time is shown. The entropy is a good measure of the total gravothermal energy of a planet because it contains both the inner energy content and the gravitational energy (volume work).  The entropy contained in our simulated planets ranges from 7.6 to 9.5 $\kB \, {baryon}^{-1}$ at young ages and from 7.0 to 8.7 $\kB \,  {baryon}^{-1}$ at 10 Myr. We have fixed the post formation luminosity. This means that the post formation entropy varies for a given mass depending on the atmospheric model \citep{marleaucumming2014}.

We also studied the effective temperature. The effective temperature varies slowly until about one Kelvin-Helmholtz timescale ($\tau_{KH}$) has passed. The initial $\tau_{KH}=E/L$ ranges from 8.96 Myr to 9.62 Myr for the 318 $\mearth$ , to 13.50 Myr to 14.52 Myr for the 5 $\mearth$ planet, depending on the atmosphere model. For example, the effective temperature of the 100~$\mearth$ planet decreases from 565 to 603 K at 0.11 Myr, to 107 to 116 K  at 1 Gyr depending on the atmospheric model.   {The difference between the temperatures from this work and from \cite{baraffechabrier2003} correspond to the differences in luminosity that were discussed before.}

The  bottom right panel in Fig. \ref{fig:resultssolarclear} shows the temporal change of the radius. It corresponds to $\tau=2/3$ for the Eddington grey atmosphere and to the coupling pressure (50 bar) for the other atmosphere grids.  {The radii from \cite{baraffechabrier2003} are constantly bigger than the radii from our simulation of the planets with an \texttt{AMES-Cond} atmospheric grid. When simulating the planets with a luminosity as was chosen in \cite{baraffechabrier2003}, and assuming a core mass of  0.5 $\mearth$ to mimic the settings in \cite{baraffechabrier2003}, we find a much better agreement in the radii. For example, at 1 Myr the radius from this work would then be 0.4\% bigger (1.9\% smaller) than the \cite{baraffechabrier2008} radius for the 1 (2) $\mj$ planet.}  We see that initially there is a non-monotonic relationship between mass and radius, with the biggest radius occurring for the 20 $\mearth$ planet with the \texttt{AMES-Cond} atmosphere. This is a consequence of the initial conditions ($L_{\text{pf}}$ and $\menve$ versus $\mcore$), which predicts for the 20 $\mearth$ planet a rather high envelope mass fraction  {relative to the total mass of the planet}.  {At later times, the radii computed with different atmosphere models converge, the radius also increases monotonically with mass, even though there is a spread of up to 0.9~$\rj$ (for the 20 $\mearth$) at early times.}

It is important to note here that the \texttt{AMES-Cond} grid was not designed for such low-mass planets. To give an overview, Fig. \ref{fig:loggTeff} illustrates the cooling of the planets in the  $\log g$--$T_{\rm eff}$ plane. The top left panel shows the cooling in the \texttt{AMES-Cond} grid. Thick lines indicate that the evolution path is inside the \texttt{AMES-Cond} grid, which starts at logg = 2.5 and an effective temperature of 100 K. As can be seen, the 636 $\mearth$ planet is evolving on the \texttt{AMES-Cond} grid, whereas all the others fall partially off the grid due to either too small surface gravities at early times or too low temperatures at later times. This means that all \texttt{AMES-Cond} results for planets below 50 $\mearth$ must be taken with caution. For the \texttt{petitCODE} and the \texttt{HELIOS} grid, the coverage in $\log g$--$T_{\rm eff}$ is in contrast good, initially, when $T_{\rm eff}$ is  {larger} than 150 K (\texttt{petitCODE}) or 100 K (\texttt{HELIOS}). These limits are shown in the $T_{\rm eff}$ panel. Again, once the planets are below these temperatures, caution must be used when employing the cooling tracks.

\subsection{Non-solar or cloudy atmospheric models}\label{subsect:nonsolarcloudy}

\begin{figure*}
\begin{center}
\begin{minipage}[t]{0.49\textwidth}
\centering\includegraphics[width=1\linewidth]{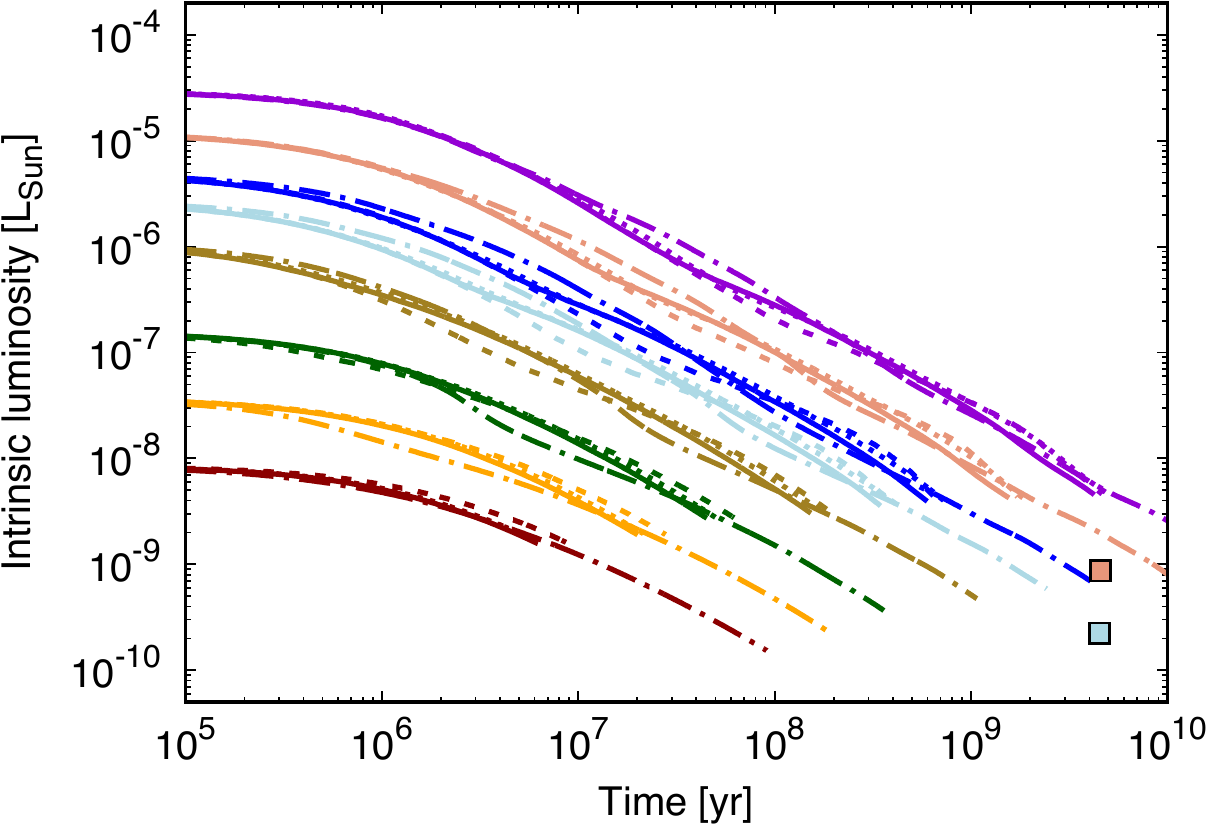}
\end{minipage}
\begin{minipage}[t]{0.49\textwidth}
\centering\includegraphics[width=1\linewidth]{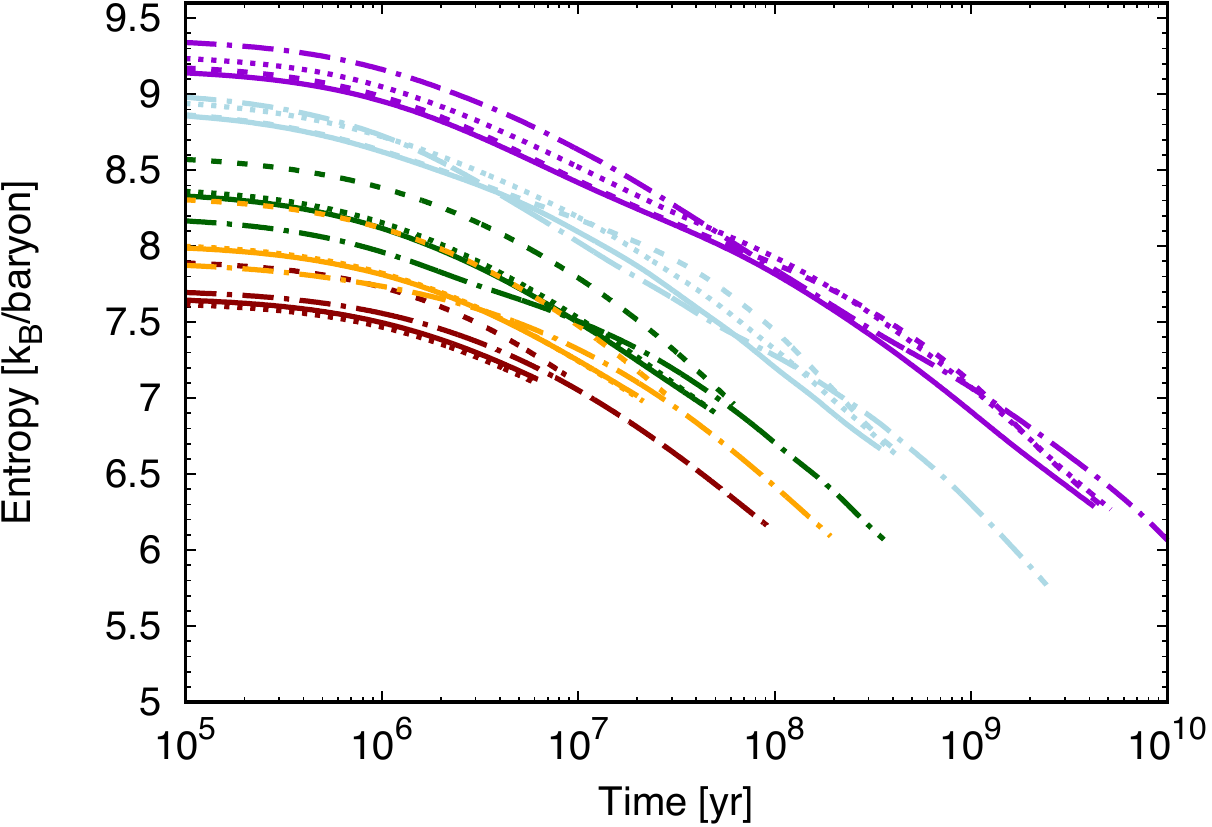}
\end{minipage}
\begin{minipage}[t]{0.49\textwidth}
\centering\includegraphics[width=1\linewidth]{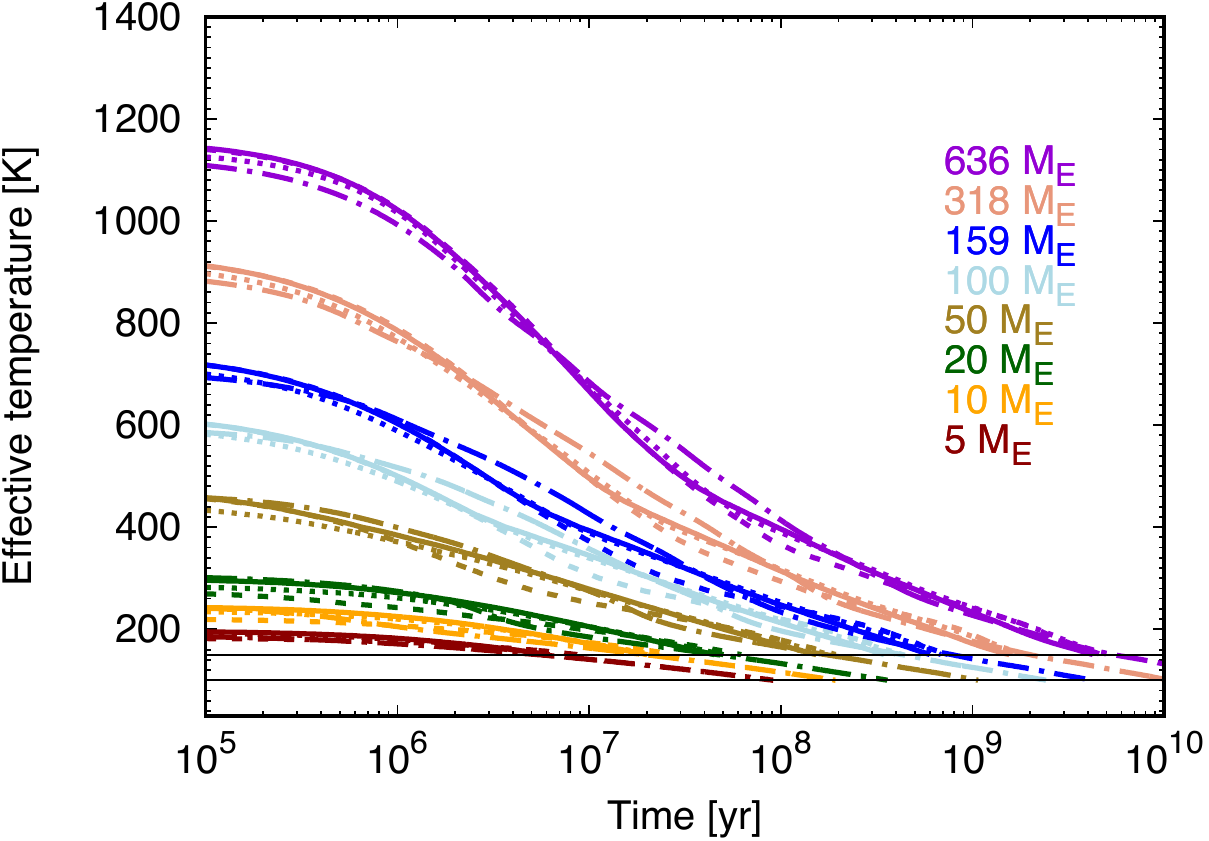}
\end{minipage}
\begin{minipage}[t]{0.49\textwidth}
\centering\includegraphics[width=1\linewidth]{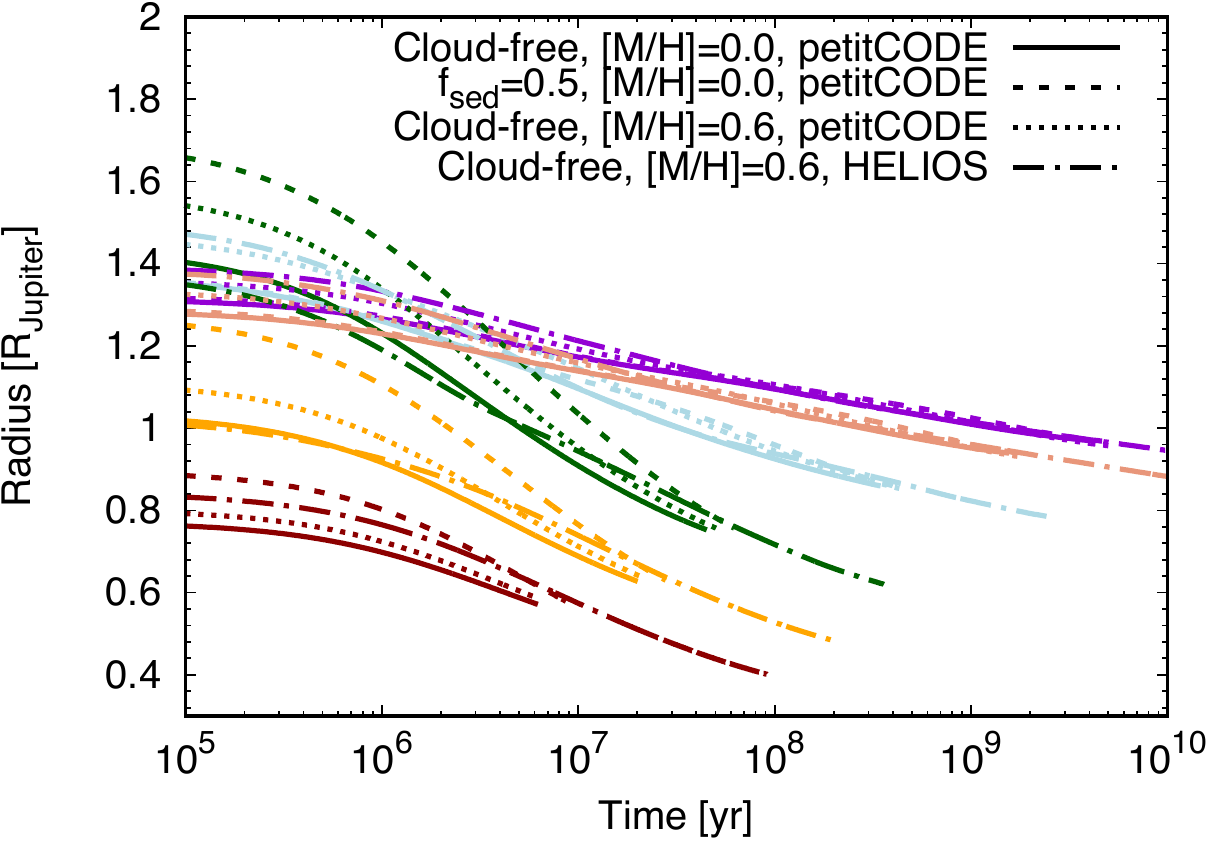}
\end{minipage}
\caption{Temporal evolution of fundamental properties of the simulated planets with different metallicities and cloudy atmospheres. The colours represent different masses, where the colour code is given in the  {bottom left} panel. The two  squares in the top left  {panel} indicate the luminosity of Jupiter and Saturn. {Solid lines} correspond to a cloud-free atmosphere with solar metallicity, {dashed lines} correspond to an atmosphere with clouds ($f_{\rm sed}$=0.5) and solar metallicity (as in Fig. \ref{fig:resultssolarclear}), {dotted lines} correspond to a cloud-free atmosphere with a metallicity [M/H]=0.6, all of them using the \texttt{petitCODE} model. {Dash-dotted lines} indicate a cloud-free atmosphere with a metallicity [M/H]=0.6 from the \texttt{HELIOS} grid. For better visibility, three { resp. two } planets  are not shown in the entropy {resp. }radius plot. The horizontal lines at 100 and 150 K in the $T_{\rm eff}$ plot show the lower limit of the \texttt{HELIOS} and \texttt{petitCODE} grid, respectively.  {The tracks are shown as long as the planet is evolving on the atmospheric grid.} }\label{fig:resultsnonsolarcloudy}
\end{center}
\end{figure*}

 {In contrast with} the former section, we now study the influence of non-solar metallicities and clouds on the evolution of the planets.
In Fig. \ref{fig:resultsnonsolarcloudy} in the top left panel, the evolution of the intrinsic luminosity from 0.1~Myr to 10~Gyr for planets from 636~$\mearth$ (2 $\mj$) down to 5~$\mearth$ is shown; again the lower mass planets are only shown as long as they are evolving not too far from the boundary of the atmospheric grid. The colour code is given in the top left panel. These evolution calculations were done with the \texttt{petitCODE} and \texttt{HELIOS} grids described in Sect. \ref{sect:atmgrids}. As a baseline model the evolution of the planets with a cloud-free solar metallicity atmosphere in the \texttt{petitCODE} grid is shown with a solid line; these are the same lines as in Fig. \ref{fig:resultssolarclear}.  We note that a $f_{\rm sed}$=0.5 is probably not realistic for the coldest planets studied here ($T_{\rm int}\lesssim 300-400~$~K), because the cloud species implemented in our simulations (Na$_2$S and KCl) form deep in the atmosphere in these cases. A low $f_{\rm sed}$ potentially mixes the clouds to locations too high up in the atmosphere. However, the evolution with a $f_{\rm sed}$=3.0 was also calculated in the \texttt{petitCODE} grid.  Since the difference to the cloud-free evolution is not visible on the scale chosen here (the cloud-free luminosity is less than 1\% fainter at 5 Myr than the luminosity calculated with an atmosphere with $f_{\rm sed}$=3.0), it is not shown. This is not surprising, as the high $f_{\rm sed}$ corresponds to a strong cloud settling or weak cloudiness.  {As mentioned before, at such low temperatures a more realistic choice for the cloud  {description} would have been to also include the effect of water clouds.} Finally, the evolution in the cloud-free atmosphere with a metallicity of [M/H]=0.6 in the \texttt{HELIOS} grid is given as a dash-dotted line.

 {In the top left panel of Fig. \ref{fig:resultsnonsolarcloudy}, the change of the bolometric luminosity over time is shown. For example, at 16 Myr the cloud-free \texttt{HELIOS} atmosphere with  [M/H]=0.6 predicts a 1.11 times brighter luminosity than the clear solar metallicity \texttt{petitCODE} atmosphere, whereas the cloudy \texttt{petitCODE} atmosphere predicts a 0.15 times fainter luminosity. }  {Analysing the figure in more detail, the  evolution of the bolometric luminosity of the planets shows average deviations (calculated as explained in Sect. \ref{subsect:solarclear}) of about 6\% for metallicities of 0.6 instead of 0.0. This is smaller than the deviations seen in Sect. \ref{subsect:nonsolarcloudy} where we compared different clear atmospheric models with solar composition, and thus  smaller than the error bars in current luminosity measurements. We thus see a non-negligible, but also not very large effect of the atmospheric model on the bolometric luminosity of the planet. We remind the reader, however, that Fig. \ref{fig:resultsnonsolarcloudy} only shows a maximum enrichment of [M/H]=0.6. For the maximum enrichment that we simulated ([M/H]=1.2), the deviation relative to the solar case is around 12\% on average (for the clear case).  For f$_{\rm sed}$=0.5, the differences to the clear solar metallicity case are similar as those between clear solar metallicity and the [M/H]=0.6 case. We caution that we have not investigated systematically the consequences of all  cloud parameters for the bolometric luminosity.  }  Again the result from the literature (e.g. \citealt{burrowsliebert1993, burrowshubbard2001, baraffechabrier2002, saumonmarley2008, dupuyliu2015}) is reproduced, as in the former section, that the choice of the atmospheric model has only a  {limited} impact on the evolution of the bolometric luminosity, at least within the models studied here.

The change of the entropy over time is given in the top right panel in Fig. \ref{fig:resultsnonsolarcloudy}. For clarity the evolution of the 318, 159, and 50 $\mearth$ mass planet is not shown. It is again important to keep in mind that the simulations start always with the same luminosity, which corresponds  to a different entropy at the beginning depending on the atmospheric model. The biggest post formation entropy spread occurs for the three lightest planets with a difference of up to 0.4 $\kB\, {baryon}^{-1}$.  {The discrepancy between the atmospheric models can be attributed to a large extent to differences in the employed opacity sources. Since \texttt{petitCODE}       and \texttt{HELIOS} use different line lists for some of the absorbing species, such as H$_2$O, NH$_3$, H$_2$S, and the alkali metals, and use different scatterers, their calculations may result in somewhat deviating atmospheric temperature profiles. This directly impacts the coupling to the deep convective adiabat and the corresponding entropy value. With increasing metallicity this discrepancy  {is exacerbated}  as the relative amount of  {absorbing} gases increases.}

The bottom left panel shows the evolution of the temperature of the planets and mirrors the luminosity evolution in the top left panel. As an example, the 100 $\mearth$  planet has a temperature of 585 to 605 K depending on the atmosphere model at 0.1 Myr and cools down to 107 to 125 K at 1 Gyr, respectively. 

Finally, the change of the radius over time can be seen in the bottom right panel. The spread in radius at early times can be up to 0.2 $\rj$ for the 10 $\mearth$ mass planet, up to 0.3 $\rj$ for the 20 $\mearth$ mass planet, and up to 0.1 $\rj$ for the 100 $\mearth$ planet. However, they all approximately converge at later times, qualitatively similar to cloud-free, solar metallicity atmospheres. 

\subsection{Varying the post formation luminosity}\label{subsect:varyLpf}

\begin{figure*}[h]
\begin{center}
\begin{minipage}[t]{0.48\textwidth}
\centering\includegraphics[width=1\linewidth]{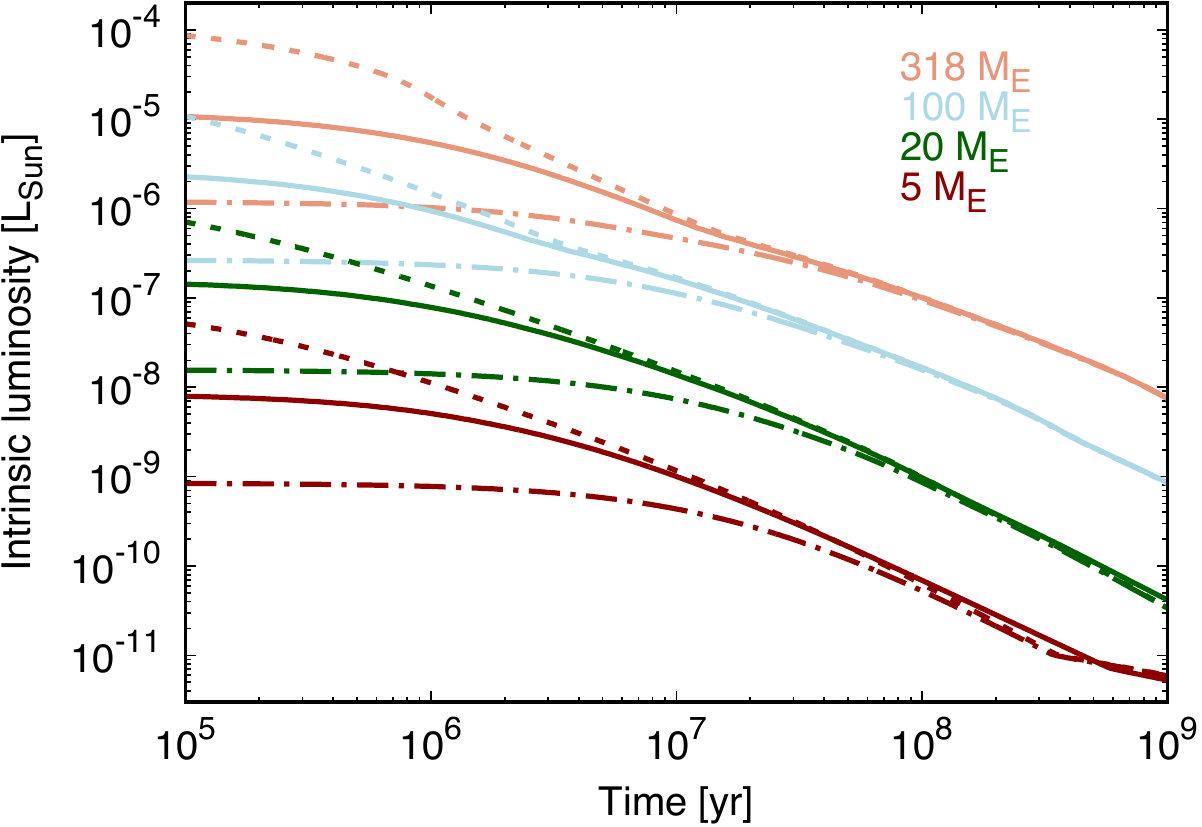}
\end{minipage}
\begin{minipage}[t]{0.48\textwidth}
\end{minipage}
\begin{minipage}[t]{0.48\textwidth}
\centering\includegraphics[width=1\linewidth]{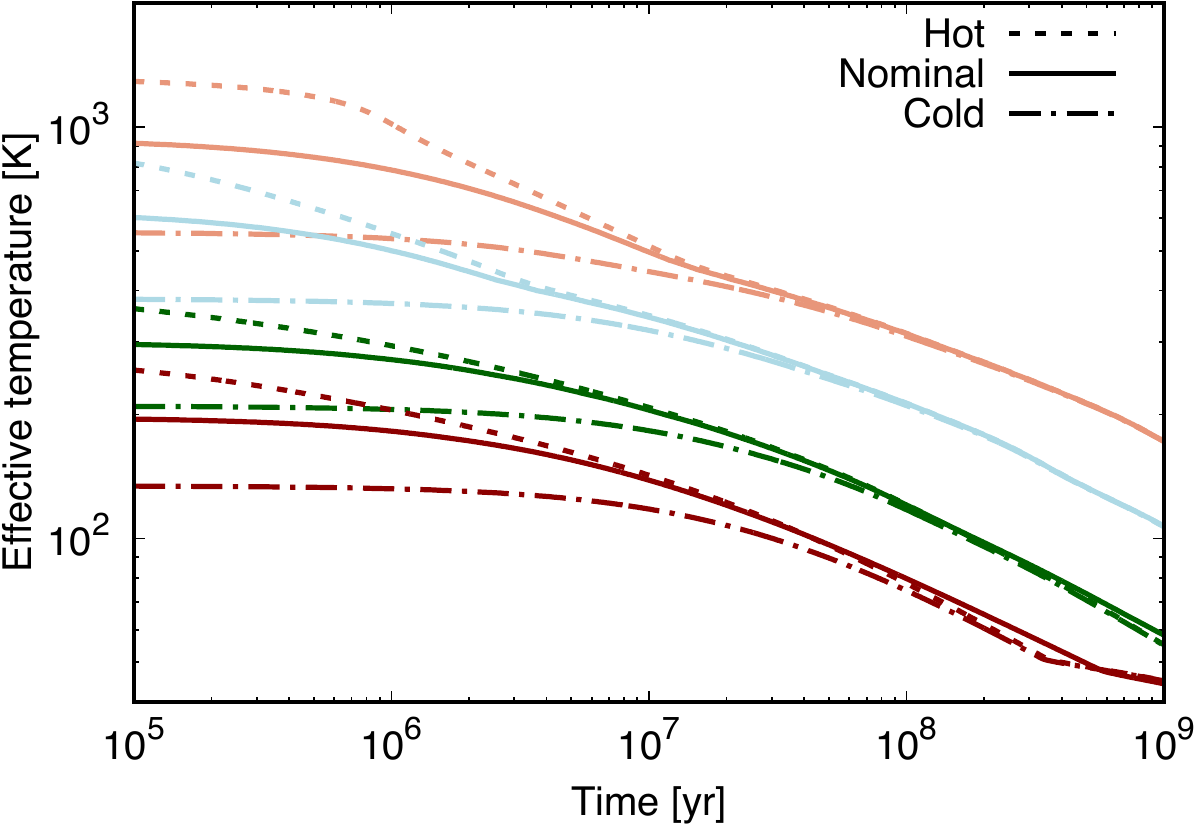}
\end{minipage}
\begin{minipage}[t]{0.48\textwidth}
\end{minipage}
\caption{Influence of varied  {post formation luminosity (}$L_{\text{pf}}$ {)} on the intrinsic luminosity and temperature for four different masses. The colour code is given in the  left  panel, the line style is given in the right panel. All  cooling curves are calculated with a cloud-free atmosphere with  solar metallicity in the \texttt{petitCODE} grid. {Solid lines} show the evolution with the nominal $L_{\text{pf}}$ of Table \ref{tab:fitexamples}. {Dashed lines} show the evolution with a ten times bigger $L_{\text{pf}}$ (hot scenario). {Dash-dotted lines} show the evolution with a ten times fainter $L_{\text{pf}}$ (cold scenario).} \label{fig:varyentr}
\end{center}
\end{figure*}

Different post formation luminosity or entropy ($S_{\text{pf}}$ or $L_{\text{pf}}$) can be used to represent different planet formation scenarios \citep{marleyfortney2007,mordasinimarleau2017}. 
 {The post formation Kelvin-Helmholtz timescale is given as}
\begin{equation}
  {{\tau}_{ {KH,pf}} {\approx \dfrac{ {G} \mtot^2}{R_{\rm  {pf}}L_{\rm  {pf}}}}}, 
\end{equation} 
 {where G is the gravitational constant, $\mtot$ is the total planetary mass, and $R_{\rm pf}$ and $L_{\rm pf}$ are the post formation radius and luminosity of the planet.}
At late times, much longer than $\boldsymbol{\tau_{\text{ {KH,pf}}}}$, the influence of the initial condition has disappeared, therefore the choice of the post formation entropy ($S_{\text{pf}}$) is considered to have only a minor influence on the late planetary evolution. However, since $\tau_{\text{KH,pf}}\propto 1/L_{\text{pf}}$, a low post formation luminosity (a very cold start) can influence the $L(t)$ over extended and observationally relevant times of up to $\sim1$ Gyr for giant planets \citep{marleyfortney2007}. In contrast, a planet with a "hotter start" will converge on the same evolutionary track as a "hot start", again because $\tau_{\text{KH,pf}}\propto 1/L_{\text{pf}}$. Not surprisingly, very low $S_{\text{pf}}$  {(implying low} $L_{\text{pf}}$ {)} can have a strong impact on the predicted magnitudes of giant planets \citep{fortneylodders2008,spiegelburrows2012}, with magnitudes that are 1-6 mag fainter than standard hot start models. The very low post formation luminosities found in the original core accretion model of \cite{marleyfortney2007} were originally thought to be a diagnostic distinguishing core accretion from other formation pathways (disc instability, turbulent fragmentation of molecular clouds). However, later core accretion models showed that core accretion can also lead to warm and even hot starts depending on the mass of the planet's core \citep{mordasini2013,marleauklahr2017, berardocumming2017}. This means that the brightness is likely not a property distinguishing strictly the formation modes, and recent population syntheses based on the core accretion paradigm \citep{mordasinimarleau2017} instead find warm starts. Therefore, because we want to make predictions about the detectability of young planets, it is important to quantify the impact of the $S_{\text{pf}}$ and also to consider for how long the impact on the evolution  of the planets remains. To study the impact of different $S_{\text{pf}}$,  the cooling of planets with a $L_{\text{pf}}$ ten times brighter (hot scenario) or fainter (cold scenario) relative to what is noted in Table \ref{tab:fitexamples} (nominal) was simulated. Such a spread in luminosity is suggested by population synthesis calculation of planet formation and can  be seen in Fig. \ref{fig:initialcond}. The resulting spread in $S_{\text{pf}}$ reaches from 8.4 to 10.5 $\kB \, {baryon}^{-1}$ (difference of 2.1 $\kB \, {baryon}^{-1}$) for the 318 $\mearth$ planet, and from 6.9 to 8.5 $\kB \, {baryon}^{-1}$ for the 5 $\mearth$ planet.

Figure \ref{fig:varyentr} shows the cooling curves for four different planetary masses (5, 20, 100, and 318 $\mearth$) and various $L_{\text{pf}}$. The evolution is calculated with a cloud-free atmosphere with solar metallicity in the \texttt{petitCODE} grid. The solid lines represent the nominal scenario and are therefore the same as in Figs. \ref{fig:resultsnonsolarcloudy} and \ref{fig:resultssolarclear}. The cooling curves with a ten times higher $L_{\text{pf}}$ are shown as dashed lines, those with a ten times lower $L_{\text{pf}}$ are given as dash-dotted lines.

The left panel in Fig. \ref{fig:varyentr} shows the evolution of the intrinsic luminosity.  It can be seen that the cooling curves for different $L_{\text{pf}}$ converge as expected (e.g. \citealt{marleyfortney2007,marleaucumming2014}). The planets of the hot scenario have a Kelvin-Helmholtz-time ($\tau_{\text{KH,pf}}$) of $\sim$1~Myr, the ones with nominal $L_{\text{pf}}$ as shown in Table \ref{tab:fitexamples} have a $\tau_{\text{KH,pf}}$ of $\sim$10~Myr and finally, the planets in the cold scenario have a $\tau_{\text{KH,pf}}$ of $\sim$100~Myr. This means that the post formation state (e.g. the initial condition) influences the planets' properties during about $\tau_{\text{KH,pf}}\sim$1,10, and 100 Myr after formation for cold, nominal, and hot scenario. These phases can be seen in Fig. \ref{fig:varyentr} as those parts of the lines that do not yet follow a $L\propto t^{-1}$ behaviour, but which are more horizontal.  As an example, at 1 Myr, the hot scenario model for the 100 $\mearth$ planet is 1.6 times brighter than the nominal evolution, and the cold scenario model is five times fainter than the nominal model.

In \citet{marleyfortney2007}, $L_{\text{pf}}$ is independent of the planetary mass because a larger mass fraction for the more massive planets went through an entropy-reducing shock, and therefore $\tau_{\text{KH,pf}}\propto \mtot^{2}$. Here we find that $\tau_{\text{KH,pf}}$ is  {approximately} independent of $\mtot$ (at least for $\mtot$~\textless~100~$\mearth$), as  {circa} $L_{\text{pf}}\propto\mtot^{2}$, and the planetary radius{, which enters linearly into $\tau_{\text{KH,pf}}$, only changes by factors of 2-3 between hot and cold scenarios.}

We note that because of the non-uniqueness of the $M-L_{\text{pf}}$ relation, there is a mass-luminosity dependency during $\sim\tau_{\text{KH,pf}}$, as already widely discussed in the literature in the context of giant planets (e.g. \cite{mordasini2013}). For example, if the age of a planet could be estimated to be 1 Myr and the planet has a brightness of $2\times10^{-7}\lsun$, then this could correspond to either a 20 $\mearth$ mass planet with a high post formation luminosity, or to a 100 $\mearth$ mass planet with a low post formation luminosity. With a further mass measurement through for example the astrometric or radial-velocity method, it is  possible to disentangle this mass-luminosity ambiguity. In principle, if the planetary temperatures can be measured precisely enough, it would also be possible to reduce the degeneracy. This should be possible for differentiating between the hot and cold scenario for planets that are heavier than 2 $\mj$ \citep{samlandmolliere2017}. Fortunately, at 10 Myr (a more likely observable age), none of the hot versus cold  {scenario} lines representing 5, 20, 100, and 318 $\mearth$ from Fig. \ref{fig:varyentr} overlap any more, so that at least a very rough mass estimate should be possible. 

The right panel in Fig. \ref{fig:varyentr} shows the evolution of the temperature over time. At 1 Myr the temperature for the 318 $\mearth$ planet in the hot, nominal, and cold scenario is 1014 K, 781 K, and 536 K, respectively. For comparison, \cite{spiegelburrows2012} have at 1 Myr a $T_{\text{eff}}$ of $\sim$850 and $\sim$550 K for their hot and cold case for a 318 $\mearth$ planet, which is similar to our result.

If planet masses and luminosities could be determined observationally at early times ($<\tau_{\text{KH,pf}}$), important constraints about hot versus cold scenarios could be made for planet formation theory, regarding for example the heating by giant impacts (e.g. \citet{anicalibert2007}), the efficiency of heat transport in the interior \citep{nettelmannhelled2013}, or the magnitude and timing of heating by planetesimal accretion \citet{mordasini2013}.

\subsection{JWST magnitudes and fluxes}\label{sect:JWSTmagnitudesandfluxes}

\begin{figure*}
\begin{center}
\begin{minipage}[t]{0.48\textwidth}
\centering\includegraphics[width=1\linewidth]{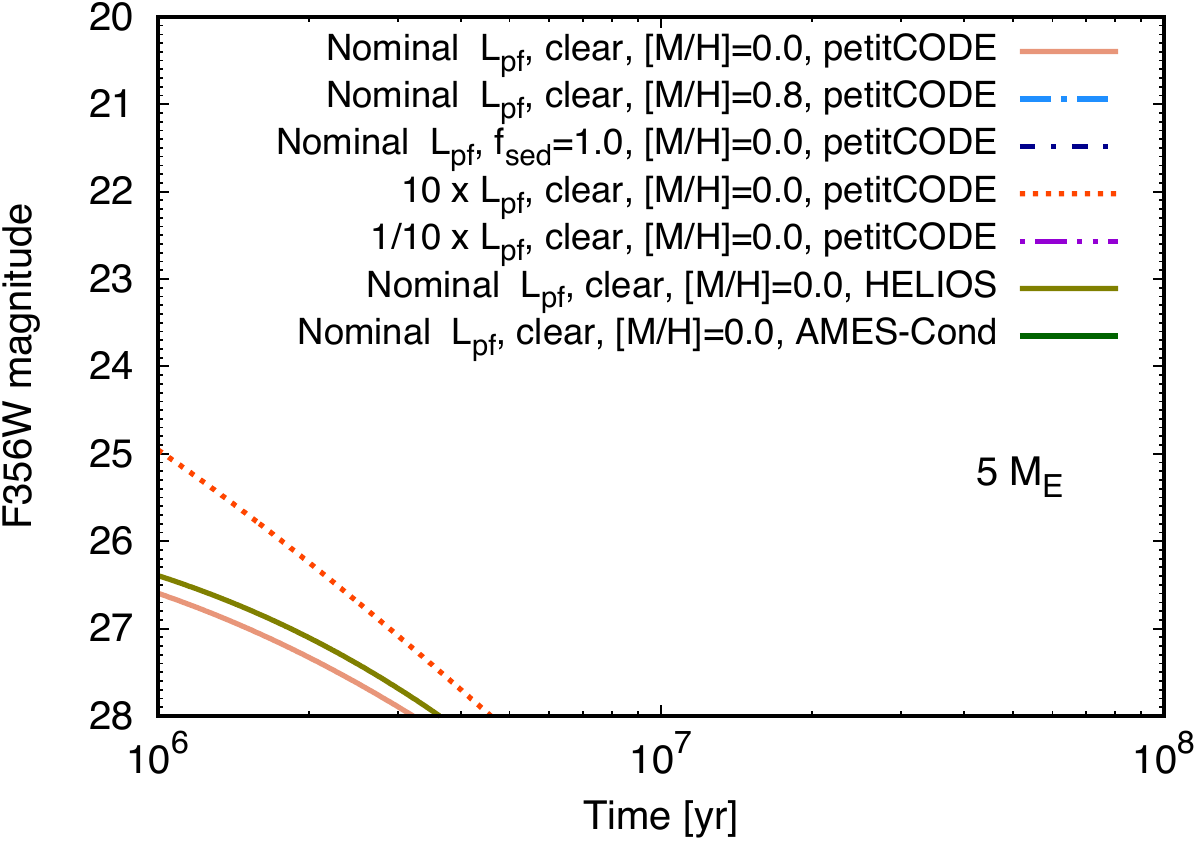}
\end{minipage}
\begin{minipage}[t]{0.48\textwidth}
\centering\includegraphics[width=1\linewidth]{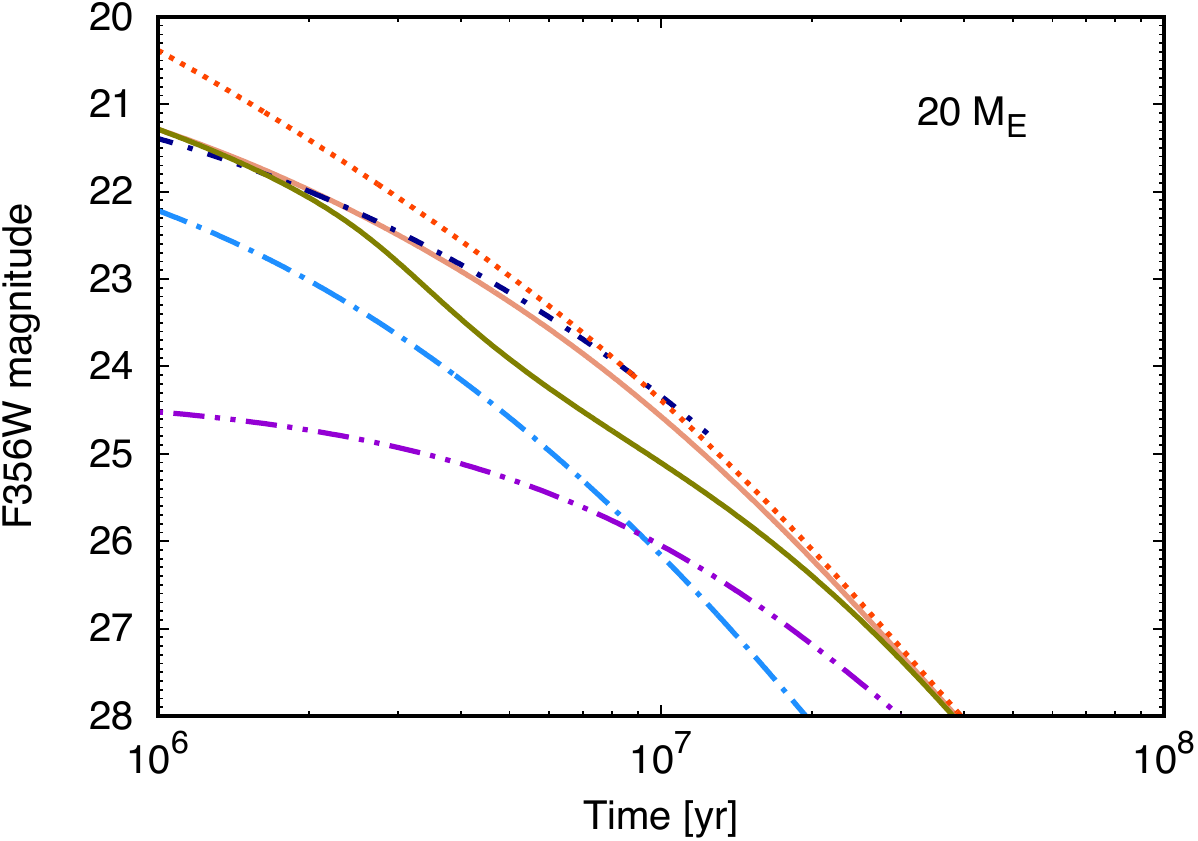}
\end{minipage}
\begin{minipage}[t]{0.48\textwidth}
\centering\includegraphics[width=1\linewidth]{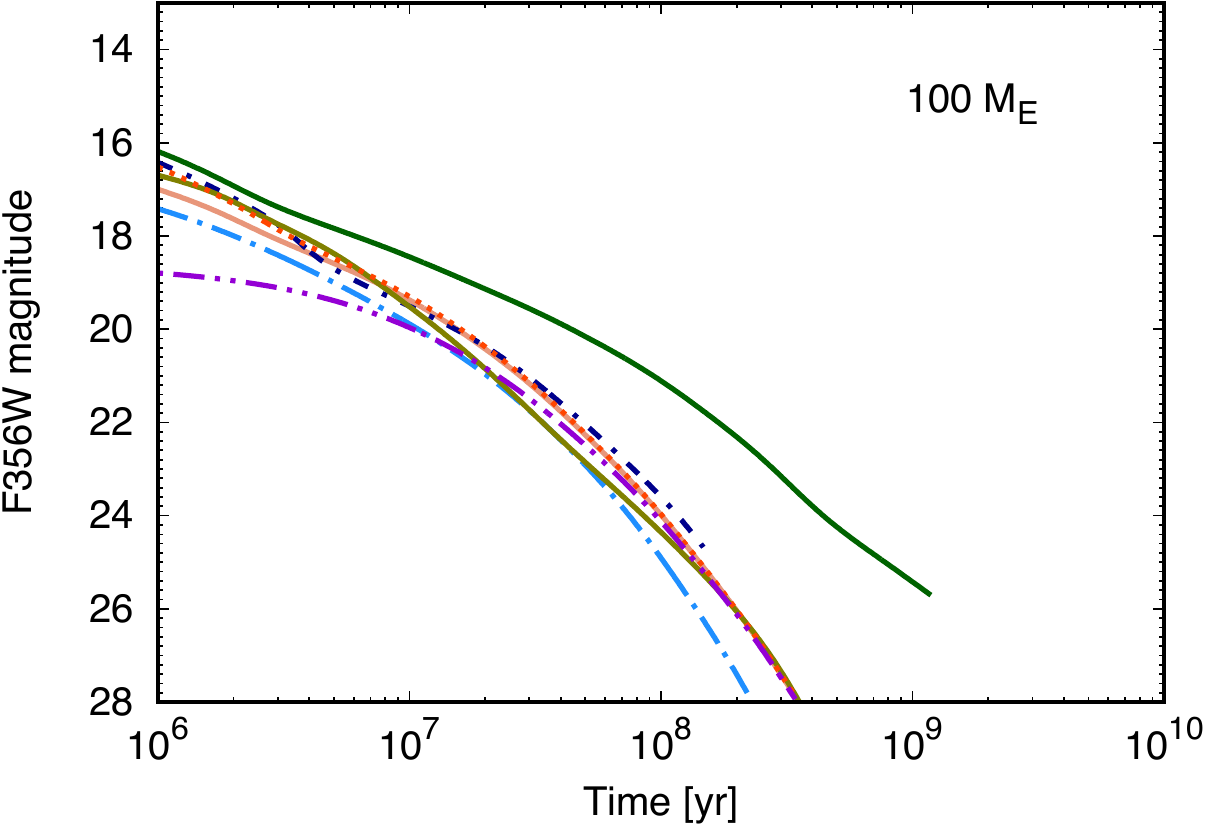}
\end{minipage}
\begin{minipage}[t]{0.48\textwidth}
\centering\includegraphics[width=1\linewidth]{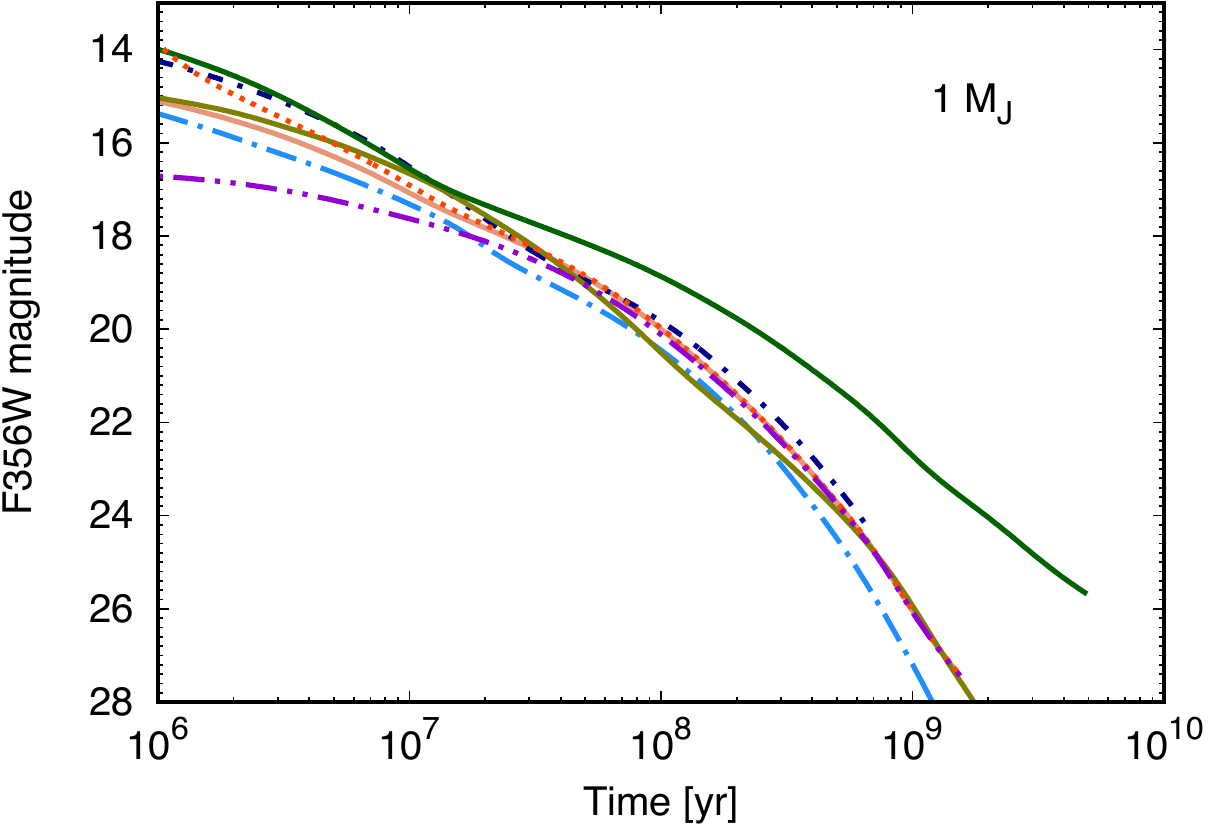}
\end{minipage}
\caption{Absolute magnitudes in{ {JWST}} filter band F356W ({JWST/NIRCam}) for four different planetary masses and various atmosphere parameters and grids  as well as a variation on post formation luminosity with a clear and solar metallicity \texttt{petitCODE} grid. The magnitudes are shown as long as they are in the atmosphere grid{, and for the cloudy models  {($f_{\rm sed}$=1.0)} as long as they are above 200 K}.} \label{fig:JWSTabsmagsF356}
\end{center}
\end{figure*}

\begin{figure*}
\begin{center}
\begin{minipage}[t]{0.48\textwidth}
\centering\includegraphics[width=1\linewidth]{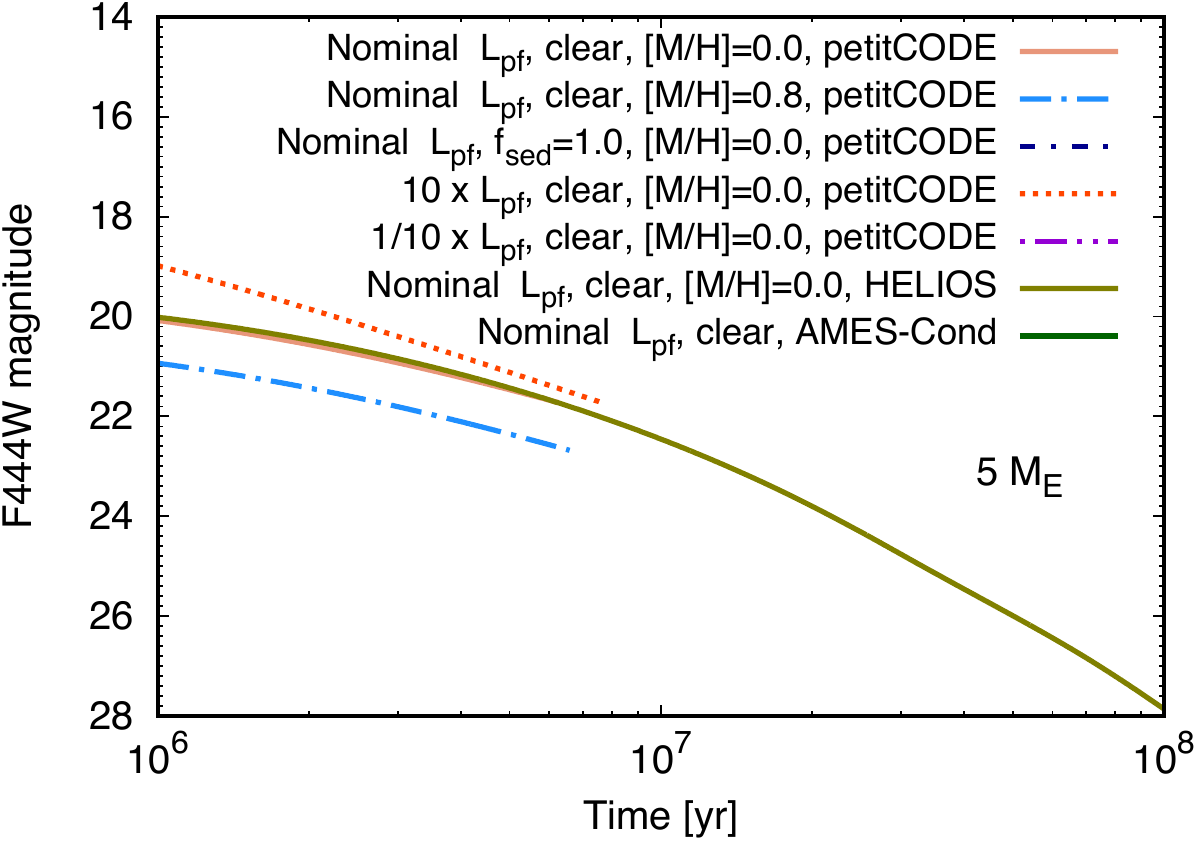}
\end{minipage}
\begin{minipage}[t]{0.48\textwidth}
\centering\includegraphics[width=1\linewidth]{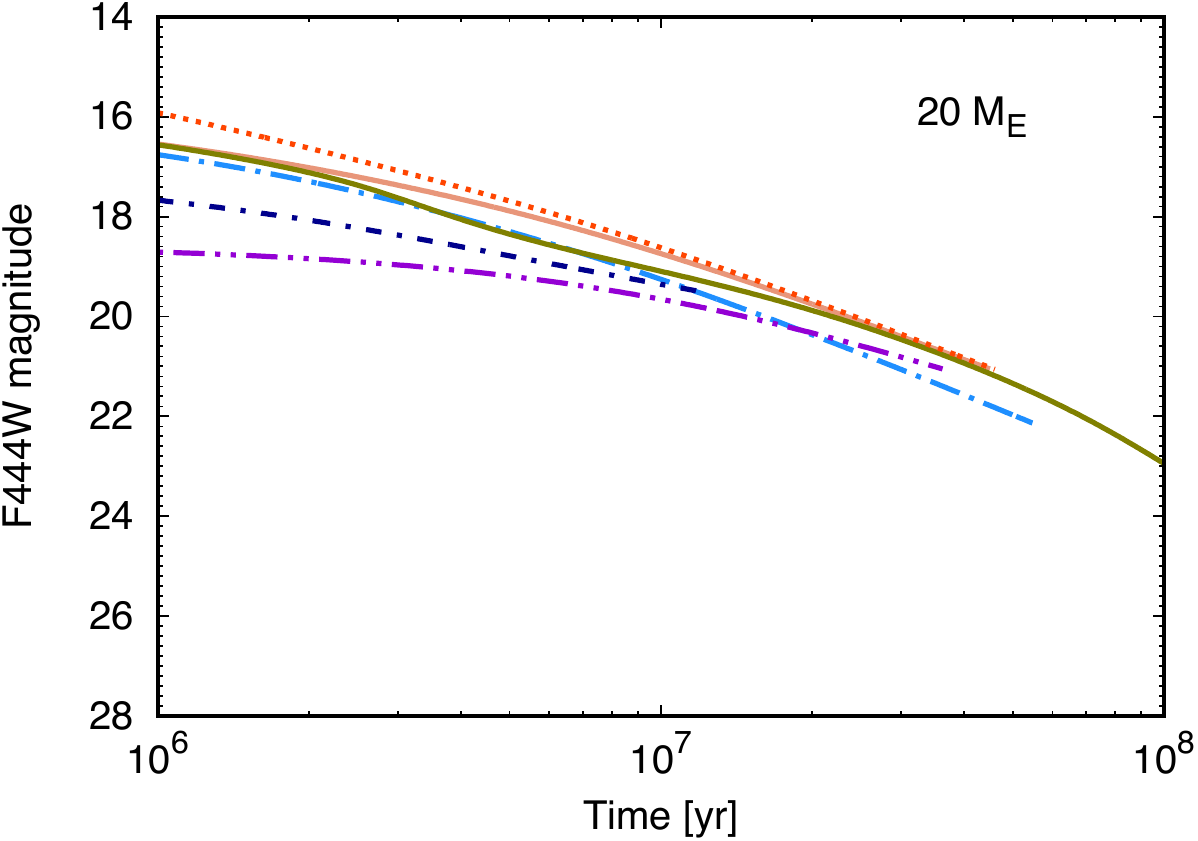}
\end{minipage}
\begin{minipage}[t]{0.48\textwidth}
\centering\includegraphics[width=1\linewidth]{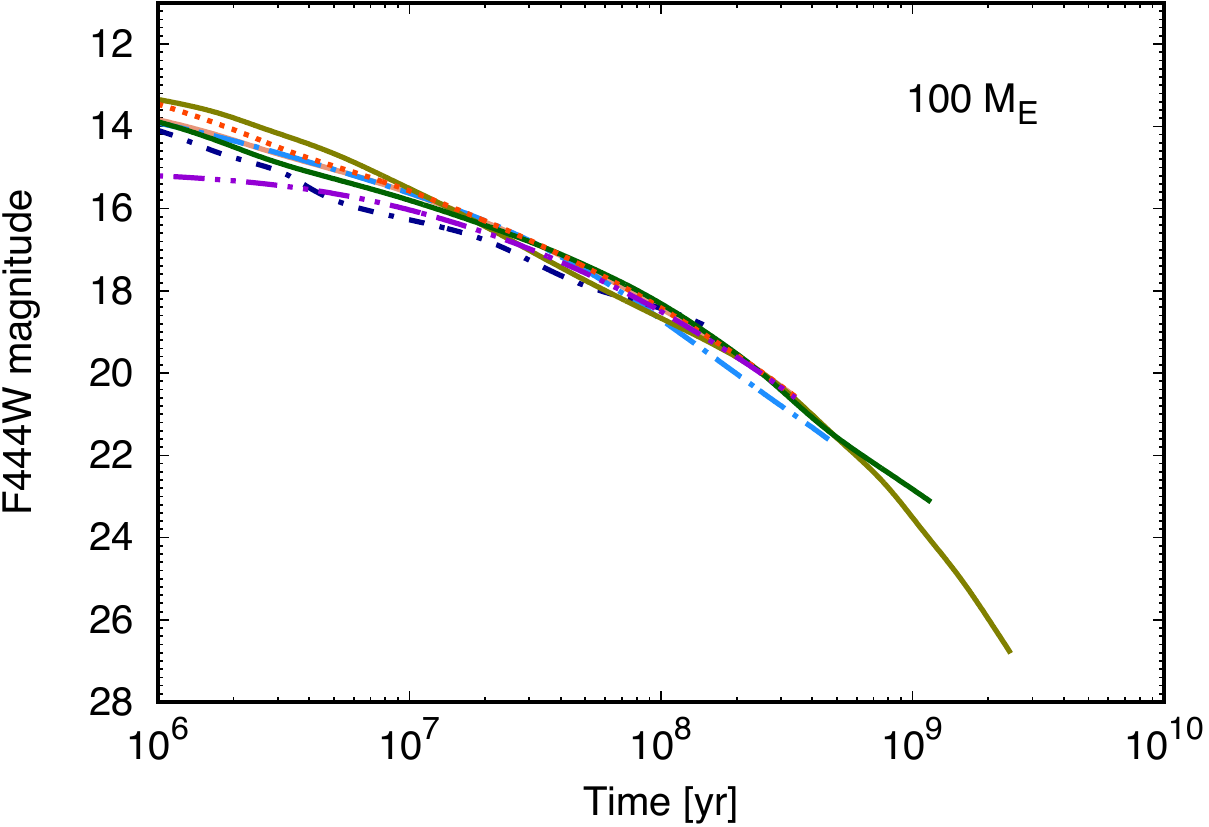}
\end{minipage}
\begin{minipage}[t]{0.48\textwidth}
\centering\includegraphics[width=1\linewidth]{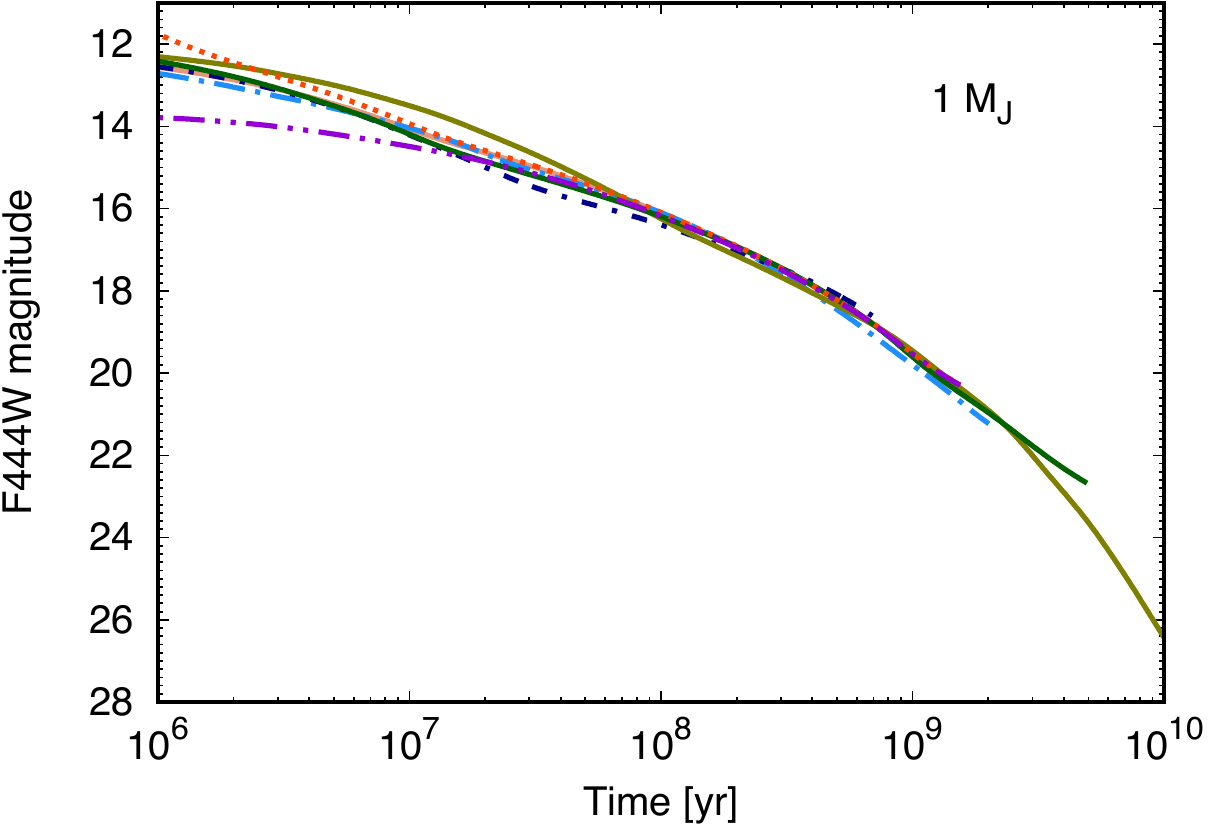}
\end{minipage}
\caption{Absolute magnitudes in{ {JWST}} filter band F444W ({JWST/NIRCam}) for four different planetary masses and various atmosphere parameters and grids  as well as a variation on post formation luminosity with a clear and solar metallicity \texttt{petitCODE} grid. The magnitudes are shown as long as they are in the atmosphere grid{, and for the cloudy models  {($f_{\rm sed}$=1.0)} as long as they are above 200 K}. } \label{fig:JWSTabsmagsF444}
\end{center}
\end{figure*}

Magnitudes for the most relevant { {JWST}} filters for exoplanet imaging, namely,
 {JWST/MIRI} (F560W, F770W, F1000W, F1280W, F1500W, F1800W, F2100W, F2550W) and 
 {JWST/NIRCam} (F115W, F150W, F200W, F277W, F356W, F444W) were calculated. The wide (W) filters were chosen as they are considered the standard imaging filters and also because we quantify the detection of planets in the background limited regime and not necessarily in the contrast limit, where certain coronagraph-filter combinations may be the preferred choice. The zero points of the magnitudes were obtained with a Vega spectrum.\footnote{From  {\url{ftp://ftp.stsci.edu/cdbs/current_calspec} and
\url{http://www.stsci.edu/hst/observatory/crds/calspec.html} as alpha\_lyr\_stis\_003.txt, accessed in 2014. We provide the spectrum we used on \url{http://www.space.unibe.ch/research/research_groups/planets_in_time/numerical_data/index_eng.html}}.} To check our magnitude calculation, the spectra from F.~Allard's website\footnote{\url{https://phoenix.ens-lyon.fr/Grids/AMES-Cond/STRUCTURES/}.} were downloaded, convolved with the filter profiles, and the resulting  magnitudes were compared with the  magnitudes given on F. Allard’s website\footnote{\url{https://phoenix.ens-lyon.fr/Grids/AMES-Cond/COLORS/}.} as well. It was found that the magnitudes are the same. The magnitudes of the modelled planets were obtained by convolving their  {spectra} with the filter transmission profiles. For this we interpolated their spectra to the desired $\log g$--$T_{\rm eff}$ combination given the grid of atmospheres. The filter profiles used in this work are available together with the magnitudes for different planetary masses and ages, metallicities, and post formation luminosities for clear or cloudy atmospheres.  {To illuminate the effects of clouds, a $f_{\rm sed}$=1.0 was chosen.} We are considering  {the most important cloud species that occur at intermediate} temperatures (Na$_2$S and KCl,  {for} T~$\gtrsim$~400~K).  {Consequently, we show the evolution} {tracks with clouds only  {down} to 200 K (instead of 150 K as is the case for the clear \texttt{petitCODE} atmosphere magnitudes).} An example  {of the tables } is given in the appendix in Table \ref{tab:magexamplefixedmass} for a fixed planetary mass as a function of time, and in Table \ref{tab:magexamplefixedtime} for a fixed time for the eight masses used in this work; all the  tables can be found at the CDS and at:  {\url{http://www.space.unibe.ch/research/research_groups/planets_in_time/numerical_data/index_eng.html}}.

As an illustration, in Figs. \ref{fig:JWSTabsmagsF356} and\ref{fig:JWSTabsmagsF444}, the absolute magnitudes in the F356W filter (centred  at 3.45 $\mu m$) and the F444W filter (centred  at 4.44 $\mu m$) are shown for the 5, 20, 100, and 318 $\mearth$ mass planets and for different atmosphere grids with different parameters, such as metallicity and $f_{\rm sed}$, as well as for the variation in post formation luminosity ($L_{\text{pf}}$) introduced in Sect. \ref{subsect:nonsolarcloudy}. These two filters were chosen for reasons discussed below. 

The clear solar metallicity line in the \texttt{petitCODE} grid is   {sometimes hard to see due to the} other lines. For the case of a ten times lower $L_{\text{pf}}$, the 5 $\mearth$ mass planet  {is} too cool to still be on the atmosphere grid and hence these magnitudes were not calculated.   {Since the 5 and 20 $\mearth$ planets are evolving outside of the \texttt{AMES-Cond} grid, the magnitudes corresponding to this atmosphere are not shown for these two low-mass planets.} 

In principle, the \texttt{AMES-Cond}, the \texttt{HELIOS}, and the \texttt{petitCODE} magnitudes for a clear and solar metallicity atmosphere should be the same. However, because of different input line lists, this is not the case. We quantified this for the 159 $\mearth$ planet. In all the  {JWST/MIRI} filters we calculated, the magnitudes calculated with these atmospheres for the evolution of the 159 $\mearth$ planet differ at most by 0.5 to 2 magnitudes. In the  {JWST/NIRCam} filters we considered, the  maximum difference ranges from 0.3 magnitudes up to 6 magnitudes in the F227W and F115W filter at a few 100 Myr. In the F356W filter, which is shown in Fig. \ref{fig:JWSTabsmagsF356}, and chosen because the impact of the line lists is very strong, the  {heavier} planets are brightest when calculated with an \texttt{AMES-Cond} atmosphere.  {For example, the 100 $\mearth$ planet is 2.9 mag brighter at 100 Myr with an   \texttt{AMES-Cond} atmosphere than with a clear solar \texttt{petitCODE} atmosphere.}
  {A strong methane absorption feature is located at this wavelength}, which makes the corresponding flux emission very sensitive to the employed methane line list.
The \texttt{petitCODE} and \texttt{HELIOS} models use the recent EXOMOL line list for CH$_4$ \citep{yurchenkoatennyson2014} turning the atmosphere more opaque than assumed in the older \texttt{AMES-Cond} models before. Hence there is more flux passing through in the \texttt{AMES-Cond} spectra, which leads to a much brighter planet at shorter wavelengths and is most prominent in some filters at shorter wavelengths. For example in the F356W filter the different methane line lists  {can} have a higher impact than a higher post formation  luminosity. A higher $L_{\text{pf}}$ influences the magnitudes only early on, by about 1 mag  {for the 20 $  {\mearth}$ for example}. In contrast,  {the same planet} with a lower $L_{\text{pf}}$  {is} up to  {3.2} magnitudes fainter in the F356W band  {early on}. For higher planetary masses, when the atmosphere becomes warmer, the effect of the methane becomes smaller as the relative abundance of methane decreases and hence its spectral effect compared to water diminishes, even in the methane bands. The F444W (Fig. \ref{fig:JWSTabsmagsF444}) filter is, in contrast to the F345W, not sensitive to the methane abundance. Hence, all atmospheres lead to similar magnitudes. The largest impact is now  due to different $L_{\text{pf}}$.

Figure \ref{fig:JWSTsenslim} shows the temporal evolution of the blackbody and non-grey spectrum for four planetary masses. The colour code gives the objects' age. The spectra are calculated for a cloud-free, solar metallicity \texttt{petitCODE} atmosphere by interpolating to the required temperature and surface gravity and are thus given as long as the planet is evolving on the atmospheric grid. 
 {W}e over-plot the sensitivity limits for the  {JWST/NIRCam} instrument  {(grey dots) and} for the {JWST/MIRI} instrument  {(black dots).}\footnote{The sensitivity limits are taken from \url{https://jwst-docs.stsci.edu/display/JTI/NIRCam+Sensitivity}, Table 1 and \url{https://jwst-docs.stsci.edu/display/JTI/MIRI+Sensitivity}, Table 1,  respectively, and they correspond to a signal-to-noise of 10 for an integration time of 10$^4$ s (page accessed 24 May 2018).} These  are background limits and do not take  the final contrast performance of the instruments into account, which will only be known after commissioning of the high-contrast imaging modes and which will also depend on the selected targets and observing strategy. The prominent features that can be seen, for example at 1-2 $\mu m$ and the peak at  {4.7} $\mu m$ , originate from the water and methane opacities. Especially the cut-off at 1.6 $\mu m$ is typical for methane. At longer wavelengths, the calculated spectral emission resembles more closely the theoretical blackbody emission. However, for certain temperatures and surface gravities, the spectral flux at shorter wavelengths of 1-2~$\mu m$ can be up to  orders of magnitude higher than the theoretical blackbody flux. As an example, the 5 $\mearth$ object shows a blackbody flux at 1 Myr and at 1.6 $\mu m$ that is  {13} orders of magnitudes lower than the one from the calculated spectrum. This can be understood because at the corresponding low temperature of 182~K there are very few absorbers in the atmosphere; water is condensed to quite high pressures in the atmosphere (0.5~bar), and we do not model water clouds here. Therefore, the depth that is probed in the atmosphere is strongly influenced by collision induced absorption (CIA) in addition to the water and methane opacities. As can be seen in Fig. \ref{fig:JWSTsenslim}, when the atmosphere of the planet becomes warmer or the planet is more massive, this mechanism is diminished.

Figure \ref{fig:JWSTsenslim} also shows the unprecedented sensitivity of { {JWST}} at thermal infrared wavelengths. For young nearby stars, such as, members of the $\beta$ Pictoris moving group with an estimated average age and  distance of 23 Myr and 15 pc, respectively \citep{mamajek2016}, planets with masses below that of Neptune seem to be within reach at  separations from the star where background limited performance is  {achieved}. This is truly uncharted territory in comparison to what has been achievable up to now with exoplanet imaging in terms of mass limits \citep[see e.g.][ for a recent review]{bowler2016}.

\subsection{Predictions for space and ground-based observations}\label{predictionsforgroundbasedobservatories}

\begin{figure*}
\begin{center}
\begin{minipage}[t]{0.49\textwidth}
\centering\includegraphics[width=1\linewidth]{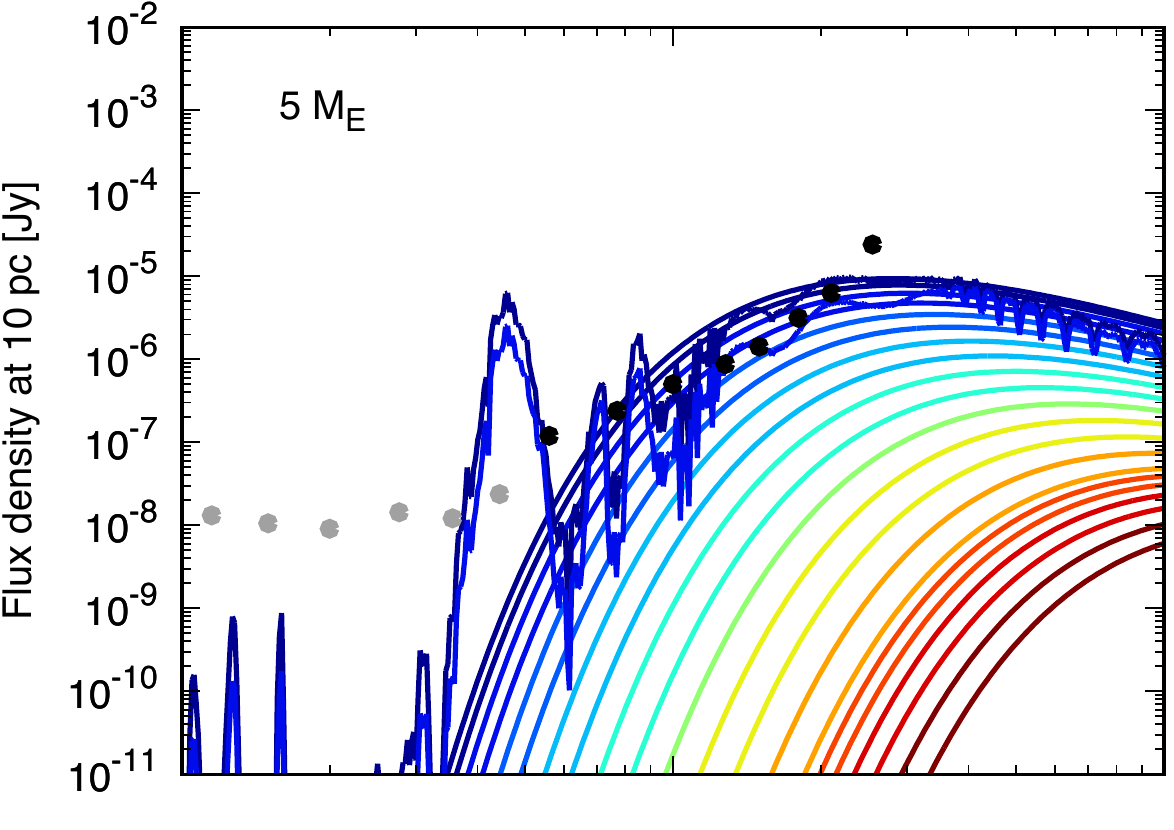}
\end{minipage}
\hspace{0.01\textwidth}
\begin{minipage}[t]{0.48\textwidth}
\centering\includegraphics[width=1\linewidth]{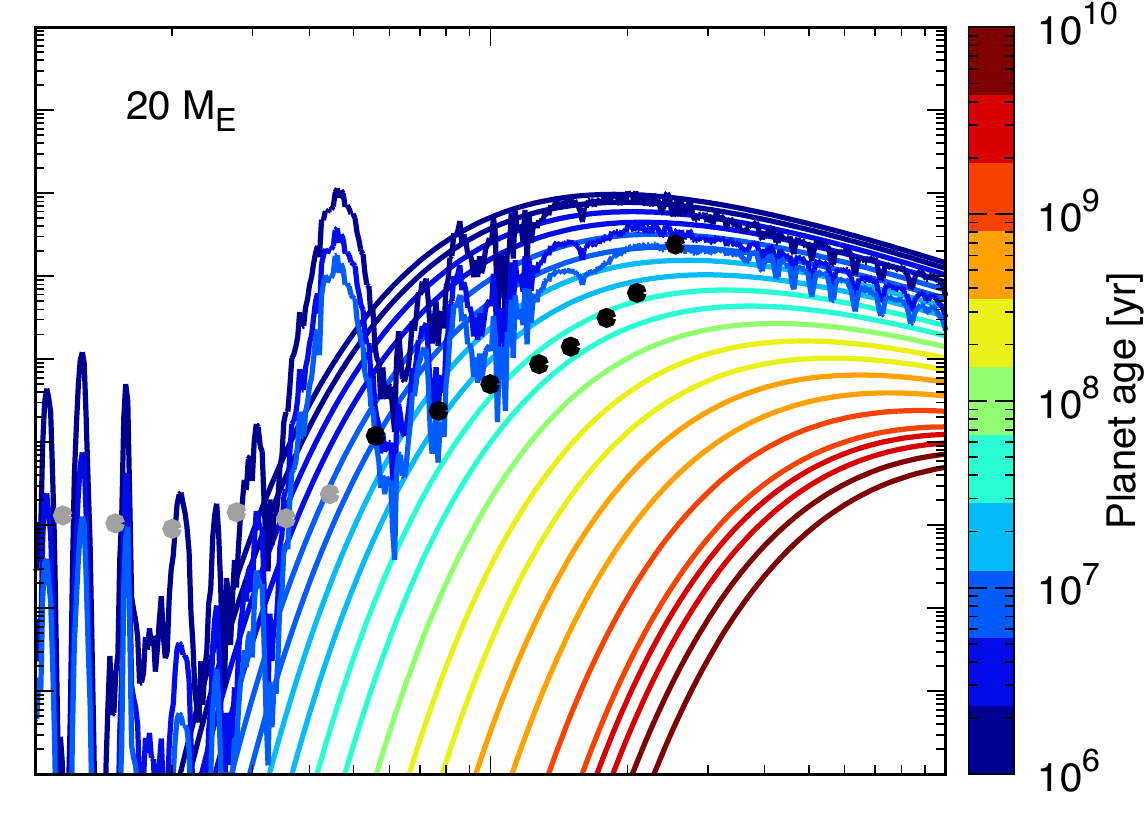}
\end{minipage}
\begin{minipage}[t]{0.50\textwidth}
\centering\includegraphics[width=1\linewidth]{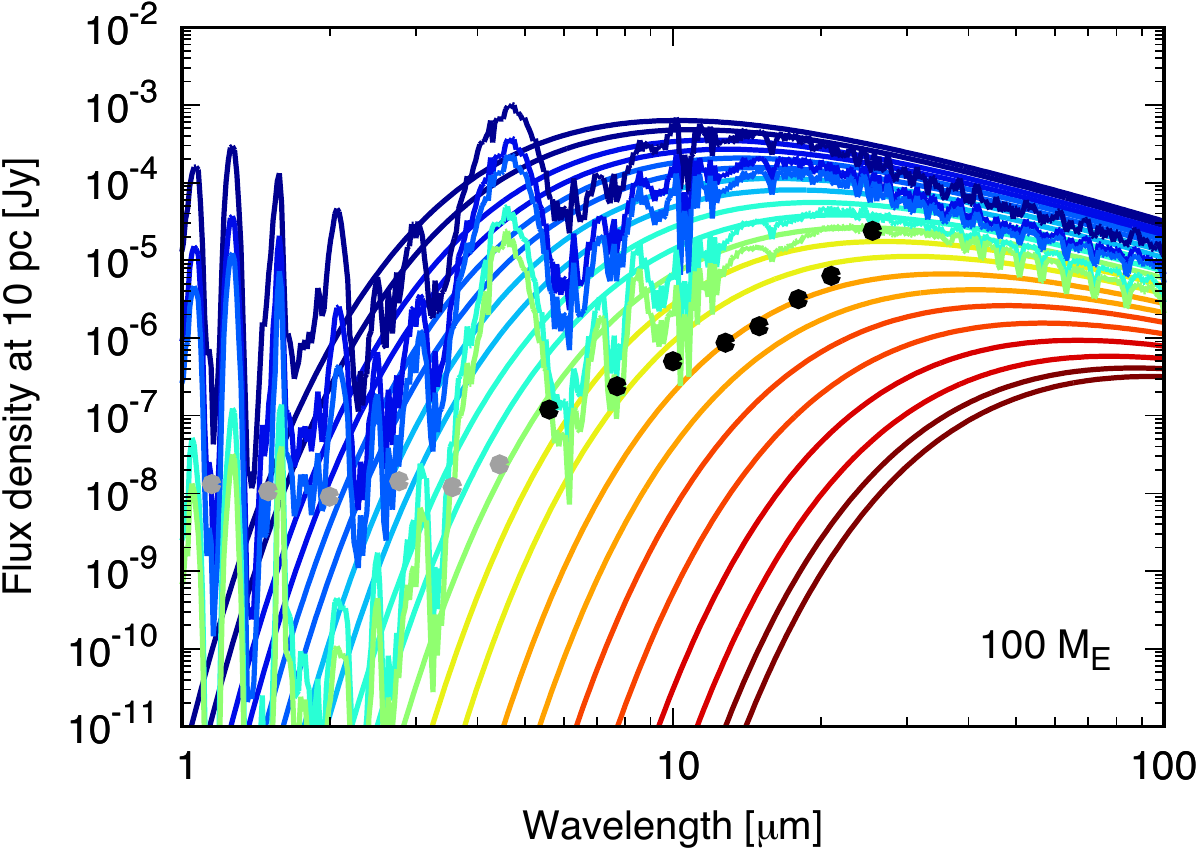}
\end{minipage}
\begin{minipage}[t]{0.48\textwidth}
\centering\includegraphics[width=1\linewidth]{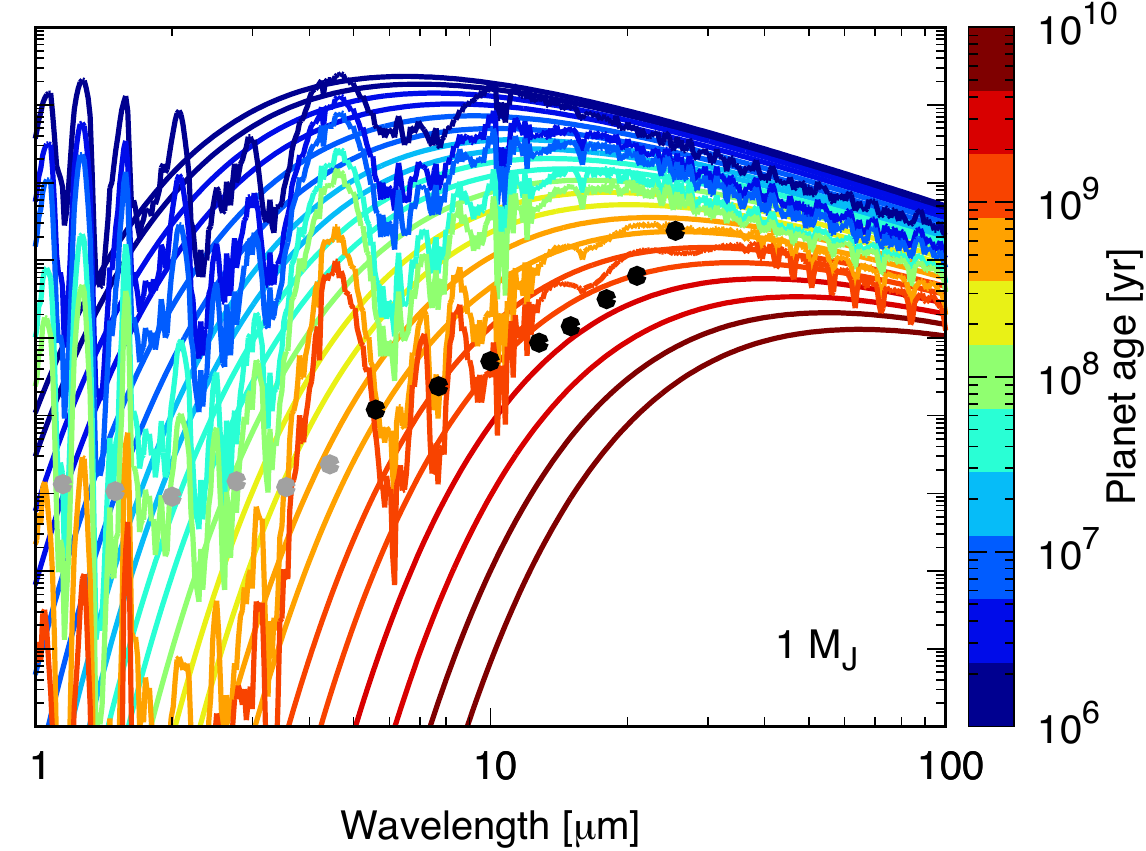}
\end{minipage}
\caption{Spectra for  cloud-free solar metallicity atmospheres from the \texttt{petitCODE} grid together with the theoretical blackbody for four planetary masses.  The age is given in color, the x- and y-axis are the same for all the figures.  {The temperature of the blackbodies corresponds to the temperature of the planet at the age given in color.} The grey dots show the background sensitivity limits for the  {JWST/NIRCam} filters included in this work, the black dots those for  {JWST/MIRI}.  {There are  2, 3, 5, and 7 spectra shown for the 5, {20}, 100, and 318 $\mearth$ planet. }} \label{fig:JWSTsenslim}
\end{center}
\end{figure*}

\begin{figure*}
\begin{center}
\includegraphics[width=\textwidth]{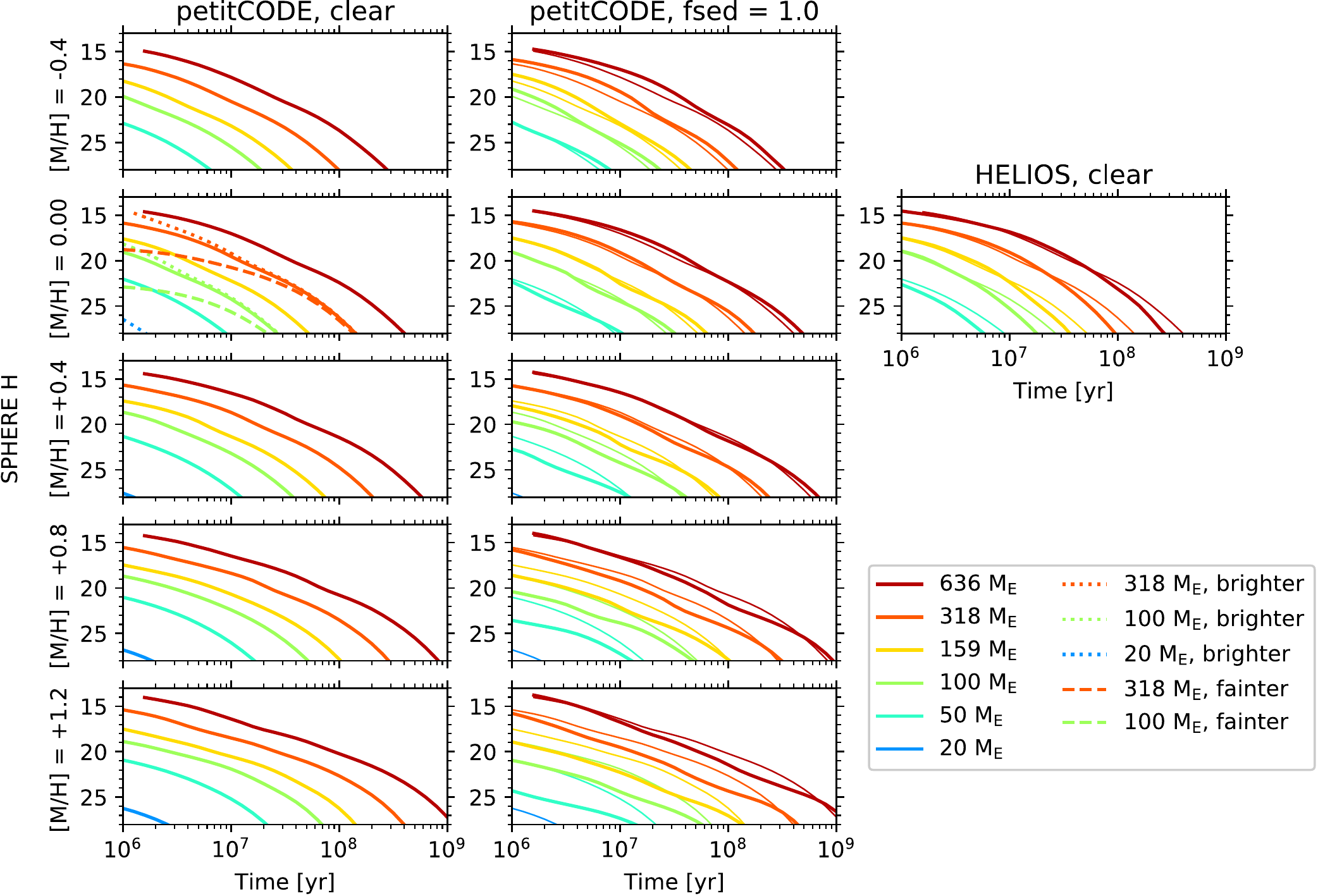}
\caption{Overview of all  evolutionary models that were calculated for this work. As an example, the  {SPHERE/IRDIS} H band magnitude is shown.   {Some of the lower mass planets might not be visible on the axis range chosen here. 
The \texttt{petitCODE} grid clear  magnitudes for the nominal post formation luminosity that are shown as thick solid lines in the first  {column}  are repeated in the other panels as thin lines for comparison between clear versus cloudy cases.
In the clear solar metallicity panel ({first panel, second row}), we also show the magnitudes corresponding to an evolution with a ten times higher post formation luminosity (brighter, {dotted}) and with a ten times lower post formation luminosity (fainter, {dashed}) than   the nominal case. This simulations were introduced in Sect. \ref{subsect:varyLpf}.
The magnitudes are only shown as long as the planets evolve in the atmospheric grid, and for the cloudy models as long as they are above 200 K. } } \label{fig:examplemag}
\end{center}
\end{figure*}

While{ {JWST}} will remain unchallenged in terms of sensitivity in the thermal infrared for many years to come, the currently operational extreme adaptive-optics (AO), high-contrast imaging near-infrared (NIR) instruments, such as  {VLT/SPHERE} or  {Gemini Planet Imager (GPI) \citep{beuzitfeldt2008,macintoshgraham2008} achieve better detection limits at small separations close to the diffraction limit, that is, in the contrast-limited regime. To put our models in the context of the exoplanet imaging surveys presently conducted,  we show in Fig. \ref{fig:examplemag} our model predictions for the  {SPHERE/IRDIS} H filter. The magnitude was calculated as described in Sect. \ref{sect:JWSTmagnitudesandfluxes}

For the solar metallicity and clear atmosphere, simulations with  {a ten times higher {and }lower post formation luminosity relative to the nominal scenario were calculated,} as introduced in Sect. \ref{subsect:varyLpf}.  {This was done for four planetary masses, namely for the} 5, 20, 100, and 318 $\mearth$  {mass planets.}
 Since the 5 $\mearth$ is too cool to still be on the atmosphere grid, ``fainter" magnitudes are not calculated for this mass. The range in post formation luminosity can lead up to a magnitude difference in the  {SPHERE/IRDIS} H band of about 10, 5, and 4 mag for the 20, 100, and 318 $\mearth$ (1 $\mj$) mass planets at 1 Myr. At 30 Myr, the different cooling paths for the 100 and 318 $\mearth$ are no longer distinguishable, which is in agreement with what was found by \cite{spiegelburrows2012} for masses of 1-2 $\mj$. 

Also note the mass-magnitude degeneracy in Fig. \ref{fig:examplemag}, similar to the mass-luminosity degeneracy already noted in Sect. \ref{subsect:varyLpf} and in the literature (e.g. \citealt{spiegelburrows2012}). For example, at 3 Myr a 20 mag object could correspond to a 159~$\mearth$ planet with nominal post formation luminosity or to a 318~$\mearth$ planet with a cold scenario post formation luminosity. From 10 Myr on, however, the lines representing the different masses and post formation luminosities no longer overlap, so that a rough mass estimate should be possible at this more likely observable age.

For comparison, the thin lines in the background show the evolution in a  clear \texttt{petitCODE} atmosphere (the same lines as in the first column).  {It is interesting to note that, depending on the metallicity, the clouds seem to have a  {dimming} (for high metallicities) or brightening (for low metallicites) effect. }

For further reference, in Fig. \ref{fig:isochrones}, isochrones for the   {VISIR} SiC magnitude are shown  for solar  {-metallicity} \texttt{HELIOS} (solid lines) and \texttt{petitCODE} (dashed lines) grid atmosphere for times starting at 1 Myr. Magnitudes are calculated as long as the planet is evolving  {within} the atmospheric grid.  {For the \texttt{HELIOS} grid, all the masses considered here evolve on the atmospheric grid from 1 to 100 Myr. On the other hand, for the \texttt{petitCODE} grid, all the masses evolve on the grid from 3 to 10 Myr. This difference comes from the different temperature coverage of the grids: the \texttt{petitCODE} grid goes from 150 K to 1000 K, whereas the \texttt{HELIOS} grid goes from 100 K up to 1200 K (see Fig. \ref{fig:loggTeff}).} At 1 Gyr,  masses from 50 $\mearth$ on are still in the \texttt{HELIOS} grid. The tables  {are} available   {for log(age/yr) = 6 to 10 in steps of 0.1 dex while the planet is}   evolving on the atmospheric grid.

Our evolutionary models together with the applied filter profiles are available  for a variety of space and ground-based filter systems: 
 {NACO} (J, H, K$_{\rm s}$, L$_{\rm p}$, M$_{\rm p}$), 
 {Cousin} (R, I), 
 Wide-field Infrared Survey Explorer (WISE1, WISE2, WISE3, WISE4), 
 {VISIR} (B87, SiC), and
 {SPHERE} (Y, J, H, K$_{\rm s}$, Y23, J23, H23, K12). 
The magnitudes were calculated as described in Sect. \ref{sect:JWSTmagnitudesandfluxes}. The available atmospheric models are shown in Fig. \ref{fig:examplemag}: \texttt{petitCODE}, clear, [Fe/H]=-0.4, 0.0, 0.4, 0.8, 1.2; \texttt{petitCODE}, $f_{\rm sed}$= {1.0}, [Fe/H]=-0.4, 0.0, 0.4, 0.8, 1.2; and \texttt{HELIOS}, clear, [Fe/H]=0.0. This gives a total of 11 atmospheric models and 35 filters, each for eight planetary masses, plus  four (three) masses in the clear solar enrichment \texttt{petitCODE} models with a higher (lower) post formation luminosity. There are three masses for the lower post formation luminosity cases because the fainter 5 $\mearth$ planet is too cold to be on the atmospheric grid.  An example of the tables is given in the appendix in Tables \ref{tab:magexamplefixedmass} and \ref{tab:magexamplefixedtime}.

\section{Summary and conclusion}\label{sect:conclusions}

In this study, we first presented (Sect. \ref{sect:model}) the extensions made to our evolution model that was originally designed for gas giants but now we applied it to core-dominated, low-mass planets. In Sect. \ref{sect:benchmarking} the updated model was then validated against the solar system gas and ice giants.  The results  {we find} are in agreement with the literature (e.g. \citealt{fortneyikoma2011,nettelmannhelled2013}).  {Comparing  {our simulations with} two independent cooling calculations (\citealt{baraffechabrier2008} and \citealt{lopezfortney2014}), we find a satisfactory agreement as well. }

We then turned to the main subject of the paper, which is the extension of classical cooling models like \cite{baraffechabrier2003} or \cite{burrowsmarley1997} to lower mass planets. For this, we  {computed initial conditions from formation models (Sect. \ref{sect:intialcond}.) and applied three different sets of atmosphere models (Sect.  \ref{sect:atmgrids})}.

The models used in this work include simple Eddington grey atmospheres, the \texttt{AMES-Cond} atmosphere, and recent \texttt{petitCODE} \citep{mollierevanboekel2017} and \texttt{HELIOS} \citep{malik17} atmospheres. Th{e clouds in the} \texttt{petitCODE} models are Na$_2$S and KCl.  {Cloud species that are important at lower temperatures  {such as} water are not (yet) included in the cloudy atmospheric models. The  surface boundary conditions and therefore the predicted magnitudes  {have the potential to be significantly a}ffected by water clouds that could form at low temperatures. When comparing the condensation curves of water with the atmospheric  {p--T} structures, one finds that planets with masses below  20 $\mearth$ could have water clouds from the start of their evolution in the outer atmosphere. For the more massive planets, water clouds could appear at about 30 (100) Myr for  {a} 50 (100) $\mearth$ planet.}

 {In this first publication we have  {not} considered different C/O ratios in the atmosphere. The C/O ratio gives  constraints on a planet's formation path (e.g. \cite{oebergmurrayclay2011,madhusudhanamin2014,mordasinivanboekel2016,laviebataille2016}), and to study the impact of a varied C/O on magnitudes will be an important next step. }

The \texttt{petitCODE} {,} \texttt{HELIOS,}  {and the \texttt{AMES-Cond}} grid{s} assume chemical equilibrium. Although we expect disequilibrium processes like turbulent mixing and photo-chemistry to have a non-negligible influence on the atmospheric composition \citep{mosesmarley2016}, we focus in this study on the treatment of clouds. We postpone the consideration of disequilibrium chemistry to future work.

The initial conditions together with the atmospheric grids and {p--T} structures were then used to calculate the evolution of a set of eight planetary masses { (Sects. \ref{subsect:solarclear}, \ref{subsect:nonsolarcloudy}, and \ref{subsect:varyLpf}). These} range from 5 $\mearth$ to 2 $\mj$.  {The atmospheric models as well as the post formation luminosities were varied.}  {Following this, magnitudes for 35 filters were calculated for  clear and cloudy atmospheres for various metallicities (Sects. \ref{sect:JWSTmagnitudesandfluxes} and \ref{predictionsforgroundbasedobservatories}). The magnitudes were calculated as described in Sect. \ref{sect:JWSTmagnitudesandfluxes}. The calculated magnitudes  together with the applied filter profiles are available at the CDS and at:  {\url{http://www.space.unibe.ch/research/research_groups/planets_in_time/numerical_data/index_eng.html}}. }  {We summarize the main findings in the following.} 

\begin{itemize}
\item When simulating hot and cold  {formation} scenarios (Sect. \ref{subsect:varyLpf}), we found that the spread in the initial (i.e. post formation) luminosity,  as suggested by formation models, has a greater influence on a planet's bolometric luminosity  than the   atmospheric  {model}, as already noted by \cite{spiegelburrows2012}.

\item In  the  {SPHERE/IRDIS} H filter  {for example}, the difference between a hot versus cold start can be up to 10 mag for the 20 $\mearth$ planet  (Sect. \ref{predictionsforgroundbasedobservatories}). We also note that there is a mass-magnitude  {indeterminacy with initial conditions} in certain filter bands at young ages, similar to what was found by \cite{spiegelburrows2012}. However, after 10 Myr a rough mass estimate should be possible (Fig. \ref{fig:examplemag}).

\item  {The atmospheres have a large impact on magnitudes in specific filter bands.} For example, in JWST/NIRCam F356W the magnitude calculated with an \texttt{AMES-Cond} atmosphere is up to  {2.9} mag  {  {brighter} at 100 Myr for the 100 $\mearth$ planet than with a clear solar \texttt{petitCODE} atmosphere (Sect. \ref{sect:JWSTmagnitudesandfluxes}). This can be explained  by the differences in  methane line lists.  {We assume that the newer line list ExoMol yields  more accurate results because it should be much more complete, especially at higher temperatures.} }

\item  {Comparing the sensitivity limits of JWST with the emergent flux from the planet, w}e find that a 20, 100, and 318 $\mearth$ (1 $\mj$) mass planet should be detectable with  {JWST/MIRI}   {in the background-limited regime} until 10, 100,  {and} 1000 Myr after formation, respectively (see  {Sect.  \ref{sect:JWSTmagnitudesandfluxes} and} Fig. \ref{fig:JWSTsenslim}).

\item  {Filters at  wavelengths around 4.7 $ {\mu m}$ seem to be favourable for exoplanet detection, as there is a prominent  {window in} the water and methane opacities (see Fig. \ref{fig:JWSTsenslim}) that enhances the emergent flux relative to a blackbody of the same temperature.}

\end{itemize}

\begin{figure}
\begin{center}
\centering\includegraphics[width=1\columnwidth]{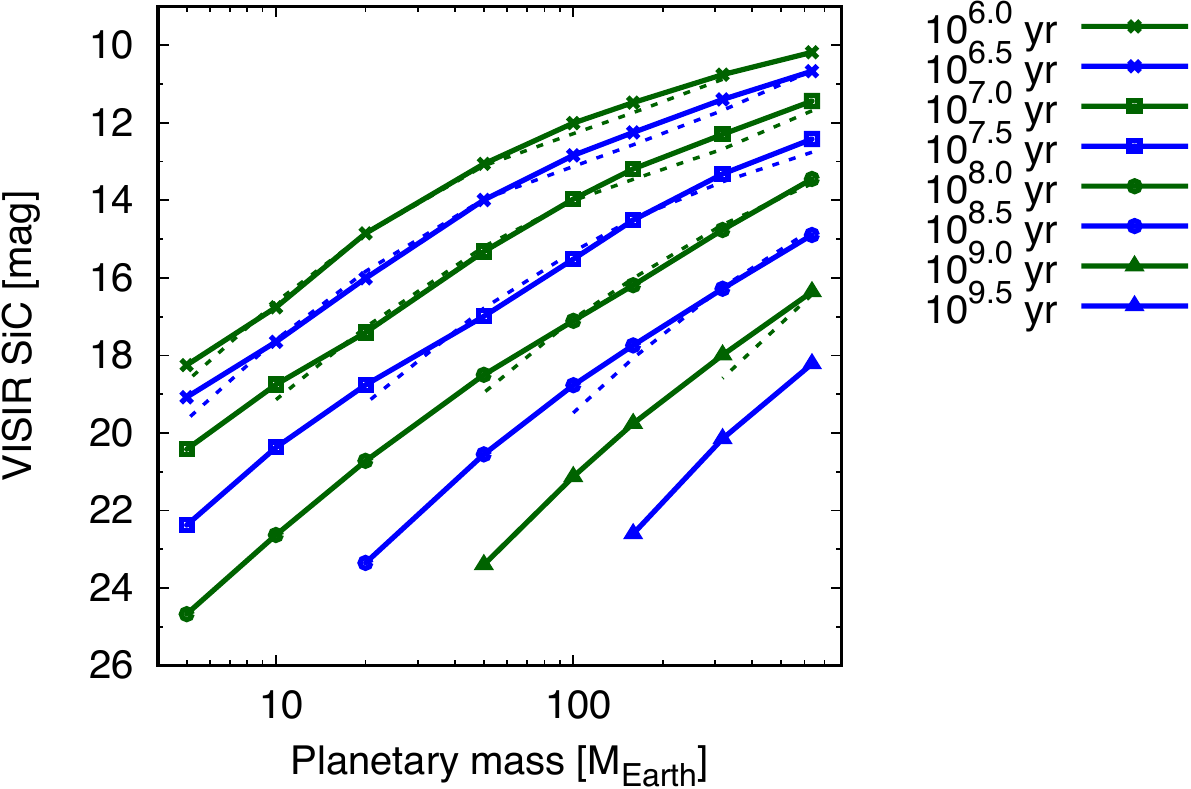}
\caption{Isochrones in the VISIR SiC filter for the \texttt{HELIOS}  {(solid)} and \texttt{petitCODE} {(dashed)} solar metallicity grid. The different lines represent times in years, where the dashed isochrones correspond to the same times as the adjacent solid isochrones. 
Masses from 5 to 636 $\mearth$ are covered. The tables  are available  {for log(age/yr) = 6 to 10 in steps of 0.1 dex while the planets are}  evolving  {within} the grid  {of the atmospheric models}.} \label{fig:isochrones}
\end{center}
\end{figure}

 {While it seems unlikely that (sub-)Jupiter mass planets are within the reach of current instruments like SPHERE or GPI for a  {large} sample of targets, there are a few special cases where our models are applicable. A good example is the nearest pre-main sequence star AP Col \citep{riedelmurphy2011}, where ground-based, high-contrast imaging can actually probe for young Jupiter analogs \citep{quanzcrepp2012}.}

 {More importantly, however, our models will be of relevance when planning  future observations with JWST and/or the next generation of 30-40 m ground-based telescopes and their exoplanet imaging instruments. {While JWST  will provide unprecedented sensitivity} at mid-infrared wavelength, future ground-based facilities will remain unchallenged in terms of spatial resolution for many years to come and provide complementary discovery space in comparison to { {JWST}}. Most notably  {METIS} \citep{brandlagocs2016} and the Planetary Camera and Spectrograph {(PCS)}  \citep[formally called  Exoplanet Imaging Camera and Spectrograph {(EPICS)};][]{kasperbeuzit2010} for the ESO  {E-ELT} will be equipped with high-contrast imaging cameras and search for exoplanets around the nearest and nearby young stars. This opens up the exciting perspective of detecting young and forming (e.g. \citealt{vanboekelhenning2017}) low-mass planets and old  gas giants instead of young giant planets only (e.g. \citealt{bowler2016}) and will put important new constraints on theoretical models of planet formation and evolution. For these future observations, this paper provides a theoretical framework for interpretation.}

\begin{acknowledgements}
We thank Eric Lopez for providing the numerical data used in Fig.~\ref{fig:lopezfortney}, and Isabelle Baraffe for informative correspondence and the data  {shown in} Fig.~\ref{fig:GJ436benergies}. We also thank them for interesting discussions. 
E.F.L thanks Marcus Wernberger Jonsson for a thorough Python introduction. 

E.F.L., C.M.,\ and G.-D.M.\ acknowledge the support from the Swiss National Science Foundation under grant BSSGI0\_155816 ``PlanetsInTime''. Parts of this work have been carried out within the frame of the National Center for Competence in Research PlanetS supported by the SNSF. 

G.-D.M. also acknowledges the support of the DFG priority programme SPP 1992 ``Exploring the Diversity of Extrasolar Planets'' (KU 2849/7-1).

We thank the referee for helpful feedback that greatly improved the manuscript.
\end{acknowledgements}

\bibliographystyle{aa} 
\bibliography{literature.bib}

\begin{appendix}

\section{First-order correction for the temperature}\label{app:coretemp}

In our previous works, the radius of a solid core (using the astrophysical, not geophysical nomenclature) is determined by numerically solving  the internal structure equation of mass conservation and hydrostatic equilibrium assuming a differentiated interior consisting of iron, silicate, and ice. As equation of state, the modified polytropic equation of state (EoS) of \citet{seagerkuchner2007} is used. This simple equation of state yields the density as a function of pressure for a wide pressure range including the degenerate limit, but neglects the change of the density with temperature. At a given moment in time, this has only little effect on the resulting radii \citep{valenciasasselov2007,grassetschneider2009}, especially for the more massive super-Earth/Neptunian and Saturnian planets considered here \citep{seagerkuchner2007}, but for the long-term cooling we need to consider it by adding a first order temperature correction of the mean density. 

From the law of thermal expansion, we estimate the variation of the mean core density $\rho$ with temperature as 
\beq
\rho=\frac{\rho_{0}}{1+\alpha (T_{\rm ceb}-T_{\rm ref})}
,\eeq
where $\rho_{0}$ is the mean density yielded by solving the structure equations with the EoS of \citet{seagerkuchner2007}. It is a function of the core mass $M_{\rm core}$, the pressure $P_{\rm ceb}$ exerted by the gaseous envelope  at the core-envelope boundary, and the ice mass fraction $\fice$ (see \citealt{mordasinialibert2012c}). The iron-silicate ratio is fixed at 2:1 in mass{,  inspired by Earth's composition and condensation models of solar-composition gas (e.g. \citealt{santosadibekyan2015})}. The other quantities are the thermal expansion coefficient $\alpha$, the reference temperature $T_{\rm ref}$ , which we set to 300 K, and the temperature at the core-envelope boundary $T_{\rm ceb}$.

The thermal expansion coefficient is found by using the ANEOS equation of state \citep{thompson1990} and the Maxwell relations as 
\beq
\alpha=\frac{1}{\rho}\left(\frac{\partial P}{\partial T}\right)_{\rho}\left(\frac{\partial \rho}{\partial P}\right)_{T}
,\eeq
where the two derivatives are an output of the EoS. We considered the ANEOS data for water ice and dunite  at higher pressures, and measured values at lower ones \citep{poirier2000} for a set of  pressure-temperature pairs representative of the Earth's mantle, the Earth's centre, Jupiter's centre, and the centre of a 10 $\mj$ super-Jupiter.  Figure \ref{fig:alphaANEOS} shows the $\alpha$ of these two materials, while the data is given in Table \ref{tab:alphaANEOS}. The plot shows that the measured  (horizontal part) and ANEOS data  can be approximated with a broken power law, so that we write
\beq 
\alpha=\min\left(a \times\left(\frac{P_{\rm ref}}{P_{\rm ceb}}\right)^{b}, \alpha_{0}\right).
\eeq
The parameters are $a=4\times 10^{-6}$ 1/K, $b=0.45$, and $\alpha_{0}=1\times 10^{-4}$~1/K for ice, and  $a=2\times 10^{-6}$ 1/K, $b=0.5$, and $\alpha_{0}=1.5\times 10^{-5}$ 1/K for dunite. The latter value is chosen to be somewhat lower than the typically used value of $2.0 \times 10^{-5}$ 1/K since we use it for the entire rocky core consisting of silicate and iron, which has a lower $\alpha$ \citep{poirier2000}.  The reference pressure $P_{\rm ref}$ is $10^{13}$ dyn/cm$^{2}$ in both cases. 

It is clear that our description represents only a simple approximation of the actual physical process, as it assumes for example a uniform expansion coefficient for the entire core. The comparisons with more complex models (see also \citealt{thomasmadhusudhan2016}) and observational data presented below nevertheless indicate a relatively good match. A simple model as a first step also seems appropriate given our goal of studying the luminosities of young extrasolar planets, and not for example the detailed internal structure of solar system planets. In future work, we will still include the more accurate description of \citet{alibert2014}. 

\begin{figure}
\begin{center}
\includegraphics[width=\columnwidth]{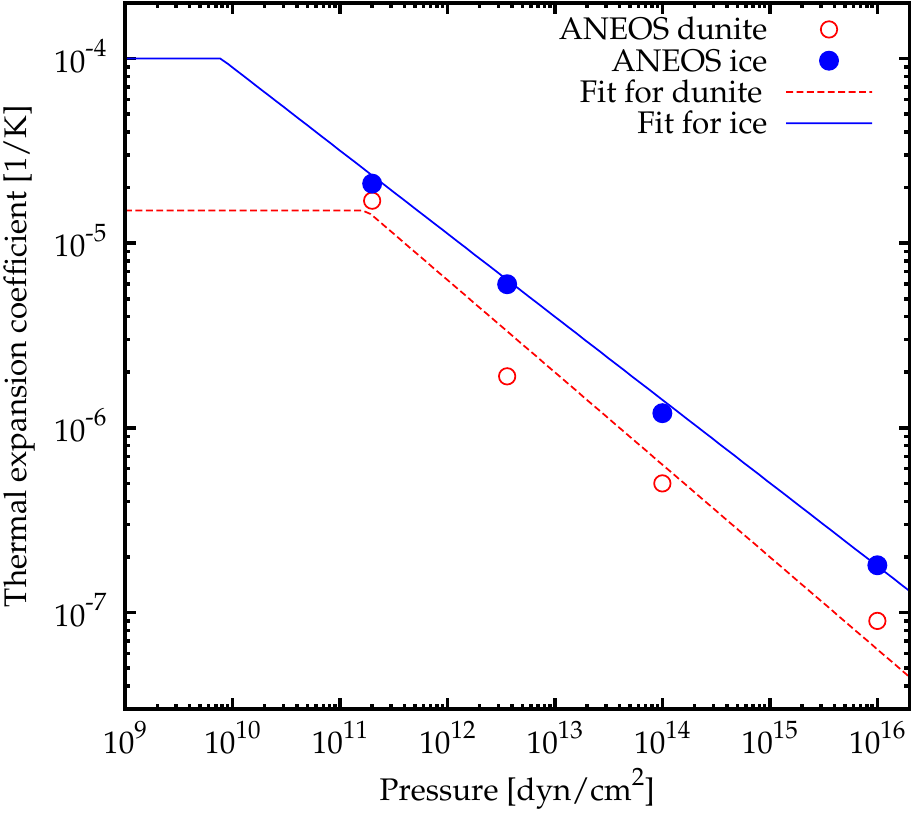}
\caption{Thermal expansion coefficients of water and dunite as predicted by ANEOS (filled and empty circles), and the broken power law approximation used in our model (solid and dashed lines).}\label{fig:alphaANEOS}
\end{center}
\end{figure}

\begin{table}
\caption{Thermal expansion coefficient of ice and dunite predicted by the ANEOS equation of state.}\label{tab:alphaANEOS}
\begin{center}
\begin{tabular}{lcccc}
\hline\hline
Location  &     $P$ [dyn/cm$^{2}$]       &      $T$ [K] &       $\alpha_{\rm ice}$   &        $\alpha_{\rm dunite}$    \\ \hline
Earth mantle & 2$\times 10^{11}$        & 2000 & 2.1$\times10^{-5}$ &  1.7$\times10^{-5}$\\
Earth centre  & 3.6$\times10^{12}$      & 5700 & 6$\times10^{-6}$ &  1.9$\times10^{-6}$  \\
Jupiter centre  &       $\sim10^{14}$   & $\sim$10$^{4}$ & 1.2$\times10^{-6}$ &  5$\times10^{-7}$\\
10 $\mj$ centre  &       $\sim10^{16}$  & $\sim$10$^{5}$ & 1.8$\times10^{-7}$ &  9$\times10^{-8}$\\  \hline
\end{tabular}
\end{center}
\label{default}
\end{table}

\begin{table}
\caption{Example of the table format that can be found at the CDS and at:  {\url{http://www.space.unibe.ch/research/research_groups/planets_in_time/numerical_data/index_eng.html}}, where the filename (BEX\_evol\_mags\_-2\_MH\_0.00\_ME\_050.dat) gives the information that it is for a cloud-free solar atmosphere in the \texttt{petitCODE} grid. This format shows the evolution of one planetary mass as a function of time.}
\begin{tiny}
\begin{center}
\begin{tabular}{llllllll}
\hline\hline
\#atmflag =           -2, &[M/H] = 0.00, &clear\\
 {\#Conversion} & {factor}  & {L\_{Jupiter} = }& {8.710e-10 x L\_{Sun}}&  {with}   {L\_{Sun} = } &  {3.846e+33 erg/s}\\
\#1: {log(Age/yr)} & 2:Mass/Mearth & 3:Radius/Rjupiter & 4:Luminosity/Ljupiter & 5:Teff/K &6:logg/cgs \\
\hline
6.0    &  50.0    & 1.300   &  397.240   &  383.446  &2.363 \\
6.1   &      50.0   &     1.274   &   346.560   &   374.332 &2.381 \\
6.2   &      50.0   &     1.247   &   299.831   &   364.946&2.400\\
...&...&...&...&...&...\\
7:NACOJ & 8:NACOH & 9:NACOKs &10:NACOLp & 11:NACOMp  &12:CousinsR \\
\hline
20.74    & 21.63    & 22.16    & 17.17    & 14.15&29.53\\
21.02   &     21.92   &     22.58   &     17.35   &     14.29&29.67\\
21.33   &     22.24   &     23.03   &     17.55   &     14.45&29.82 \\
...&...&...&...&...&...\\
13:CousinsI   & 14:WISE1   & 15:WISE2   & 16:WISE3   & 17:WISE4 &18:F115W  \\
\hline
25.77    & 19.60    & 14.50    & 13.19    & 11.62&20.76 \\
25.93   &     19.81   &     14.64   &     13.32   &     11.71&21.05 \\
26.10   &     20.05   &     14.80   &     13.47   &     11.81&21.37 \\
...&...&...&...&...&...\\
19:F150W   & 20:F200W   & 21:F277W   & 22:F356W   & 23:F444W  &24:F560W \\ 
\hline
21.97    & 22.66    & 21.61    & 18.45    & 14.81&15.81 \\
22.26   &     23.07   &     21.92   &     18.64   &     14.96 &15.99\\
22.57   &     23.52   &     22.26   &     18.85   &     15.11 &16.18\\
...&...&...&...&...&...\\
25:F770W   & 26:F1000W   & 27:F1280W   & 28:F1500W   & 29:F1800W  &30:F2100W \\
\hline
15.27    & 13.78    & 13.00    & 12.18    & 11.77&11.67\\
15.43   &     13.96   &     13.15   &     12.30   &     11.88 &11.76\\
15.60   &     14.15   &     13.31   &     12.43   &     12.00&11.86\\
...&...&...&...&...&...\\
31:F2550W   & 32:VISIRB87   & 33:VISIRSiC   & 34:SPHEREY   & 35:SPHEREJ &36:SPHEREH \\
\hline
11.60    & 14.04    & 13.11    & 21.42       &    20.74& 21.54\\
11.68   &     14.20   &     13.26   &  21.74   &    21.02  &21.83 \\
11.78   &     14.37   &     13.42   &    22.08   &   21.33 &22.14 \\
...&...&...&...&...&...\\
37:SPHEREKs   &  {38:SPHEREJ2} &   {39:SPHEREJ3} &   {40:SPHEREH2} &   {41:SPHEREH3} &   {42:SPHEREK1} \\
\hline
 22.03       &       {25.80}       &    {19.15} &  {19.96}       &  {25.83} &    {21.24}\\
22.43   &     {26.11}   &      {19.43}   &      {20.23}   &    {26.12} &  {21.65}\\
22.87   &    {26.45}   &    {19.73}   &    {20.52}   &    {26.44} &  {22.09}\\
...&...&...&...&...\\
 {SPHEREK2}  \\
\hline
 {25.92}\\
 {26.29}\\
 {26.69}\\
...\\
\hline
\end{tabular}
\end{center}
\end{tiny}
\label{tab:magexamplefixedmass}
\end{table}

\begin{table*}
\caption{Example of the second table format that can be found at the CDS and at:  {\url{http://www.space.unibe.ch/research/research_groups/planets_in_time/numerical_data/index_eng.html}}, showing the magnitudes at  certain  {times} for all the planetary masses considered here if they are still in the atmosphere grid at that time. This example table is for a non-solar metallicity of [M/H]=0.8 and a cloudy atmosphere with $f_{\rm sed}$= {1.0} in the \texttt{petitCODE} grid (filename: BEX\_evol\_mags\_-2\_MH\_0.80\_fsed\_1.00.dat).}
\begin{tiny}
\begin{center}
\begin{tabular}{lllllllll}
\hline\hline
\#atmflag = -2,& [M/H] = 0.80, &fsed= {1}.00\\
 {\#Conversion} & {factor}  & {L\_{Jupiter} = }& {8.710e-10 x L\_{Sun}} &  {with} &  {L\_{Sun} = } &  {3.846e+33 erg/s}\\
\#1: {log(Age/yr)} & 2:Mass/Mearth & 3:Radius/Rjupiter & 4:Luminosity/Ljupiter & 5:Teff/K &6:logg/cgs &7:NACOJ\\
\hline
 {6.0}   &    {10.0}   &    {1.075}   &  {25.805}       &  {212.903}   &  {1.830}  &   {44.10}\\
 {6.0}   &    {20.0}   &    {1.536}   &  {97.297}       &  {245.974}   &  {1.805}   &   {39.58} \\
 {6.0}  &    {50.0}   &     {1.487}   &  {389.188}      &  {356.610}   &  {2.246}   &  {31.42}\\
 {6.0}  &    {100.0}  &     {1.357}   &  {1110.963}   &  {485.242}   &  {2.627}  &   {26.55}\\
 {6.0}  &    {159.0}   &    {1.297}   &  {2191.610}   &  {588.325}   &  {2.868}  &  {23.32} \\
 {6.0}  &    {318.0}   &    {1.285}   &  {6642.524}   &  {779.730}   &  {3.177}  &   {18.29} \\
 {6.1}   &    {10.0}   &    {1.046}   &  {23.487}       &  {210.758}   &  {1.853}   &  {44.31} \\
 {6.1}   &    {20.0}   &    {1.513}   &  {87.422}       &  {243.416}   &  {1.834}   &  {39.97}  \\
...&...&...&...&...&...&...\\
  8:NACOH & 9:NACOKs &10:NACOLp & 11:NACOMp  &12:CousinsR  & 13:CousinsI   & 14:WISE1 \\
\hline
   {39.05}   &    {33.54}   &  {24.04}          &  {19.89}   &  {64.15}   & {59.40}   &    {26.96}  \\
      {34.80}   &    {29.43}   &  {21.30}       &  {17.88}   &  {57.75} &   {53.26}   &    {23.80}  \\
     {27.61}   &     {23.66}   &  {17.86}       &  {15.46}   &  {45.63}   &  {41.59}  &    { 19.73}\\
      {23.37}  &     {20.42}   &  {15.92}   &  {14.05}   &  {38.24}    &  {34.58}  &    {17.41} \\
      {20.98}   &    {18.60}   &  {14.78}   &  {13.22}   &  {34.17}    &  {30.37}  &    {16.08 } \\
      {17.31}   &    {15.88}   &  {13.09}   &  {12.13}   &  {28.57}    &  {24.41}  &    {14.15} \\
      {39.28}   &    {33.86}   &  {24.23}       &  {20.01}   &  {64.39}   &  {59.64}   &    {27.19}\\
   {35.03}   &    {29.64}   &  {21.47}          &  {18.00}   &  {58.08}    &  {53.56}   &    {24.00}  \\
...&...&...&...&...&...\\
  15:WISE2   & 16:WISE3   & 17:WISE4 &18:F115W  & 19:F150W   & 20:F200W   & 21:F277W\\
\hline
      {20.29}   &  {15.91}      &  {13.07}   &  {44.78} &    {39.50}   &    {34.11}   &    {31.24} \\
      {18.23}   &  {14.24}      &  {11.89}   &  {40.34}    & {35.40}   &    {29.98}   &    {27.13} \\
        {15.72}   &  {12.65}    &  {11.09}   &  {32.30}  &  {28.31 }  &    {24.14}   &     {21.35}  \\
    {14.26}   &  {11.75}   &  {10.59}   &  {27.43}    &  {24.08}  &    {20.80}   &    {18.28}\\
      {13.40}   &  {11.23}   &  {10.26}   &  {23.94}    & {21.69}  &    {18.92}   &    { 16.69}\\
      {12.27}   &  {10.52}   &  {9.87}   &  {18.52}    &  {18.02}   &    {16.21}   &    {14.68} \\
       {20.41}   &  {16.03}     &  {13.16}   &  {44.99 } &   {39.72}   &    { 34.43}   &    {31.51}   \\
    {18.35}   &  {14.37}        &  {11.99}   &  {40.55}    &  {35.61}   &    {30.19}   &    {27.43}   \\
...&...&...&...&...&...&...\\
   22:F356W   & 23:F444W  &24:F560W& 25:F770W   & 26:F1000W   & 27:F1280W   & 28:F1500W\\ 
\hline
    {25.52}     &  {20.73}   &  {21.14}  &  {18.92}   &    {18.33}   &    {16.72}   &  {14.35}  \\
    {22.58}     &  {18.65}   &  {18.85}    &  {16.95}   &    {15.72}   &    {1.536}   &  {97.297} \\
    {18.90}     &  {16.07}   &  {15.82}    &  {14.61}  &    {13.15}   &     {12.49}   &  {11.71}\\
     {16.83}   &  {14.55}   &  {13.92}    &  {13.20}  &    {11.91}  &     {11.62}   &  {11.06} \\
     {15.61}   &  {13.65}   &  {12.94}    &  {12.46}  &    {11.31}   &    {11.10}   &  {10.63} \\
    {13.78}     &  {12.41}   &  {11.78}    &  {11.46}  &    {10.51}   &    {10.29}   &  {10.15}  \\
  {25.73 }      &  {20.85}   &  {21.28}    &  {19.06}   &    {18.53}   &    {16.89}   &  {14.46} \\
 {22.77}        &  {18.78}   &  {19.01}    &  {17.11}   &    {15.92}   &    {14.65}   &  {12.97} \\
...&...&...&...&...&...\\
    29:F1800W  &30:F2100W &31:F2550W   & 32:VISIRB87   & 33:VISIRSiC   & 34:SPHEREY   & 35:SPHEREJ \\
\hline
          {13.66}   &  {13.24}  &   {12.86}   &    {17.83}   &    {16.49}   &  {46.48}    &  {43.33} \\
          {245.974}   &  {1.805}&    {11.77}   &    {15.72}   &    { 14.45}   &  { 42.48}           &  {39.03}  \\
          {11.31}   &  {11.16}    &  {11.04}  &    {13.42}   &     {12.53}   &  {34.57}    &  {31.31}  \\
    {10.76}   &  {10.64}    &  {10.53}  &    {12.20}  &     {11.60}   &  {29.28}   &  {26.51} \\
    {10.42}   &  {10.31}    & {10.21}  &    {11.61}   &    {11.06}   &  {25.00}   &  {23.33} \\
   {9.98}   &  {9.90}    &  {9.82}  &    {10.85}   &    {10.28}   &  {18.97}   &  {18.29} \\
          {13.76}   &  {13.33} &   {12.95}   &    {17.99}   &    {16.64}   &  {46.72}    &  {43.54}  \\
          {12.38}   &  {12.10}   &  {11.86}   &    {15.88}   &    {14.61}   &  {42.76}    &  {39.28}  \\
...&...&...&...&...&...&...\\
36:SPHEREH &37:SPHEREKs   &  {38:SPHEREJ2} &   {39:SPHEREJ3} &   {40:SPHEREH2} &   {41:SPHEREH3} &   {42:SPHEREK1}\\
\hline
    {39.01}  &    {33.27}   &    {42.48}   &    {42.06}   &  {37.52}    &  {40.08}   &  {32.42}  \\
    {34.84}    &  {29.20}   &    {38.62}   &    {37.73}   &  {33.52}    &  {35.54 }   &  {28.39} \\
  {27.72}    &  {23.56}  &    {31.55}   &     {30.35}   &  {26.78 }     &  {28.27}   &  {22.89}\\
  {23.49 }    &  {20.41}  &    {27.13}  &     {25.75}   &  {22.87}   &  {23.89}   &  {19.86} \\
    {21.10}    &  {18.60}  &    {24.28}   &    {22.47}   &  {20.65 }   &  {21.36}   &  {18.13}  \\
    {17.36}    &  {15.80}  &    {19.32}   &    {17.35}   &  {17.02}   &  {17.20}   &  {15.44} \\
    {39.23}    & {33.59}   &    {42.72}   &    {42.28}   &  {37.73}     &  {40.35}   &  {32.73}\\
   {35.06}    &  {29.40}   &    {38.89}   &    {37.97}   &  {33.72}     &  {35.79}   &  {28.58}  \\
...&...&...&...&...&...&...\\
 {SPHEREK2}  \\
\hline
 {33.00}\\
 {26.29}\\
 {25.79}\\
 {21.90}\\
 {19.72}\\
 {16.44}\\
 {38.64}\\
 {33.35}\\
...\\
\hline
\end{tabular}
\end{center}
\end{tiny}
\label{tab:magexamplefixedtime}
\end{table*}

\end{appendix}

\end{document}